\pdfoutput=1
      [1999/12/01 v1.4c Il Nuovo Cimento]
\documentclass{cimento}

\usepackage[latin1]{inputenc}
\usepackage[english]{babel}
\usepackage[final]{graphicx}
\usepackage{amsmath,amsfonts,amssymb}
\usepackage{latexsym}

\issue{31}{4}{2008}
\title{Very High Energy Gamma Astrophysics}
\author{A.~De Angelis\from{ins:difa}\from{ins:inaf}\from{ins:infn}\from{ins:ist},
O.~Mansutti\from{ins:difa}\from{ins:infn},
\atque\ETC\
M.~Persic\from{ins:inaf}\from{ins:infn}
}
\shortauthor{A.~De Angelis, O.~Mansutti, \atque M.~Persic}
\instlist{%
  \inst{ins:difa} Dipartimento di Fisica, Universit\`a di Udine, via delle Scienze 208, I-33100 Udine, Italy
  \inst{ins:inaf}INAF, Sezione di Trieste, via G.B.\ Tiepolo 11, I-34143 Trieste, Italy
  \inst{ins:infn}INFN, Sezione di Trieste and Gruppo Collegato di Udine, Italy
  \inst{ins:ist}Instituto Superior T\'ecnico, Av.\ Rovisco Pais, 1049-001 Lisboa, Portugal%
}

\PACSes{\PACSit{95.55.Vj}{Neutrino, muon, pion, and other elementary particle detectors; cosmic ray detectors}
\PACSit{95.85.Pw}{Gamma-ray}
\PACSit{96.50.S}{Cosmic rays}
}

\begin{document}

\maketitle

\begin{abstract}
High-energy photons are a powerful probe for astrophysics and for fundamental physics under extreme conditions.
During the recent years, our knowledge of the most violent phenomena in the Universe has impressively progressed thanks to the advent of new detectors for high-energy $\gamma$-rays.
Observation of $\gamma$-rays gives an exciting view of the high-energy universe 
thanks to satellite-based telescopes (AGILE, GLAST) and to ground-based detectors
like the Cherenkov telescopes (H.E.S.S.\ and MAGIC in particular), which recently discovered more than 60 new very-high-energy sources.
The progress achieved with the last generation of Cherenkov telescopes is comparable to the one drawn by EGRET with respect to the previous $\gamma$-ray satellite detectors.
This article reviews the present status of high-energy gamma astrophysics, with emphasis on the recent results and on the experimental developments.
\end{abstract}

\tableofcontents

\section{Introduction}

A flow of high-energy particles reaches the Earth.
About one century ago, two pioneering works by Victor Hess~\cite{vhess} (who for this was awarded the Physics Nobel Prize in 1936), and by Domenico Pacini~\cite{pacini} (who unfortunately died before 1936: his fundamental work is less known, although it was performed at the same time as Hess', independently, and with different techniques) proved that such particles were of extraterrestrial origin: they were thus called {\em cosmic rays}~\cite{rossi}.
Since their energies far exceed the temperatures ordinarily encountered in astronomical objects, cosmic rays are messengers of the non-thermal universe.

Excluding neutrinos, cosmic rays mainly consist of charged particles, such as protons ($\sim$90\%), helium nuclei ($<$10\%), ionized heavier elements ($<$1\%), and electrons ($<$1\%), while only 0.1\%--1\% of the total radiation consists of photons with energy $>$1\,MeV; for historical reasons these are called \mbox{{\em $\gamma$-rays}}.

The energy of cosmic rays covers more than 10 orders of magnitude, from tens of MeV up to 10$^{20}$ eV and higher%
\footnote{%
	A theoretical upper limit on the energy of cosmic rays from distant sources exists.
	This limit was computed in 1966 by Greisen, Kuzmin and Zatsepin~\cite{gzk}, and it is called today the GZK cutoff.
	Protons with energies above a threshold of about 10$^{20}$ eV suffer a resonant interaction with the cosmic microwave background photons to produce pions ($p$+$\gamma$$\rightarrow$$\Delta$$\rightarrow$$N$$+$$\pi$).
	This continues until their energy falls below the production threshold.
	Because of the mean path associated with the interaction, extragalactic cosmic rays from distances larger than 50 Mpc from the Earth and with energies greater than this threshold energy should not be observed on Earth, and there are no known sources within this distance that could produce them.
	The Auger experiment recently confirmed the existence of the GZK cutoff~\cite{auger0}.%
}.
The dependence of the flux on the energy $E$ of the particle can be approximated by a power law
$$
\frac{dN}{dE} \propto E^{-\alpha} \, ,
$$
where the {\em spectral index} $\alpha$ has typical values between 2.5 and 3.
The energy dependence can be explained in first approximation due to shock acceleration mechanisms~\cite{fermiacc}.

\begin{figure}
\centering
\includegraphics[width=.75\textwidth]{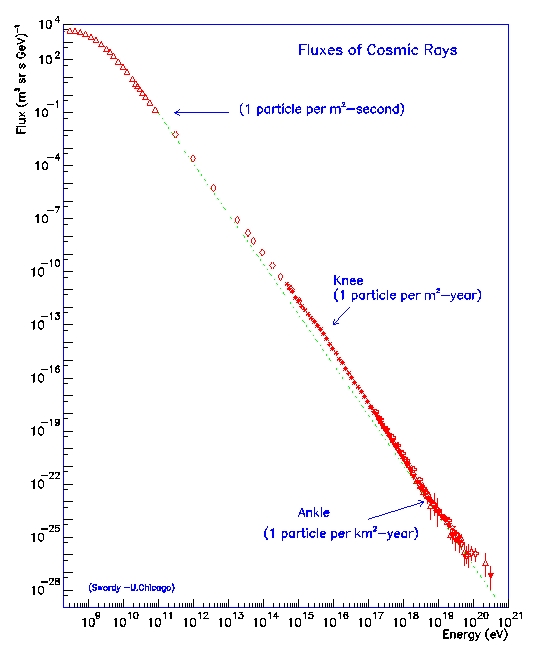}
\caption{\label{fig:cr} 
Energy spectrum of cosmic rays.}
\end{figure}

Fig.~\ref{fig:cr} shows the energy spectrum for cosmic rays.
The region around $10^{15.5}$ eV is referred as the ``knee'', for its spectral steepening (from a spectral index of about 2.7 to about 3), while the region around $10^{18.5}$ eV, where a spectral flattening occurs, is called the ``ankle'' of the spectrum.
It is generally believed that cosmic rays below the knee have a galactic origin and that they have been confined inside our Galaxy for at least $10^7$ years (this can be expected from the value of the magnetic field in the galaxy, which is of the order of 1 $\mu$G)~\cite{magfield}.
The particles above $10^{17}$ eV are believed to be mostly of extragalactic origin, since the galactic magnetic field is not able to trap them in our Galaxy.
Recently the Auger collaboration has announced that cosmic rays of energies above $10^{19}$ eV appear to come from active galactic nuclei (AGN) located within 75 Mpc from our Galaxy~\cite{auger07}.
The clustering of arrival directions around the sky positions of those AGNs has enabled to indicate a value of the order of 0.5 nG for the extragalactic magnetic field, at least within a radius of 75 Mpc from the solar system~\cite{emagfield}.

Although protons are the most abundant component of cosmic rays, their origin is very difficult to determine, since the deflection radius for a proton of energy $E$ crossing a magnetic field of intensity $B$ is:
$$
\frac{R}{\rm{1\,pc}} \simeq 0.01
\frac%
{E/{\rm{1\,TeV}}}%
{B/1\,\mu{\rm G}} \, .
$$
The directional information of a photon coming from the galactic centre, at a distance of 8 kpc from the Earth, is thus lost below an energy of 300 PeV.

Magnetic fields do not deflect photons, so that they point with very good approximation to the position of the emission source.
Since in a hadronic cascade photons can reach some 10\% of the energy of the proton originating it, high-energy photons can be a good instrument to study the production of cosmic rays also of hadronic origin.

\begin{figure}
\centering
\includegraphics[width=.7\textwidth,angle=0.5]{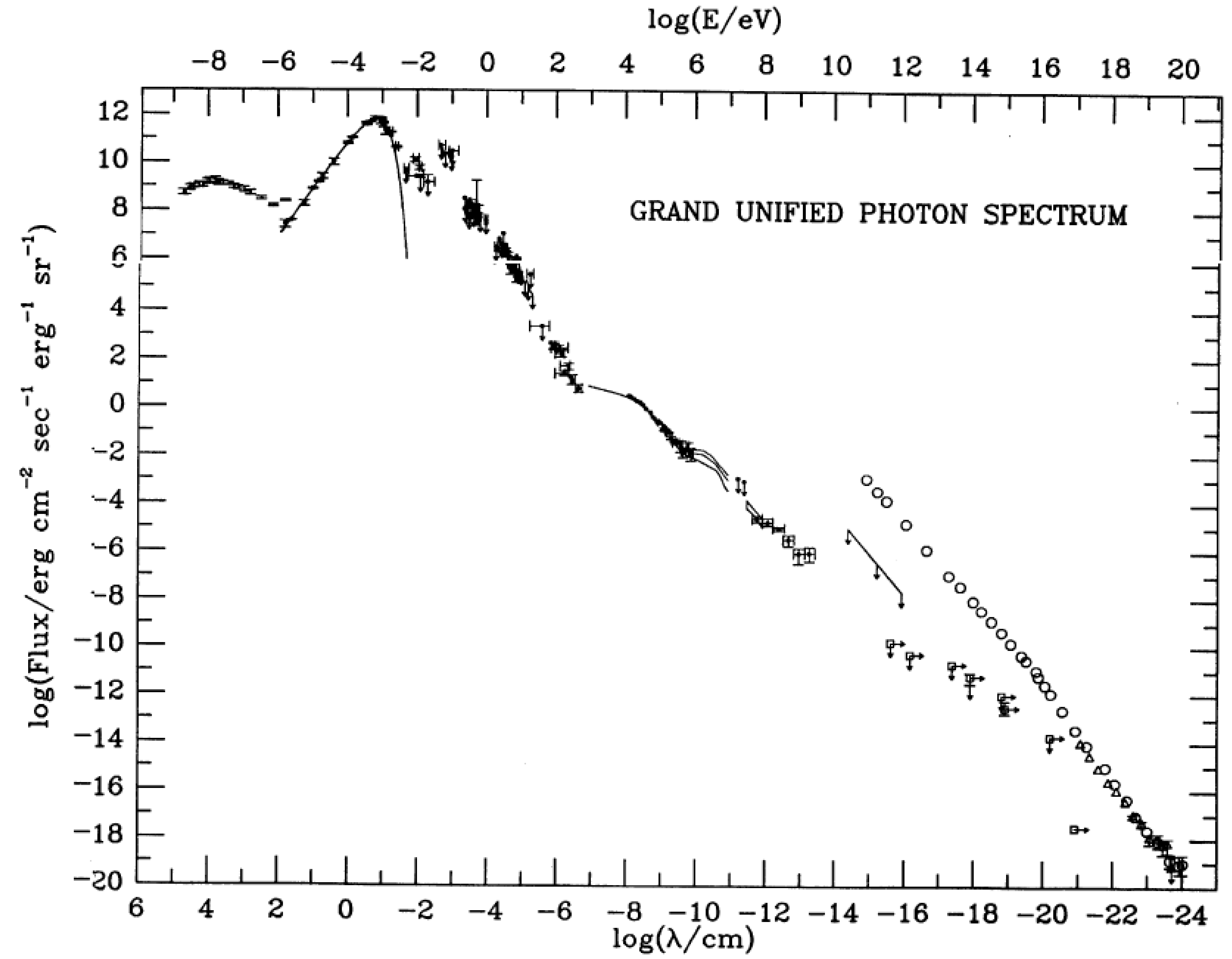}
\caption{\label{fig:Complem} 
Flux of diffuse extra-galactic photons~\cite{ressel-turner1989}.
The cosmic microwave radiation peak is visible.}
\end{figure}

$\gamma$-rays are probably the most interesting part of the spectrum of photon emission from astrophysical sources (Fig.~\ref{fig:Complem}).
During the recent years, a new window has been opened in the observation of $\gamma$-rays above 20 MeV (essentially, the photons above the threshold for pair production, plus some energy phase space), thanks to the availability of new photon detectors coming from technologies typical of experimental particle physics.
We shall call such photons, i.e., the photons above 20 MeV, high-energy (HE) $\gamma$-rays.
Excellent reviews of high energy gamma-ray astrophysics are available; see for example Ref.~\cite{aharobook}, Ref.~\cite{fleury}, Ref.~\cite{hinton}, and Ref.~\cite{hinton2}.

Although arbitrary, the definition of high-energy $\gamma$-rays reflects profound 
astrophysical and experimental arguments:
\begin{itemize}
\item[-]
the emission is {\em non-thermal}, and dominated by the conversion of gravitational energy into electromagnetic energy;
\item[-]
it is impossible to concentrate the photons, which leads to telescopes radically different form the ones dedicated to observations of larger wavelengths;
\item[-]
large quantity of background events are produced by charged cosmic particles.
\end{itemize}
High-energy $\gamma$-rays today are further classified; such a classification is of course arbitrary.
In this paper the energy ranges associated to the definition of low-medium energy, high-energy (HE), very-high-energy (VHE), ultra-high-energy (UHE) and extremely-high-energy (EHE) $\gamma$-rays, inspired with small modifications from~\cite{weekes1989}, are the following:
\begin{center}
\begin{narrowtabular}{3.5cm}{l r@{ }c@{ }r}
low-medium energy       &  0.1 MeV & - &  20 MeV
\\
HE        &  20 MeV & - &  30 GeV
\\
VHE       & 30 GeV & - &   30 TeV
\\
UHE       &  30 TeV & - &  30 PeV
\\
EHE       &  30 PeV & - & no limit
\end{narrowtabular}
\end{center}
The upper limits on EHE $\gamma$-rays come from the heavy reduction of the fluxes as the energy of the observed photons increases (at 30 PeV you expect no more than 1~photon/km$^2$ per day from the most powerful sources).

High-energy $\gamma$-rays can be detected even if they come from very far distances, since the detection efficiency is good and the mean free path at those energies is large:
regions of the sky which are opaque in other energy bands can be transparent to X- and $\gamma$-rays, providing us a key to some fundamental physics phenomena.

\section{Production processes of gamma-rays}

The source of high-energy photons from astrophysical objects is mainly gravitational energy released by collapses towards a central massive object.
In the presence of angular momentum, the dynamics of such a collapse can manifest itself in an accretion disk, with the presence of jets of plasma outflowing the accretion plane.
In addition, one could have characteristic photon signals also from annihilation/decay of heavy particles.

At the origin of the production of gamma-rays there is mostly the photon radiation off charged particles, in general electrons.
Such a radiation might happen due to
\begin{itemize}
\item
Bremsstrahlung.
An electron in an external field radiates due to bremsstrahlung.
The characteristic spectrum of bremsstrahlung is proportional to $1/E$, 
where $E$ is the energy of the emitted photon.
The fractional energy loss due to bremsstrahlung after crossing a distance $d$ 
is $(1-e^{-d/X_0})$, where $X_0$ is called the radiation length of the material.
$1/X_0$ is proportional to the density of the material.
\item
Synchrotron radiation.
An electron moving in a magnetic field radiates due a synchrotron radiation 
strongly beamed into a cone of angle $\alpha \sim {m_e}/{E}$, where $m_e$ is the electron mass.
Relativistic electrons in a typical galactic magnetic field emit synchrotron radiation at $E_\gamma$$\simeq$0.05$\bigl($$\frac{E_e}{{\rm TeV}}$$\bigr)^2$$\bigl($$\frac{B}{3\,\mu{\rm G}}$$\bigr)$\,eV.
For ultra-relativistic electrons (typically encountered in astrophysical situations), the emission spectrum is power-law.
High-energy synchrotron radiation can only originate in regions of very strong magnetic field, e.g.\ close to a neutron star surface (where $B$$\gtrsim$$10^{12}$\,G).
\end{itemize}

Alternatively, primary photons can originate from nuclear transition, or decay of $\pi^0$ in a hadronic environment.


\subsection{Photons from gravitational collapses}

The basic interpretation for the production of high-energy photons from 
gravitational collapses is the so-called self-synchrotron Compton (SSC) mechanism.
Synchrotron emission from ultra-relativistic electrons accelerated in a magnetic field generates photons with an energy spectrum peaked in the infrared/X-ray range.
Such photons in turn interact via Compton scattering with their own parent electron population: since electrons are ultrarelativistic
(with a Lorentz factor $\gamma_e$\,$\sim$\,10$^{4-5}$), the energy of the upscattered photon gets boosted by a factor 
$\lesssim$\,$\gamma_e^2$.

The upscattering of low-energy photons by collisions with high-energy electrons is the inverse Compton (IC) scattering.
This mechanism is very effective for increasing the photon energy (for this reason it is called ``inverse''), and is important in regions of high soft-photon energy density and energetic-electron number density.

For a power-law population of relativistic electrons with spectral index $q$ and a blackbody population of soft photons at a temperature $T$, mean photon energies and energy distributions can be calculated for electron energies in the Thomson regime and in the relativistic Klein-Nishina regime:
\begin{eqnarray}
{<E_\gamma>}
\, \simeq \, 
& \frac{4}{3} \gamma_e^2 \, <\eta> ~
&
{\rm for\ } \gamma_e \eta \ll m_ec^2 \ {\rm (Thompson\ limit)}
\\
\, \simeq \, 
& \frac{1}{2} <E_e> ~~~
&
{\rm for\ } \gamma_e \eta \gg m_ec^2 \ {\rm (Klein-Nishina\ limit)} 
\label{eq:compton1}
\end{eqnarray}
\begin{eqnarray}
\frac{{\rm d}N_\gamma}{{\rm d}E_\gamma}
\, \propto \,
&
E_{\gamma}^{-\frac{q+1}{2}}       ~~~~~~~~ ~
&
{\rm for\ } \gamma_e \eta \ll m_ec^2 \ 
{\rm (Thompson\ limit)}
\\
\, \propto \,
&
E_{\gamma}^{-(q+1)} \ln(E_\gamma)          ~
&
{\rm for\ } \gamma_e \eta \gg m_ec^2 \ 
{\rm (Klein-Nishina\ limit)}  
\end{eqnarray}
where where $E_\gamma$ denotes the scattered photon's energy,
$E_e$ denotes the parent electron's energy, and $\eta$ denotes the seed photon's 
energy.
A useful approximate relation linking the electron's energy and the Comptonized photon's energy is given by:
$E_\gamma \simeq 6.5 \bigl( \frac{E_e}{{\rm TeV}} \bigr)^2 \bigl( \frac{\eta}{{\rm meV} } \bigr)$\,GeV.

\begin{figure}
\centering
\includegraphics[width=.9\textwidth]{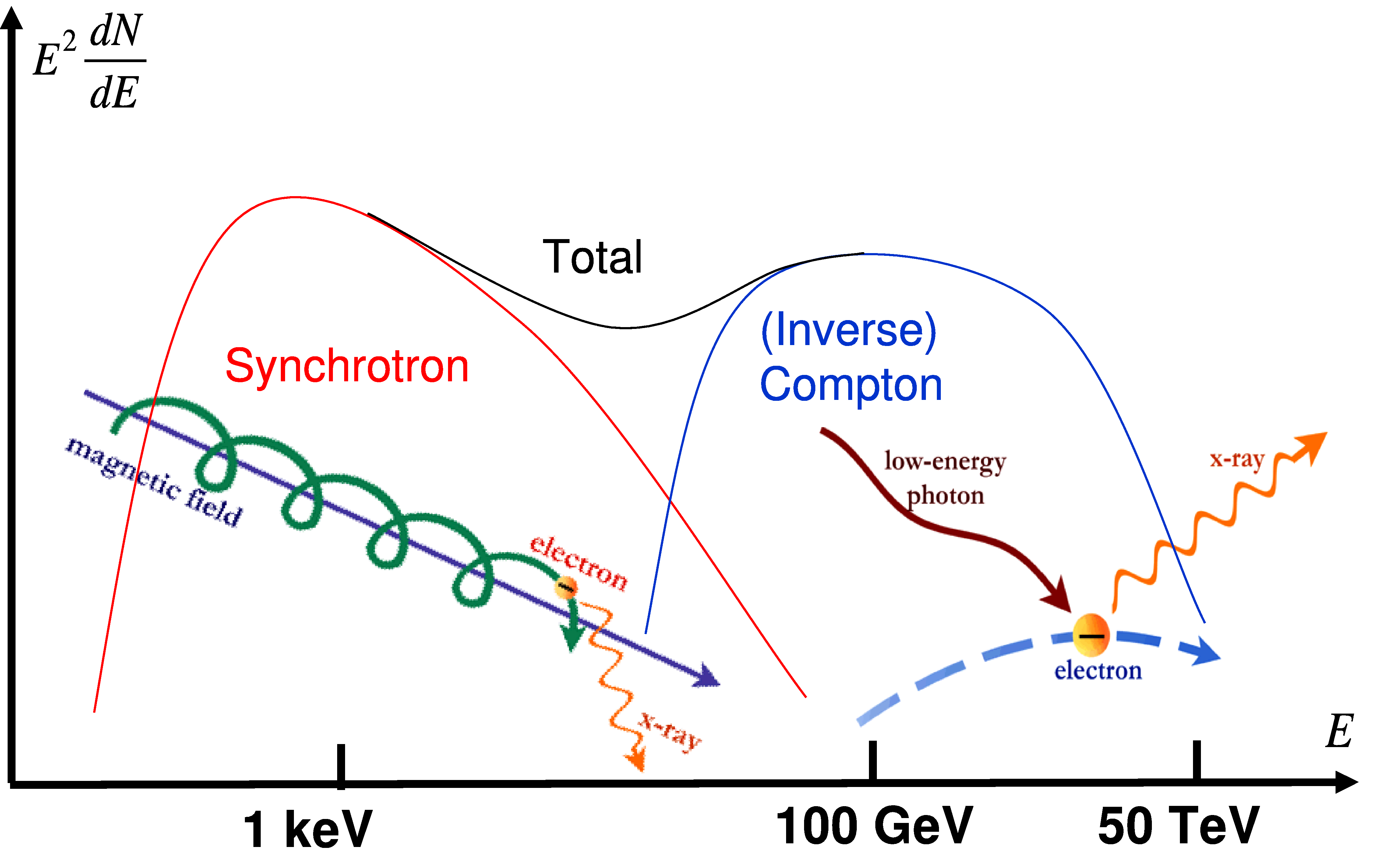}
\caption{\label{fig:StandardEmission} 
Differential energy spectrum of photons in the SSC model.}
\end{figure}

The Compton component can peak at GeV--TeV energies; the two characteristic synchrotron and 
Compton peaks are clearly visible on top of a general $E_\gamma^{-2}$ dependence. 
Fig.~\ref{fig:StandardEmission} shows the resulting energy spectrum. 
This behavior has been verified with high accuracy on the Crab Nebula, a steady VHE gamma emitter in the Milky Way which is often used to calibrate VHE gamma instruments. If in a given region the  photons  by the synchrotron radiation can be described by a power law  with spectral index $p$, in a first approximation the tails at the highest energies from {both} the synchrotron { and} the Compton mechanisms will have a spectral index  $p$.

Alternative and complementary models of VHE emission involve two electron populations, one 
--primary-- accelerated within the jet and the other --secondary-- generated by cascades initiated by primary protons/nuclei that had been accelerated in the jet~\cite{mannheim1993};
or a population of extremely energetic protons~\cite{aharonian2000}.
In such cases, the energy of the primary protons is expected by the physics of hadronic cascades to be one or two orders of magnitude larger than the energy of gammas, since the dominant mechanism for photon production is the decay of secondary $\pi^0$-mesons into~$\gamma \gamma$.
The study of $\gamma$-rays can thus provide insights on the production of charged cosmic rays.

\begin{figure}
\centering
\includegraphics[width=.5\textwidth]{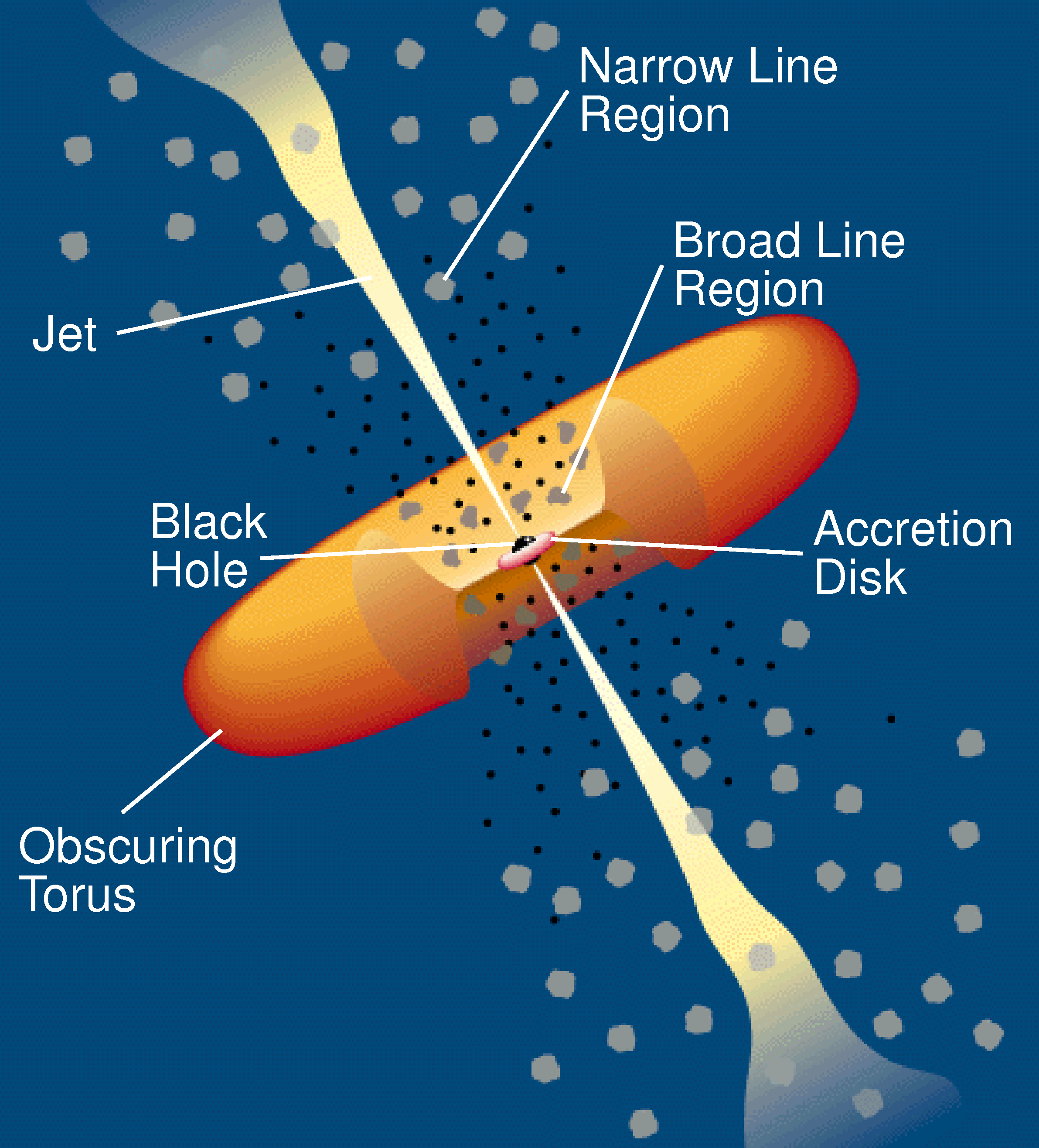}
\caption{\label{fig:agn} 
Schematic diagram for the emission by an AGN~\cite{urry-padovani1995}.}
\end{figure}

The case of AGN is particularly relevant.
Supermassive black holes of $\sim$10$^6$--10$^9$ solar masses ($M_\odot$) and beyond reside in the cores of most galaxies; their fueling by infalling matter produces a spectacular activity.
The black hole is surrounded by an accretion disk and by fast-moving clouds, which emit Doppler-broadened lines~(\cite{urry-padovani1995, padovani2007}).
In about 10$\%$ of all AGN, the infalling matter turns on powerful collimated jets that shoot out in opposite directions, likely perpendicular to the disk, at relativistic speeds 
(see Fig.~\ref{fig:agn}).

If a relativistic jet is viewed at small angle to its axis the observed jet emission is amplified by relativistic beaming%
\footnote{%
	Defining the relativistic Doppler factor as
	$\delta \equiv [\gamma_j (1 - \beta_j \cos \theta_j)]^{-1}$
	(with $\beta_j = v_j/c$ the jet speed normalized to the speed of light,
	$\gamma_j = 1/ \sqrt{(1-\beta_j^2)}$, and $\theta_j$ the angle between the jet's direction and the line of sight), the observed and intrinsic luminosities at a given frequency $f$ are related by $L_f^{\rm obs} = \delta^p L_f^{\rm em}$ with $p \sim 2-3$, and the variability timescales are related by $\Delta t_{\rm obs} = \delta^{-1} \Delta t_{\rm em}$.
	For $\theta \sim 0 ^{\circ}$ and $\delta \sim 2 \,\gamma_j$ the observed luminosity can be amplified by factors $\sim 400$~--~$10^4$ 
	(for, typically, $\gamma_j \sim 10$ and $p \sim 2-3$); 
	whereas $\theta_j \sim 1/\gamma_j$ implies $\delta \sim \gamma_j$, with a luminosity amplification of $\sim 10^2$~--~10$^3$.%
}
and dominates the observed emission; in this case the ``quasi-stellar'' (quasar) source is called blazar.

Given the compactness of blazars (as suggested by their observed short variability timescales, which can be as low as a couple of minutes for doubling the flux), all GeV/TeV photons would be absorbed through pair-producing $\gamma\gamma$ collisions with target X-ray/IR photons.
Beaming ensures the intrinsic radiation density to be much smaller than the observed one, so that $\gamma$-ray photons encounter a much lower $\gamma\gamma$ opacity and hence manage to leave the source: reversing the argument, $\gamma$-ray detection is a proof of strongly anisotropic (e.g., beamed) emission.

\begin{figure}[t]
\centering
\includegraphics[width=.7\textwidth]{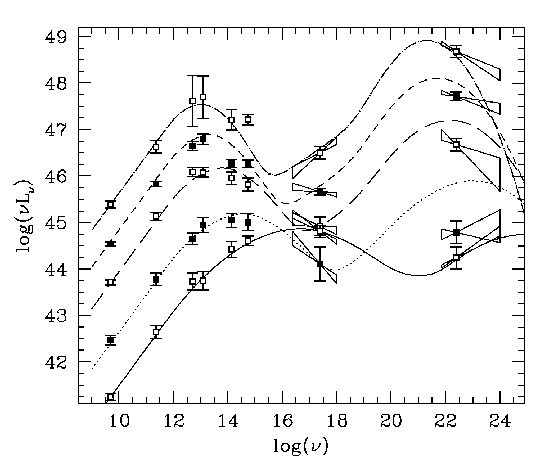}
\caption{\label{fig:AGNsequence}
The active galactic nuclei sequence~\cite{fossati1998}.}
\end{figure}

The spectral energy distributions of blazars are generally characterized by two broad humps, peaking at, respectively, infrared/X-ray and GeV-TeV frequencies~\cite{ulrich1997}.
Analyses of the spectral energy distribution of blazars~\cite{fossati1998, ghisellini1998} have suggested (see Fig.~\ref{fig:AGNsequence}) that:
\begin{enumerate}
\item
higher/lower-luminosity objects have both humps peaking at lower/higher frequencies (they are called, respectively, LBLs and HBLs);
\item
the luminosity ratio between the high- and low-frequency humps increases with luminosity;
\item
at the highest luminosities the $\gamma$-ray output dominates the total luminosity.
\end{enumerate}

Depending on the relative efficiency of the relativistic particles' cooling through scattering with photon fields that are internal to jet or external to it, the synchrotron and Compton components can also peak at IR/optical and MeV--GeV energies (external-IC, or EIC, scheme, see Ref.~\cite{dermer-schlickeiser1993}).
Hybrid SSC/EIC models have also been proposed~\cite{ghisellini1998}.

The emitting particles are accelerated within the relativistic jets which carry energy from the central black hole outwards~\cite{rees1967}.
In the SSC framework this process is approximated with a series of relativistically moving homogeneous regions (blobs), where particle acceleration and radiation take place (e.g., Ref.~\cite{maraschi1992}).
The X-ray and $\gamma$-ray emissions, with their extremely fast and correlated multi-frequency variability, indicate that often a single region dominates the emission;
such a region appears to be close to the central engine.



A cornerstone prediction from a pure SSC model is a definite correlation between the yields from synchrotron radiation and from IC during a flare (it would be difficult to accommodate in the theory an ``orphan flare'', i.e., a flare in the IC region not accompanied by a flare in the synchrotron region).
Evidence of neutrinos in a flare would be a smoking gun for the presence of a hadronic component.

Finally, another very important extragalactic source of $\gamma$-rays is given by gamma-ray bursts (GRBs).
GRBs are the most luminous events occurring in the universe since the Big Bang.
The duration of a GRB is typically a few seconds, but it can range from a few milliseconds to minutes, and the initial burst is usually followed by a longer-lived ``afterglow''.
Most of them are thought to come from a hypernova, or core-collapse supernova, a cataclysmic event resulting from the internal collapse and violent explosion of a massive star (the mass of the star needs to be at least 8 times the solar mass to undergo this process).
There is a prevailing consensus that the basic mechanism of GRB emission is an expanding relativistic fireball~\cite{rees-meszaros1992, meszaros-rees1993, sari1998}, with the beamed radiation due to internal/external shocks (prompt/afterglow phase, respectively).

In the fireball shock framework, several models have predicted VHE emission during both the prompt and afterglow phases of the GRB (e.g., Ref.~\cite{meszaros2006}).
This can occur as a result of electron self-IC emission from the internal shock or the external forward/reverse shock.
Seed photons can be produced locally (through synchrotron, or be the leftover of the initial radiation content responsible for the acceleration of the fireball) or can be produced in, e.g., the shell of a previously exploded supernova.
In the latter case, the photons emitted by the supernova may also act as targets for the $\gamma \gamma$ absorption, and in this case the VHE emission could be severely dimmed.
If the emission processes are indeed synchrotron and IC, then a blazar-like spectral energy distribution is predicted, with a double-peak shape extending into the VHE band.
In such theoretical freedom, VHE observations of GRBs could help constraining GRB models.

\subsection{Production from self-annihilation of dark matter}

Evidence for the departure of cosmological motions from the predictions of Newtonian dynamics based on visible matter, interpreted as due to the presence of dark matter, is well established
-- from galaxy scales (e.g.,~\cite{vanAlbada1985}) to galaxy-cluster scales 
(e.g.,~\cite{sarazin1986}) to cosmological scales (e.g.,~\cite{spergel2003}).

Dark matter particle candidates should be weakly interacting with ordinary matter (and hence neutral), otherwise they would have been already found.
The theoretically favored ones are heavier than the proton, and they are called weakly interacting massive particles (WIMPs); the present experimental limits from accelerators indicate a minimum mass of about 50~GeV~\cite{pdg}.
WIMPs should be long-lived enough to have survived from their decoupling from radiation in the early universe into the present epoch.

The main field of research for dark matter in the $\gamma$-ray energy range is related to the detection of photons emitted by the self-annihilation of WIMPs.
In particular, in supersymmetric models the lightest supersymmetric neutral particle, the neutralino, is predicted to be a Majorana particle, and is thus a natural candidate for such a WIMP.

\begin{figure}
\centering
\includegraphics[width=.7\textwidth]{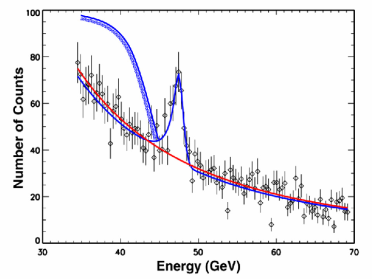}
\caption{\label{fig:WimpSignature}
Simulated signals of the WIMPs self-annihilation: a narrow peak or a continuum below the mass 
of the WIMP~\cite{GLASTscienceBrochure}.}
\end{figure}

The self-annihilation of a heavy WIMP $\chi$ can generate photons 
(Fig.~\ref{fig:WimpSignature}) in three main ways:
\begin{itemize}
\item[a)]
directly via annihilation into a photon pair ({\sl $\chi \chi \rightarrow \gamma \gamma$}) or 
into a photon - $Z$-boson pair (\mbox{$\chi \chi \rightarrow \gamma Z$}) with 
$E_\gamma = m_\chi$ or $E_\gamma = (m_\chi - m_Z)^2/(4 m_\chi)$ respectively;
these processes give a clear signature at high energies, being the energy monochromatic, but 
the process is suppressed at one loop, so the flux is expected to be very faint;
\item[b)]
via annihilation into a quark pair which produces jets emitting in turn a large number of 
$\gamma$ photons ({\sl $qq \rightarrow$ jets $\rightarrow$ many photons});
this process produces a continuum of gammas with energies below the WIMP mass;
the flux can be large but the signature might be difficult to detect;
\item[c)]
via internal bremsstrahlung~\cite{recentberg}; also in this case one has an excess of low 
energy gammas with respect to a background which is not so well known.
\end{itemize}

The $\gamma$-ray flux from the annihilation of dark matter particles of mass $m_{DM}$ can be 
expressed as the product of a particle physics component times an astrophysics component:
\begin{equation}
\frac{dN}{dE}\,=\frac{1}{4\pi}\,\underbrace{\frac{\langle
\sigma
v\rangle}{m^2_{DM}}\,\frac{dN_{\gamma}}{dE}}_{Particle\,
Physics}\,\times\,\underbrace{\int_{\Delta\Omega-l.o.s.} dl(\Omega) \rho^2_{DM}}_{Astrophysics} \ .
\end{equation}
The particle physics factor contains $\langle \sigma v\rangle$, the velocity-weighted 
annihilation cross section (there is indeed a possible component from cosmology in $v$), and 
$dN_{\gamma}/dE$, the differential $\gamma$-ray spectrum summed over the final states with 
their corresponding branching ratios.
The astrophysical part corresponds to the squared density of the dark matter distribution 
integrated over line of sight ({\it l.o.s.}) in the observed solid angle.

It is clear that the expected flux of photons from dark matter annihilations, and thus its 
detectability, depends crucially on the knowledge of the annihilation cross section $\sigma$ (which even within SUSY has 
uncertainties of one-two orders of magnitude for a given WIMP mass) and of $\rho_{DM}$, which 
is even more uncertain, and enters in the calculation squared.
A cuspy dark matter profile, or even the presence of local clumps, could make the detection 
easier by enhancing~$\rho_{\rm DM}$.
The experimental data about the centre of the Galaxy (rotation curves, microlensing, etc.) 
give however presently no evidence for a cuspy dark matter density profile, but the issue is 
still debated.

The targets for dark matter searches should be not extended, with the highest density, with no 
associated astrophysical sources, and close by to us, and possibly with some indication of 
small luminosity/mass ratio from the stellar dynamics.

\begin{figure}
\centering
\includegraphics[width=.9\textwidth]{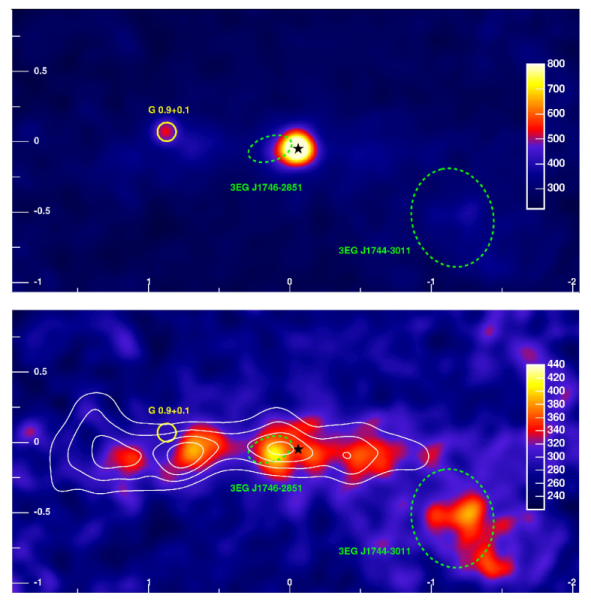}
\caption{\label{fig:GalacticCentre}
Image of the galactic centre by the H.E.S.S.\ telescope~\cite{aharonian2006g}.
Upper panel: point-like $\gamma$-ray sources at the galactic centre;
lower panel: the galactic centre in the $\gamma$-rays after subtraction of the two point-like 
sources: an extended source of $\gamma$-rays shows up.}
\end{figure}

The galactic centre is at a distance of about 8~kpc from the Earth.
A black hole of about $3.6 \times 10^6$ solar masses, Sgr~A$^{\star}$, lies there.
Because of its proximity, this region might be the best candidate for indirect searches of 
dark matter.
Unfortunately, there are other astrophysical $\gamma$-ray sources in the field of view (e.g., 
the supernova remnant Sgr~A~East) and the halo core radius makes it an extended rather than a 
point-like source (see Fig.~\ref{fig:GalacticCentre}).

Since distance dilution of the signal hinders detection, galaxies candidate for indirect dark 
matter detection should be chosen among nearby objects.
The best observational targets for dark matter detection outside the Galaxy are the Milky 
Way's dwarf spheroidal satellite galaxies (e.g., Draco, Sculptor, Fornax, Carina, Sextans, 
Ursa Minor).
For all of them (e.g., Draco), there is observational evidence of a mass excess 
with respect to what can be estimated from luminous objects, i.e., a high $M/L$ ratio.

\subsection{Top-down mechanisms}

Finally, some gamma-ray emission could originate in decays of exotic particles of very large mass possibly produced in the early Universe.
Such long-lived heavy particles are predicted in many models (e.g., technicolor models or the R-parity - violating SUSY model), and the energy distribution of particles coming from their decay should be radically different from what predicted by the standard emission models from astrophysical sources~\cite{GLASTscienceBrochure}.

\section{Propagation of $\gamma$-rays}

\begin{figure}
\centering
\includegraphics[width=.9\textwidth]{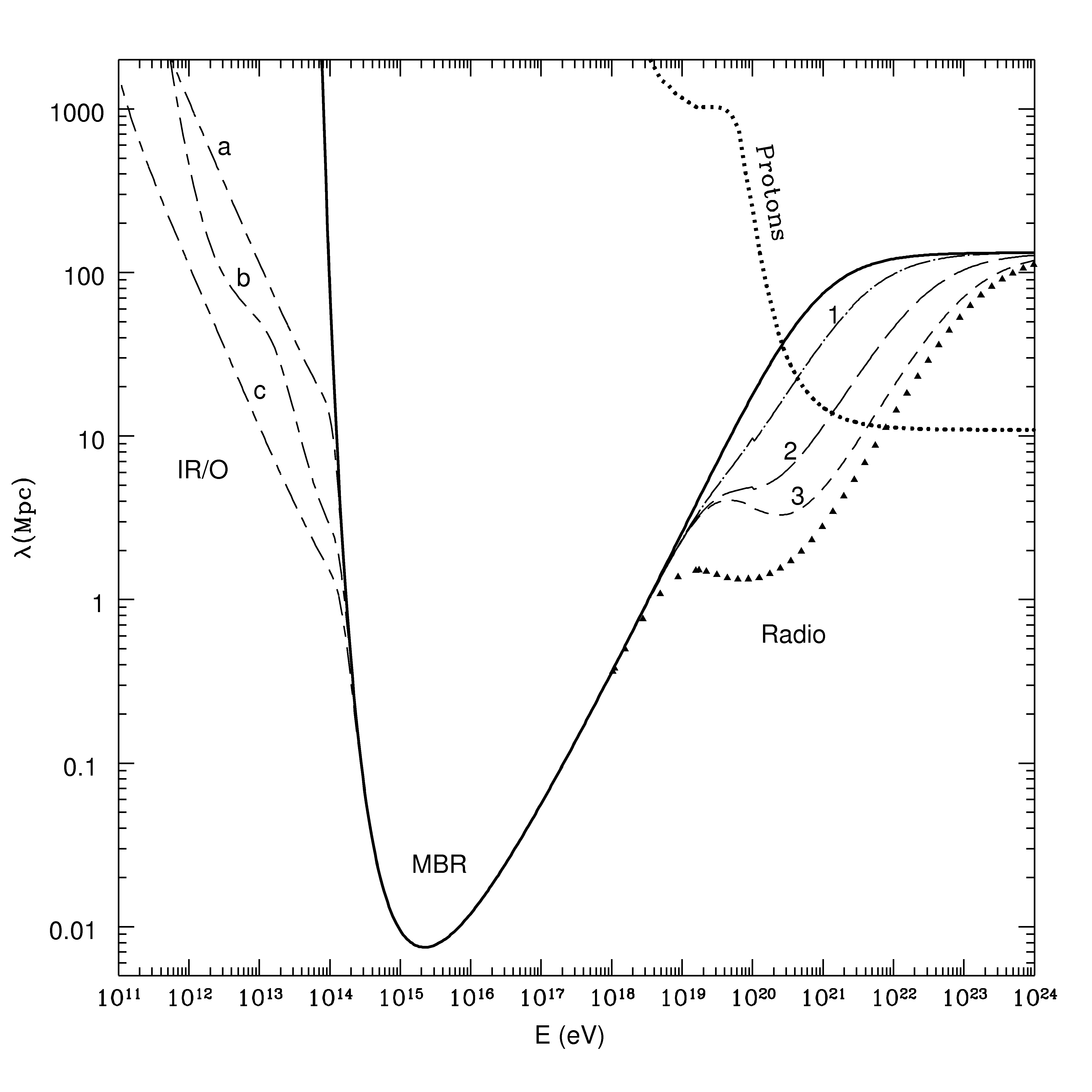}
\caption{\label{fig:CoppiAharonian}
Mean free path as a function of the photon energy~\cite{CoppiAharonian}.}
\end{figure}

Electron-positron $(e^-e^+)$ pair production in the interaction of beam photons off extragalactic background photons is a source of opacity of the Universe to $\gamma$-rays whenever the corresponding photon mean free path (Fig.~\ref{fig:CoppiAharonian}) is smaller than the source distance.

The dominant process for the absorption is the pair-creation process
$\gamma + \gamma_{\,\texttt{\scriptsize background}} \xrightarrow{} e^+ + e^-$, 
for which the cross-section is described by the Bethe-Heitler formula~\cite{heitler}:
\begin{equation}\label{eq.sez.urto}
\sigma(E,\epsilon) \simeq 1.25 \cdot 10^{-25} (1-\beta^2) \cdot \left[2 \beta ( \beta^2 -2 )
+ ( 3 - \beta^4 ) \, {\rm ln} \left( \frac{1+\beta}{1-\beta} \right)
\right] {\rm cm}^2
\end{equation}
where
$\beta = \sqrt{1-\frac{(m_e c^2)^2}{E\,\epsilon}}$, $m_e$ being the value of the electron mass,
$E$ is the energy of the (hard) incident photon and
$\epsilon$ is the energy of the (soft) background photon.
Notice that only QED, relativity and cosmology arguments are involved in the previous formula.

\begin{figure}[tb]
\centering
\includegraphics[width=.7\textwidth]{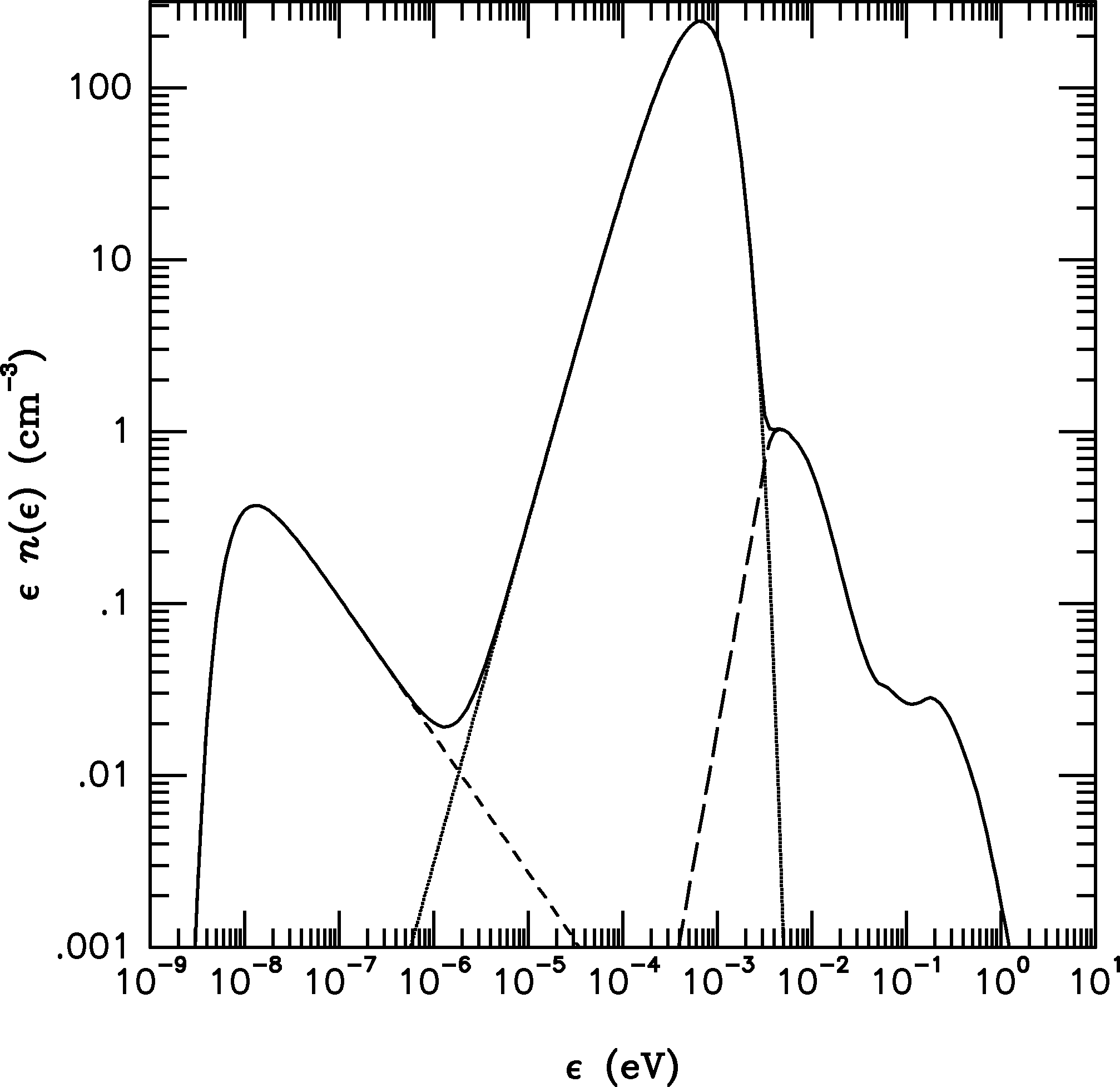}
\caption[]%
{\label{fig:density}
Photon number density of the intergalactic radiation field at $z=0$ as developed by 
Lee~\cite{lee98}, composed by the radio background (short dashed line), the cosmic microwave background (dense 
dotted line), and the infrared/optical/ultraviolet background (EBL) (long dashed line).}
\end{figure}

The cross section in Eq.~6 is maximized when $\epsilon \simeq \frac{500 \,{\rm{GeV}}}{E}$~eV.
Hence if $E = 1\,$TeV the interaction cross section is maximal if $\epsilon \simeq 0.5\,$eV 
(corresponding to a near-infrared soft photon).
In general, for very high energy photons the $\gamma \gamma \rightarrow e^+ + e^-$ interaction becomes 
important with optical/infrared photons, whereas the interaction with the cosmic microwave 
background becomes dominant at $E~\sim$~1~PeV.
Therefore, the background component relevant for interaction with VHE photons is the 
optical/infrared background radiation.
This is called extragalactic background light~(EBL)~\cite{gould:1967a}.

The EBL consists of the sum of starlight emitted by galaxies throughout their whole cosmic 
history, plus possible additional contributions, like, e.g., light from hypothetical first 
stars that formed before galaxies were assembled.
Therefore, in principle the EBL contains important information both the evolution of baryonic 
components of galaxies and the structure of the Universe in the pre-galactic era.

The attenuation suffered by observed VHE spectra can thus be used to derive constraints on the 
EBL density~\cite{stecker2001}.
First limits on the EBL were obtained in Ref.~\cite{stecker1992}, while recent determinations 
from the detection of distant VHE sources are reported in~\cite{aharonian2006i,mazin}.
Fig.~\ref{fig:density} shows the estimated photon number density of the background photons as composed by the radio background, the cosmic microwave background, and 
the infrared/optical/ultraviolet background~(EBL).

Specifically, the probability for a photon of observed energy $E$ to survive absorption along 
its path from its source at redshift $z$ to the observer plays the role of an attenuation 
factor for the radiation flux, and it is usually expressed in the form:
\begin{equation} \label{eq:flux.tau}
e^{-\tau(E,z)} \, .
\end{equation}
The coefficient $\tau(E,z)$ is called {\em optical depth}. 

To compute the optical depth of a photon as a function of its observed energy~$E$ and the 
redshift $z$ of its emission one has to take into account the fact that the energy $E$ of a 
photon scales with the redshift~$z$ as $(1+z)$~\cite{peebles}; thus when using 
Eq.~\ref{eq.sez.urto} we must treat the energies as function of~$z$ and evolve 
$\sigma\big(E(z),\epsilon(z),\theta\big)$ for 
$E(z)= (1+z)E$ and $\epsilon(z)=(1+z)\epsilon$,
where $E$ and $\epsilon$ are the energies at redshift $z=0$.
The optical depth is then computed~\cite{gould:1967a} by convolving the photon number density 
of the background photon field with the cross section between the incident $\gamma$-ray and 
the background target photons, and integrating the result over the distance, the scattering 
angle and the energy of the (redshifted) background photon:
\begin{equation}  \label{eq.comp.tau}
\tau(E,z) = 
\int_{0}^{z} dl(z)\
\int_{-1}^{1}\ d\cos \theta \frac{1-\cos \theta}{2}
\int_{\frac{2(m_e c^2)^2}{E(1-\cos\theta)}}^{\infty} d\epsilon(z)\
n_{\epsilon}\big(\epsilon(z),z\big) \ \sigma(E(z),\epsilon(z),\theta)
\end{equation}
where
$\theta$ is the scattering angle,
$n_{\epsilon}\big(\epsilon(z),z\big)$ is the density for photons of energy $\epsilon(z)$ at 
the redshift $z$, and $l(z) = c\ dt(z)$ is the distance as a function of the redshift, defined 
by~\cite{padmanabhan}
\begin{equation}
\label{eq:padmanabhan-diff}
\frac{dl}{dz} \ = \ \frac{c}{H_0} 
\frac{1}{(1+z) \left[ (1+z)^2 (\Omega_M\,z+1) - \Omega_{\Lambda}\,z(z+2) \right]^{\frac{1}{2}} } \, .
\end{equation}
In the last formula $H_0$ is the Hubble constant, $\Omega_M$ is the matter density (in units 
of the critical density, $\rho_{\rm c}$) and $\Omega_{\Lambda}$ is the ``dark energy'' density (in 
units of $\rho_{\rm c}$); therefore, since the 
optical depth depends also on the cosmological parameters, its determination constrains the 
values of the cosmological parameters~\cite{blanch-vari} if the cosmological emission of galaxies 
is known.


The energy dependence of $\tau$ leads to appreciable modifications of the observed source 
spectrum (with respect to the spectrum at emission) even for small differences in~$\tau$, due to the 
exponential dependence described in Eq.~(\ref{eq:flux.tau}).
Since the optical depth (and consequently the absoption coefficient) increases with energy, 
the observed flux results steeper than the emitted one.

\begin{figure}
\centering
\includegraphics[width=.7\textwidth]{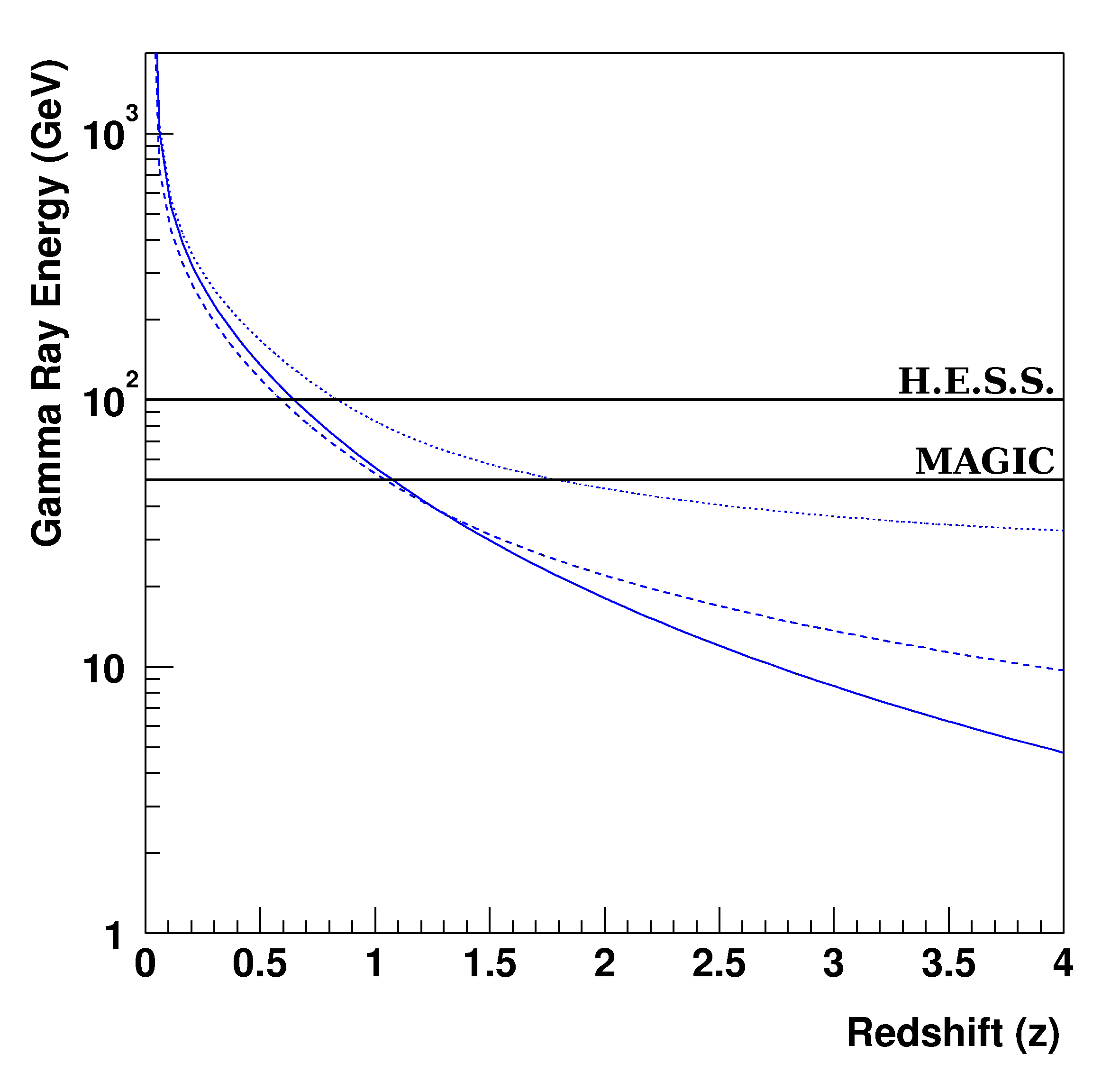}
\caption[]%
{\label{fig:gr-horizon}
Gamma-ray horizon compared with the lower energy limit of some gamma detectors: 
the MAGIC and H.E.S.S.\ telescopes;
the curves of the photon energy versus horizon are computed for different background evolution 
models by Blanch \& Martinez in the second citation in Ref.~\cite{blanch-vari}.
}
\end{figure}

The {\em horizon} (e.g.\ Ref.~\cite{blanch-vari,mannheim1999}) or {\em attenuation edge} 
(e.g.\ Ref.~\cite{primack2001}) for a photon of energy $E$ is defined as the distance 
corresponding to the redshift~$z$ for which $\tau(E,z)=1$, that gives an attenuation by 
a factor $1/e$ (see Fig.~\ref{fig:gr-horizon}).

\begin{figure}
\centering
\includegraphics[width=.7\textwidth]{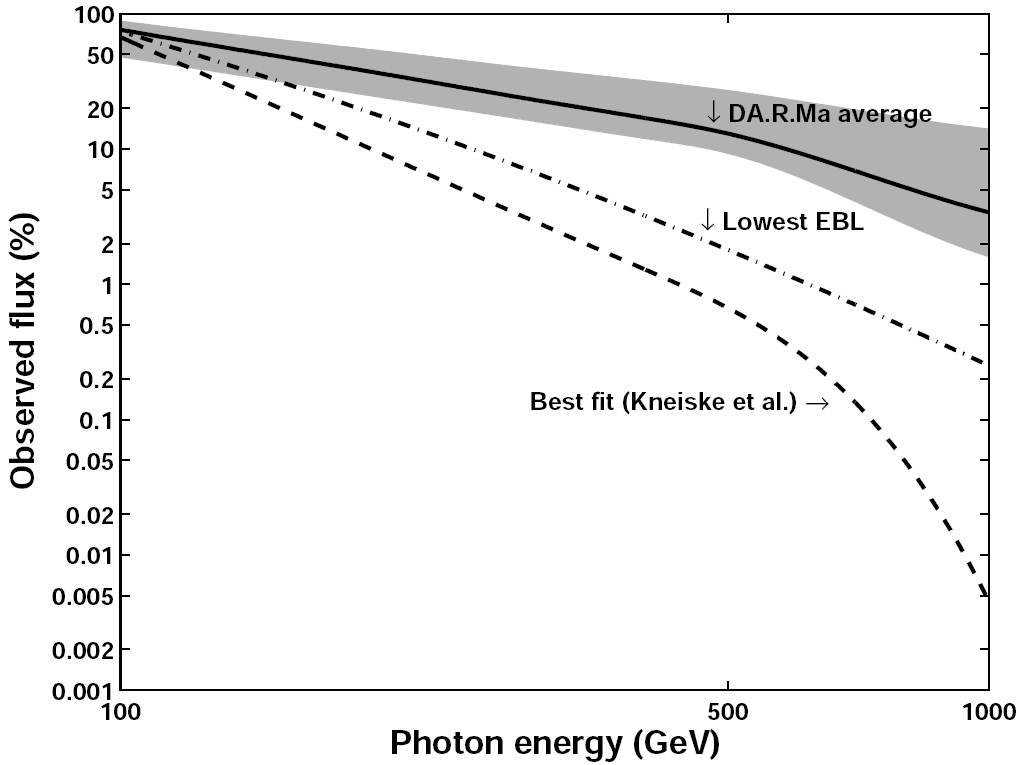}
\caption[]%
{\label{fig:darma}
The two lowest lines give the fraction of photons surviving from a source at the same distance 
of 3C\,279 without the oscillation mechanism, for the ``best-fit model'' of EBL (dashed line) 
and for the minimum EBL density compatible with cosmology~\cite{kneiske2004}.
The solid line represents the prediction in the DARMA model~\cite{noi2}, with an uncertainty represented by the gray band.
}
\end{figure}

Other interactions than the one just described might change our picture of the attenuation 
of $\gamma$-rays, and they are presently subject of thorough studies, since the present data 
on the absorption of photons are hardly compatible with the pure QED picture.
For example, $\gamma$-rays might interact with (possibly quintessential) very light axion-like 
particles, which might change the absorption length~\cite{noi1,noi2}.
In particular, in the DARMA model~\cite{noi2}, such contribution might enhance the photon flux via a regeneration mechanism (see Fig.~\ref{fig:darma}).
Such an interaction would be mediated by the (intergalactic) magnetic fields~\cite{emagfield}.
A similar mechanism invkes the conversion of photons into axion-like particles at the emission source~\cite{hooperserpico}.

Finally, mechanisms in which the absorption is changed through violation of the Lorentz 
invariance as in Ref.~\cite{kifune} are also under test; such models are particularly 
appealing within scenarios inspired to quantum gravity~\cite{amelino}.

\section{Detection techniques}

The detection of high-energy photons is complicated by the absorption by the atmosphere, 
and by the faintness of the signal, in particular when compared to the corresponding charged 
particles of similar energy.

\subsection{Atmospheric transparency and processes of interaction}

\begin{figure}
\centering
\includegraphics[width=\textwidth]{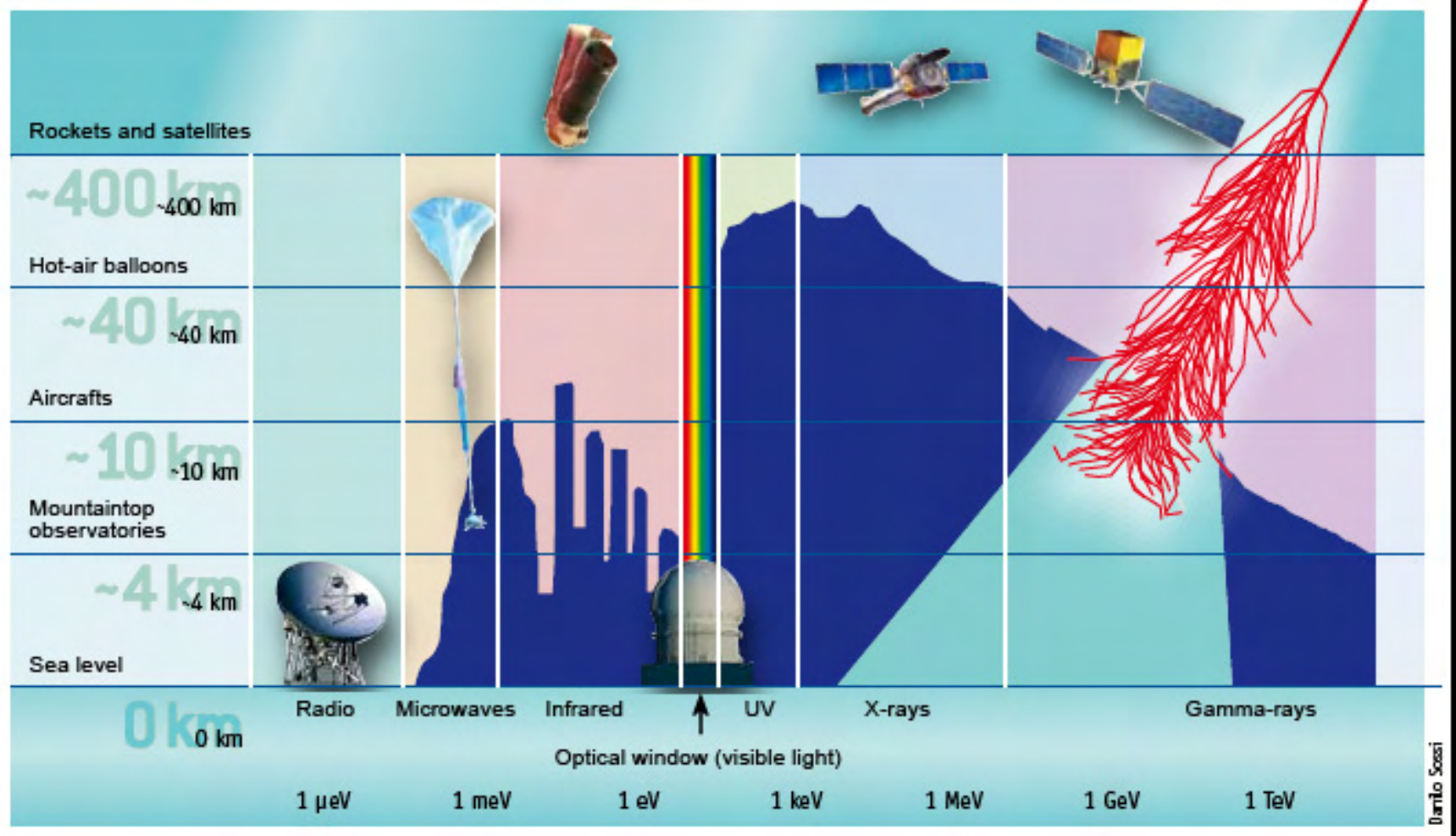}
\caption{\label{fig:atmOpac}
Transparency of the atmosphere for different photon energies and possible detection techniques~\cite{SuW}.}
\end{figure}

Photons above the ultra-violet (UV) region are shielded by the Earth's atmosphere 
(see Fig.~\ref{fig:atmOpac}).

Photons interact with matter mostly due to the Compton mechanism and to the photoelectric effect 
at energies up to about 20~MeV, while $e^+e^-$ pair production dominates above about 
20~MeV.
Above about 50~GeV the production of atmospheric showers takes place, dominated by the pair 
production and the bremsstrahlung mechanisms: an energetic photon scatters on an atmospheric 
nucleus and produces a pair, which emits secondary photons via bremsstrahlung;
such photons produce in turn a $e^+e^-$ pair, and so on, giving rise to a shower of 
charged particles and photons.
The process is described exg.\ in~\cite{rossi1941,rossigreisen}.

\begin{figure}
\centering
\includegraphics[width=.7\textwidth]{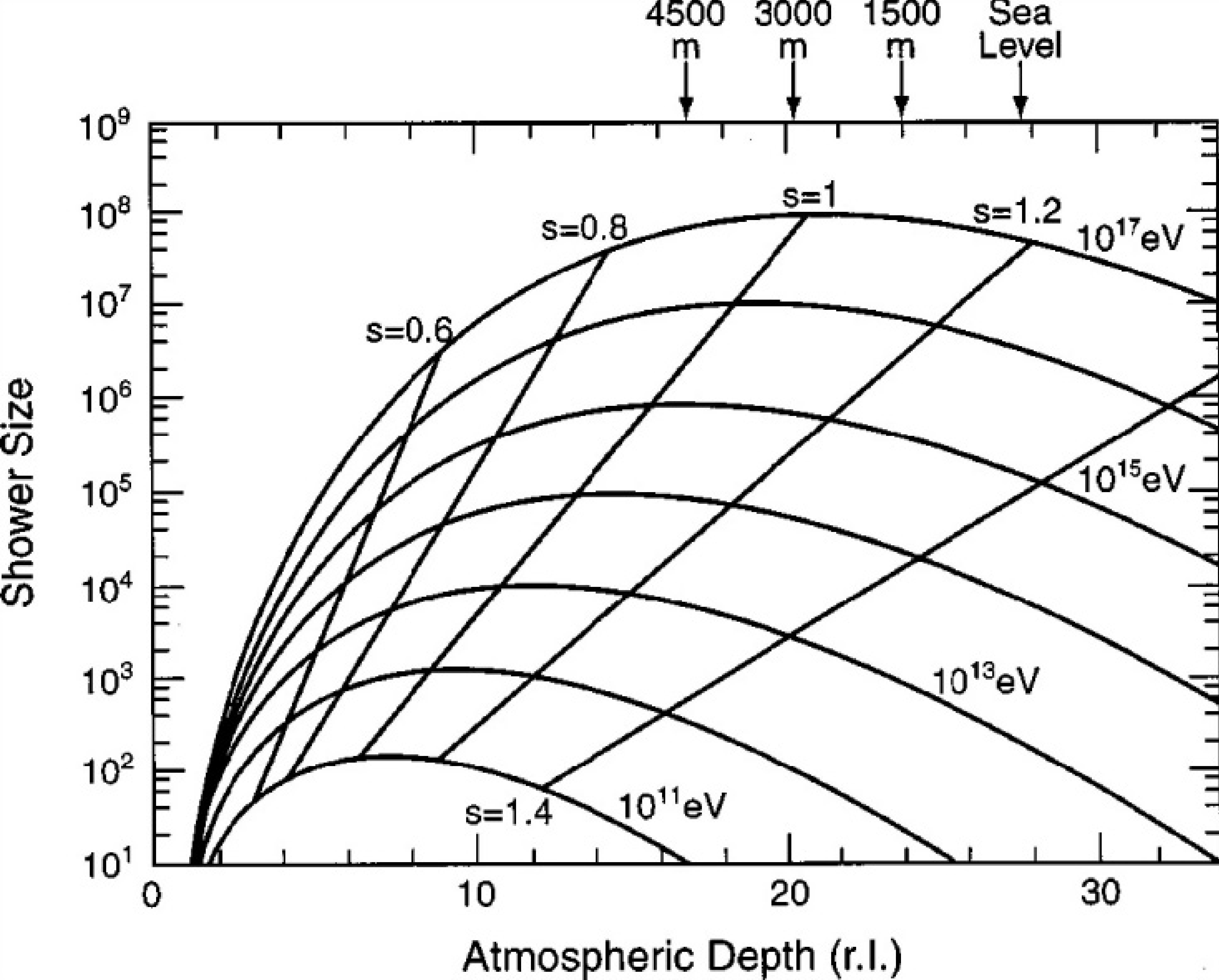}
\caption{\label{fig:longdev}
Longitudinal shower development~\cite{fleury}.
The parameter $s$ describes the shower age, being 0 at the first interaction, 1 at the maximum and 2 at the death~\cite{rossi1941}.}
\end{figure}

The longitudinal development of typical photon-induced extensive air showers is shown in 
Fig.~\ref{fig:longdev} for different values of the primary energies.
The maximum shower size occurs approximately $\ln(E/\epsilon_0)$ radiation lengths%
\footnote{%
The radiation length $X_0$ is both the mean distance over which a high-energy electron 
loses all but $1/e$ of its energy by radiation, and 7/9 of the mean free path for 
pair production by a high-energy photon.
The radiation length for air is about 37 g/cm$^{2}$.%
}
into the atmosphere, generally well above ground (the critical energy $\epsilon_0$, 
about 80 MeV in air, is the energy at which the ionization energy loss starts dominating 
the energy loss by bremsstrahlung).
Anyway, a large number of shower particles may reach the ground, especially at 
mountain altitudes.

\begin{figure}
\centering
\includegraphics[width=.9\textwidth]{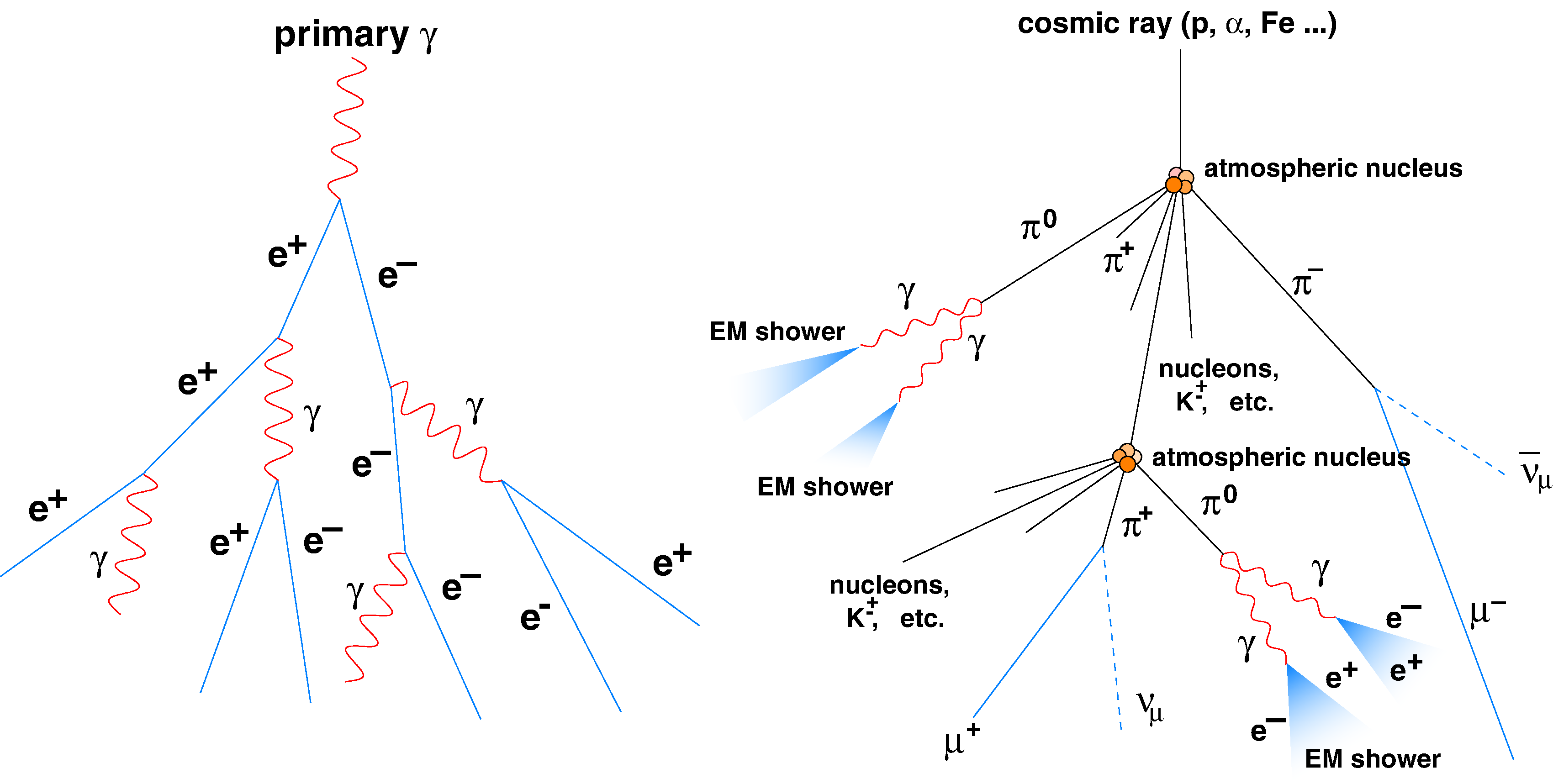}
\caption{\label{fig:showers}
Schematic representation of two atmospheric showers initiated by a photon~(left) 
or by a cosmic ray (right)~\cite{wagnertesi}.}
\end{figure}

The dominant high-energy hadrons,
protons and nuclei, also interact high in the atmosphere.
The process characterizing hadronic showers is not dissimilar (Fig.~\ref{fig:showers}).
The hadronic interaction length in air is about 61 g/cm$^2$ for protons,
being shorter for heavier nuclei.
The transverse development of hadronic showers is in general wider than for 
electromagnetic showers, and fluctuations are larger.

At sea level the thickness of the atmosphere corresponds to about 28 radiation lengths.
This means that only satellite-based detectors can detect primary X/$\gamma$-rays.
Since the fluxes of high-energy photons are low and decrease rapidly with increasing energy, 
VHE ad UHE gammas can be detected only from the atmospheric showers they produce, i.e., by 
means of ground-based detectors; such detectors should be placed at high altitudes, where 
atmospheric dimming is lower.

Let us now examine the characteristics of the satellite-based and ground-based detectors for 
high-energy photons.

\subsection{Satellites}

\begin{figure}
\centering
\includegraphics[width=.9\textwidth]{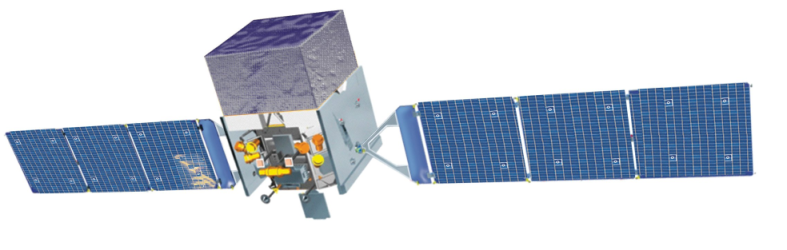}
\caption{\label{fig:Glast}
Image of the GLAST satellite~\cite{GLASTimage}.}
\end{figure}

Main figures of merit for a satellite-borne detector are its effective area (i.e., the product of the area 
times the detection efficiency), the energy resolution, the space or angular resolution 
(called as well point-spread function, or PSF), and the time resolution.
Satellite HE gamma telescopes such as EGRET, AGILE and GLAST (Fig.~\ref{fig:Glast}) detect 
the primary photons at energies lower than ground-based telescopes.
They have a small effective area, of order of 1 m$^2$, which yields a low sensitivity.
They have a large duty cycle, since they are not constrained by night operation, and they 
suffer a low rate of background events, but they have a large cost.

\begin{figure}
\centering
\includegraphics[width=.5\textwidth]{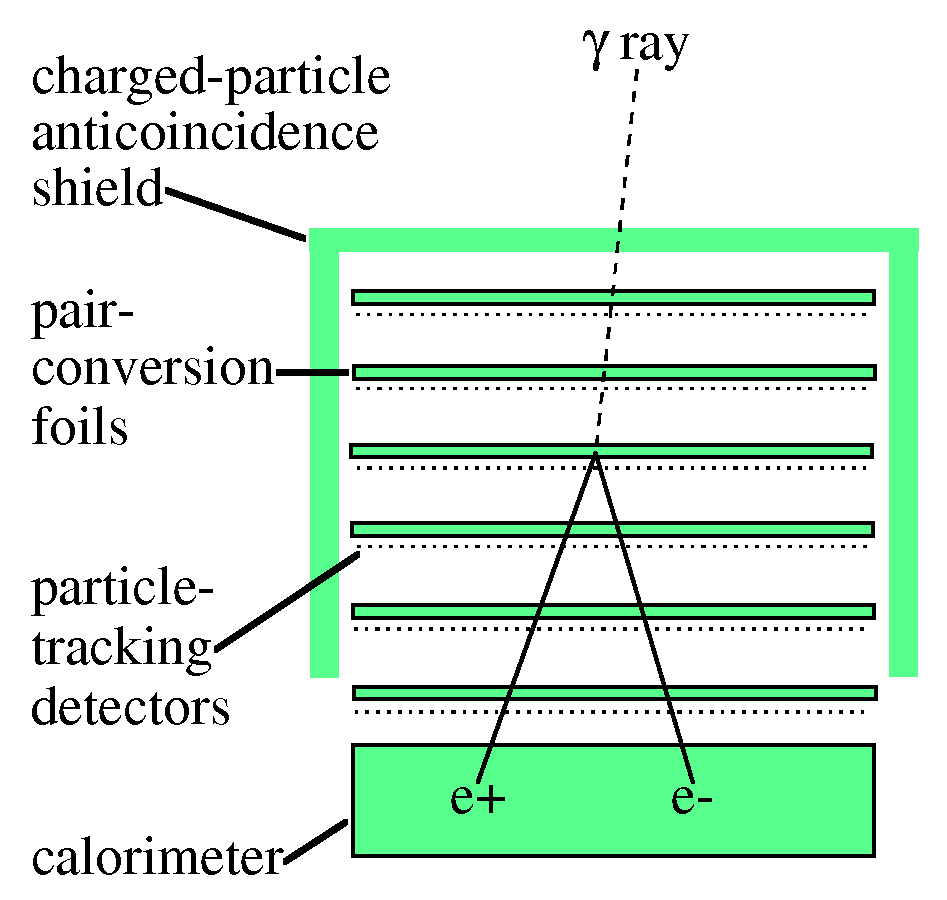}
\caption{\label{fig:TrackerCalor}
Pair-production in the typical structure of a satellite detector after an incident 
$\gamma$-ray enters the telescope~\cite{satelliteDetector}.}
\end{figure}

The technology of AGILE and GLAST has been inherited from the EGRET instrument (Fig.~\ref{fig:TrackerCalor}),
 operational in 1991--2000 on the Compton Gamma-Ray Observatory.
The direction of an incident photon is mostly determined by the geometry of its conversion into 
an $e^+e^-$ pair in foils of heavy materials which compose the instrument, and detected by 
planes of silicon detectors.
The presence of an anticoincidence apparatus realizes a veto against unwanted incoming charged 
particles.
The angular resolution of these telescopes is limited by the opening angle of the 
$e^+e^-$ pair, approximately~${m_e}/{E} \, \ln({E}/{m_e})$, and especially by the 
effect of multiple scattering.

To achieve a good energy resolution, in this kind of detectors a calorimeter in the bottom 
of the tracker is possibly used, depending on the weight that the payload is planned to 
comply with.
Due to weight limitations, however, it is difficult to fit in a calorimeter that completely holds the showers; this leakage downgrades the energy resolution.
Since at low energies the multiple scattering is the dominant process, the optimal detector design is a tradeoff between small radiation length (which decreases the conversion efficiency) and large number of samplings (which increases the power consumption, limited by the problems of heat dissipation in space).

\begin{figure}
\centering
\includegraphics[width=.9\textwidth]{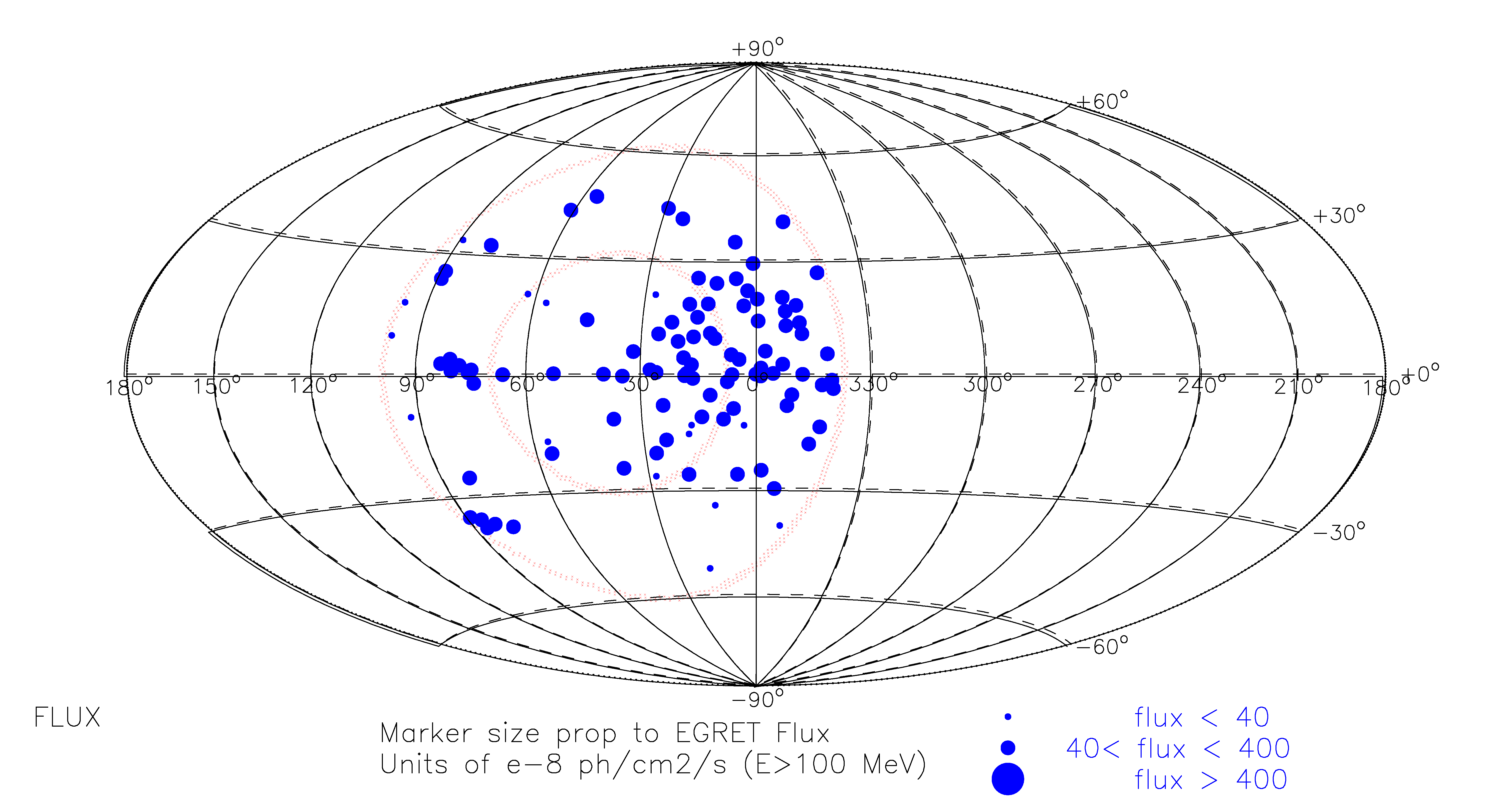}
\caption{\label{fig:AGILEbeginning}
Example of a region of space explored by AGILE in a single pointing; 
the field of view (FoV) is indicated~\cite{agile-sito}.
}
\end{figure}

AGILE is a completely Italian satellite launched 
in April 2007.
Its structure is very similar to GLAST, but its effective area is about one order of magnitude smaller.
The physics runs of AGILE (see Fig.~\ref{fig:AGILEbeginning}) started on September 1$^{\rm st}$, 2007, but many remarkable physics results were observed already during commissioning.
Several sources, some of which new, have already been found at $>$1\,GeV.

\begin{figure}
\centering
\includegraphics[width=.35\textwidth]{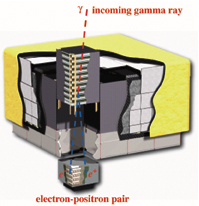}
\hspace*{1em}
\includegraphics[width=.35\textwidth]{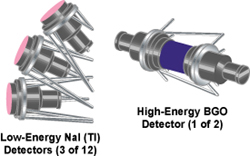}
\\[1ex]
\includegraphics[width=.45\textwidth]{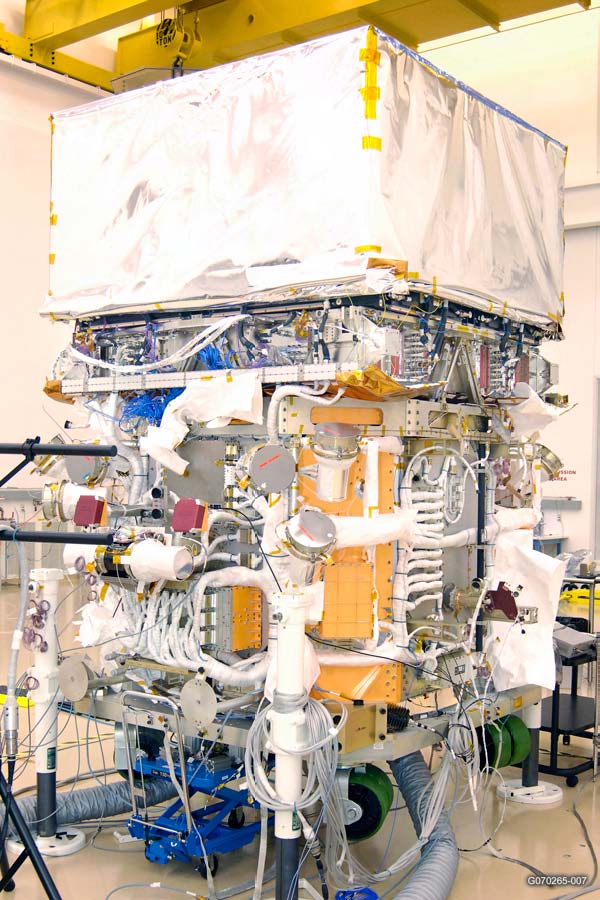}
\\[1ex]
\includegraphics[width=.7\textwidth]{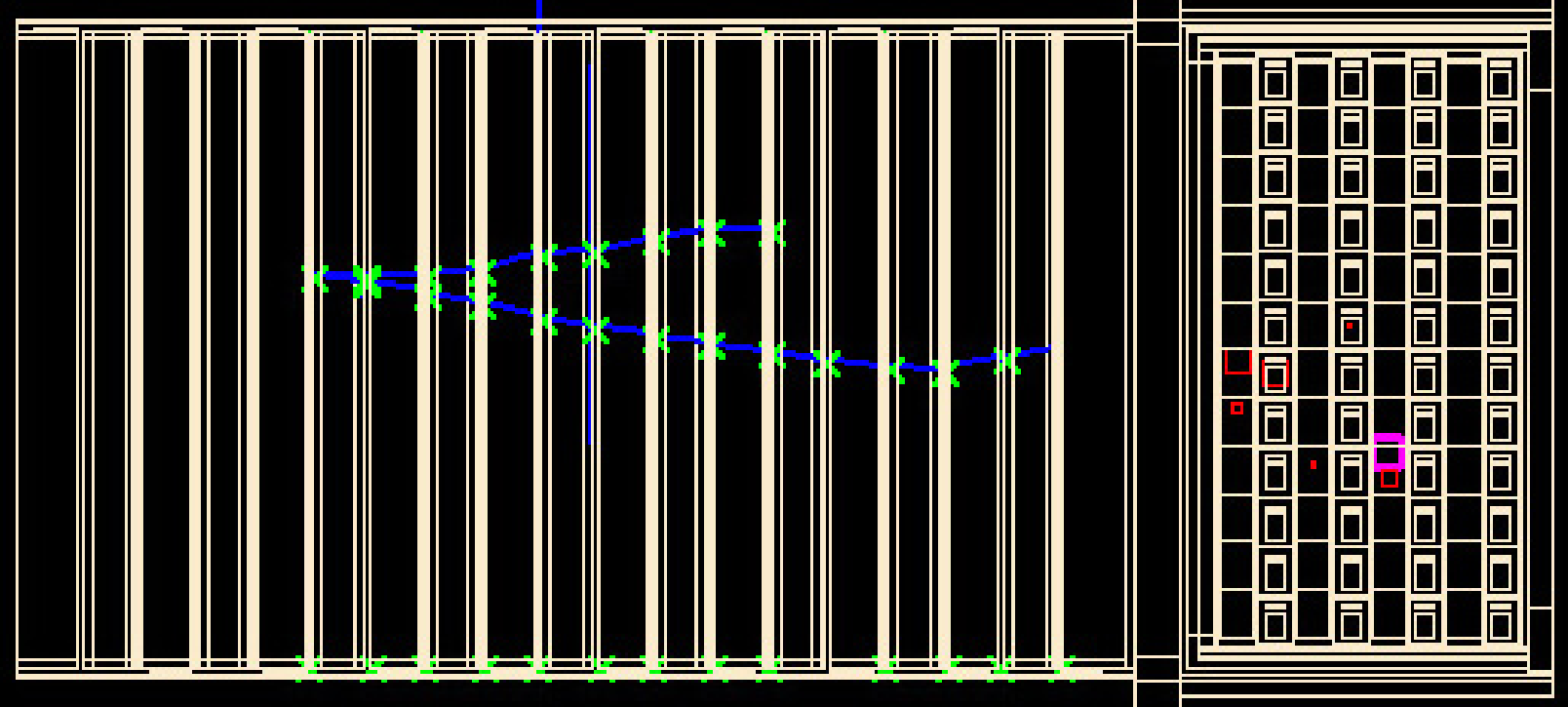}
\caption{\label{fig:Glast-LAT-GBM}
On the first  row, the two GLAST instruments: the Large Area Telescope (LAT) and the 
GLAST Burst Monitor (GBM)~\cite{lat-gmb}.
On the second row, a picture of the GLAST satellite~\cite{GLASTintero}.
On the third  row, the first gamma conversion detected by GLAST~\cite{GLASTimage}.}
\end{figure}

\begin{figure}
\centering
\includegraphics[width=.9\textwidth]{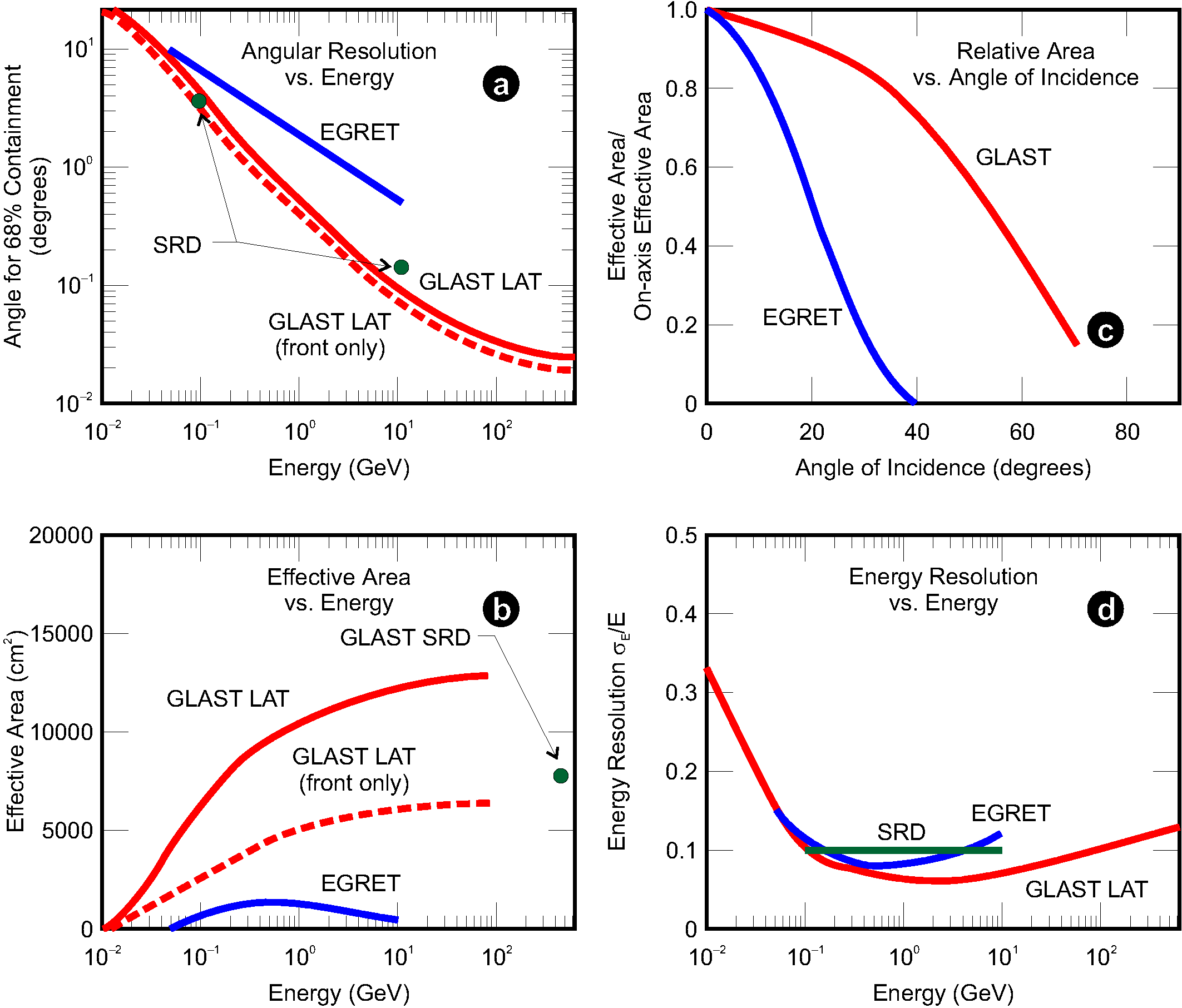}
\caption{\label{fig:GLASTperformance}
GLAST LAT performance compared with EGRET~\cite{GlastPerformance}.}
\end{figure}

\begin{figure} 
\centering
\includegraphics[width=.9\textwidth]{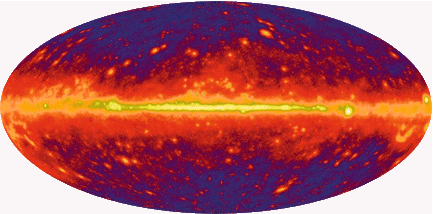}
\caption{\label{fig:GLAST1ySky}
One-year simulation of the sources that GLAST will detect~\cite{GLASTimage};
the red color corresponds to photons at an energy in the range \mbox{0.1 -- 0.4 GeV};
green corresponds to the energy range \mbox{0.4 -- 1.6 GeV} and
blue to more than 1.6~GeV.}
\end{figure}

The GLAST observatory, 
launched in June 2008, is composed by the spacecraft and by two instruments: the Large Area Telescope (LAT) and the GLAST Burst Monitor (GBM).
The two instruments are integrated and they work as a single observatory.
The structure of the LAT consists mainly in a tracker, an anticoincidence apparatus and a calorimeter (see Fig.~\ref{fig:Glast-LAT-GBM}).
Its energy range goes from 20~MeV to about 300~GeV, while the energy range explored by the GMB is 10~keV~--~25~MeV.
The GLAST LAT 
outperforms EGRET by two orders of magnitude thanks to its effective area that approaches 1~m$^2$ (to be compared to 0.15~m$^2$ from EGRET) and its time resolution of 10~$\mu$s; a comparison is shown in Fig.~\ref{fig:GLASTperformance}.
Fig.~\ref{fig:GLAST1ySky} shows a simulation of the HE $\gamma$-ray sky emerging from one year of scanning-mode observation by GLAST.

\subsection{Ground-based detectors}

Ground-based VHE telescopes such as MILAGRO, ARGO, CANGAROO, H.E.S.S., MAGIC and VERITAS 
detect the secondary particles of the atmospheric showers produced by primary photons and 
cosmic rays of energy higher than the primaries observed by satellites.
The two kinds of detectors are complementary (see Fig.~\ref{fig:Complem}).
Such ground-based detectors have a huge effective area, so their sensitivity is high;
they detect a huge amount of background events, and they have low cost.

\begin{figure}
\centering
\includegraphics[width=.7\textwidth]{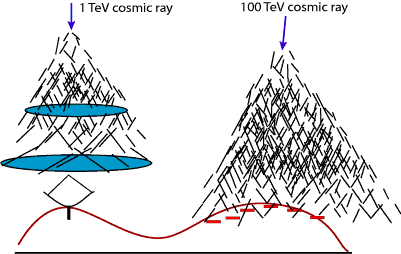}
\caption{\label{fig:EASvsIACT}
Sketch of the operation of Cherenkov telescopes and of EAS~\cite{sciamiVeritas}.}
\end{figure}

There are two main classes of ground based HE gamma detectors: the Extensive Air Shower arrays 
(EAS) and the Cherenkov telescopes (see Fig.~\ref{fig:EASvsIACT}).

\subsubsection{EAS detectors}

The EAS detectors, such as MILAGRO and ARGO, are made by a large array of detectors sensitive to charged secondary particles generated by the atmospheric showers.
They have high duty cycle and a large FoV, but a low sensitivity.
Since the maximum of a photon-initiated shower at 1 TeV typically occurs at 8 km a.s.l., the energy threshold of such detectors is rather large.

\begin{figure}
\centering
\includegraphics[width=\textwidth]{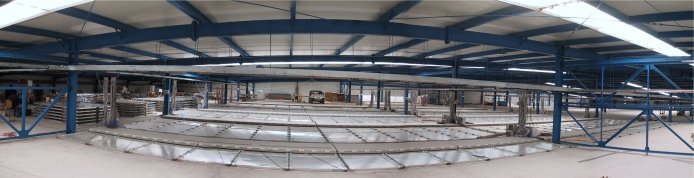}
\caption{\label{fig:ARGO}
View of the laboratory of the ARGO detector~\cite{sitoArgo}.}
\end{figure}

\begin{figure}
\centering
\includegraphics[width=.7\textwidth]{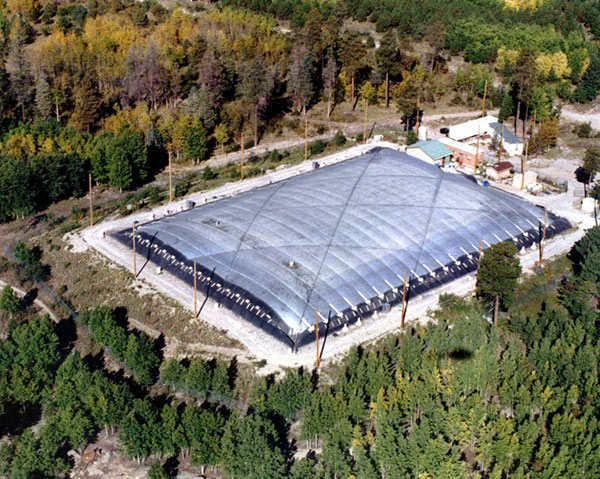}
\caption{\label{fig:Milagro}
The MILAGRO detector~\cite{sitoMilagro}.}
\end{figure}

Direct sampling of the charged particles in the shower can be achieved:
\begin{itemize} 
\item
Either by using an array of sparse scintillator-based detectors, as for example in the Tibet AS instrument (located at 4100 m a.s.l.\ to reduce the threshold).
For an energy of 100~TeV there are about 50~000 electrons at mountain-top altitudes, so sampling is possible;
\item
Or by effective covering of the ground to ensure efficient collection and hence lower energy threshold.
\begin{itemize}
\item
The ARGO-YBJ detector (see Fig.~\ref{fig:ARGO},~\cite{sitoArgo}) at the Tibet site follows this approach.
It is made of an array of resistive plate counters.
Its energy threshold lies in the 0.5 TeV-1 TeV range.
The first results show that ARGO can detect the Crab Nebula with a significance of about 
5\,standard deviations ($\sigma$) in 50~days of observation.
\item
MILAGRO (see Fig.~\ref{fig:Milagro}) is a water-Cherenkov based instrument near Los Alamos 
(about 2600~m altitude).
It is made of photomultipliers in water.
It detects the Cherenkov light%
\footnote{%
The Cherenkov light is a radiation produced in a medium by charged particles which travel faster than the 
speed of light in that medium; since its differential probability is proportional to 
($1/\lambda$) in the visible, it peaks on the blue.
Cherenkov photons are emitted at an angle $\theta_C$ such that $\cos(\theta_C)$$=$$\frac{1}{\beta n}$,
where $n$~is the refractive index of the medium in which the phenomenon takes place,
and $\beta$ is the speed of light in units of $c$.%
}
produced by the secondary particles of the shower when they pass through the water.
MILAGRO can detect the Crab Nebula with a significance of about 5\,$\sigma$ in 100~days of 
observation, at a median energy of about 20~TeV.
\end{itemize}
\end{itemize}

The energy threshold of EAS detectors is at best in the 0.5 TeV-1 TeV range, and it also depends on where the first interaction of the atmospheric shower occurred, so such detectors are built to detect UHE photons as well as the most energetic VHE gammas.
At such energies fluxes are small, so they need to have large surfaces, of order of~10$^4$~m$^2$.
EAS detectors are possibly provided with a muon detector devoted to hadron rejection; otherwise the discrimination from the background can be done based on the reconstructed shower shape.
The direction of the detected primary particles is computed by taking into account their arrival times, and the angular precision is about 1~degree.
Energy resolution is also poor.
The calibration can be performed by studying the shadow in the reconstructed directions caused by the presence of the Moon.

\subsubsection{Cherenkov telescopes}

Imaging Atmospheric Cherenkov Telescopes (IACTs), such as CANGAROO~III, H.E.S.S., MAGIC and 
VERITAS, detect the Cherenkov photons produced in air by charged, locally superluminal 
particles in atmospheric showers.
For reasons explained below, they have a low duty cycle and a small FoV, but they have a high 
sensitivity and a low energy threshold.

At sea level, the value of the Cherenkov angle $\theta_C$ in air for $\beta = 1$ is about 
1.3$^{\rm o}$, while at 8~km~a.s.l.\ it is about 1$^{\rm o}$.
The energy threshold at sea level is 21~MeV for a primary electron and 44~GeV for a primary 
muon.

Half of
the emission occurs within 21 m of the shower axis (about 70 m
for a proton shower).

\begin{figure}
\centering
\includegraphics[width=.8\textwidth]{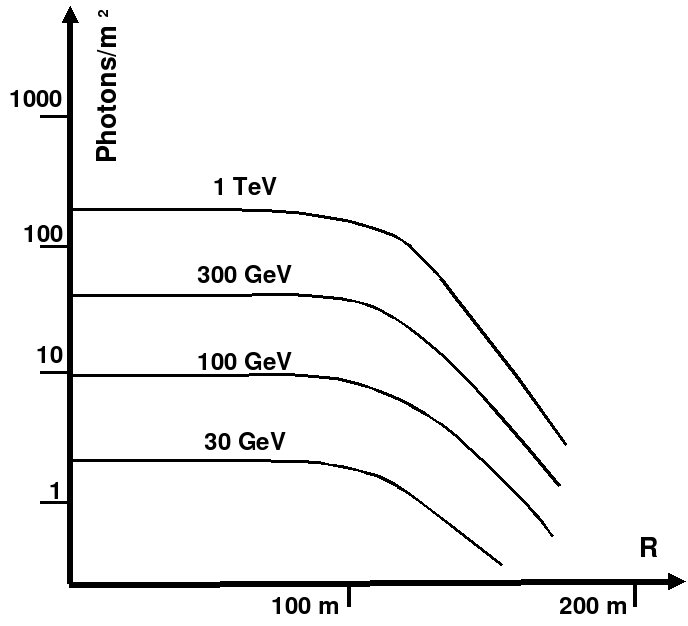}
\vspace*{-3mm}
\caption{\label{fig:yield}
Density per square meter of Cherenkov photons between 300 and 600 nm 
as a function of distance $R$ from the shower impact point for various 
photon energies as seen at 2 km a.s.l.\ for vertical showers.}
\end{figure}

Since the intrinsic angular spread of the charged particles in an electromagnetic shower is 
about 0.5 degrees, the opening of the light cone is dominated by the Cherenkov angle.
As a consequence, the ground area illuminated by Cherenkov photons from a shower of 1 TeV 
(the so-called ``light pool'' of the shower) has a radius of about 120 m.
The height of maximal emission for a primary of 1~TeV of energy is approximately 8 km a.s.l., 
and about 150 photons per m$^2$ arrive at 2000 m a.s.l. in the visible frequencies.
This dependence is not linear, being the yield of about 10 photons per square meter at 100 GeV 
(Fig.~\ref{fig:yield}).
The shower has a duration of a few (about 2 to 3) ns at ground; this duration is maintained 
by an isochronous (parabolic) reflector.

\begin{figure}
\centering
\includegraphics[width=.7\textwidth]{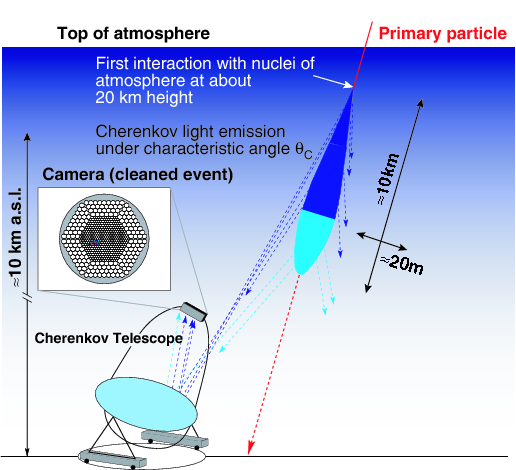}
\caption{\label{fig:IACTtecnique}
The observational technique adopted by the Imaging Atmospheric Cherenkov Telescopes 
(IACTs)~\cite{wagnertesi}.}
\end{figure}

The observational technique used by the IACTs is to project the Cherenkov light collected by 
a large optical reflecting surface onto a camera made by an array of photomultiplier tubes, 
with typical quantum efficiency of about 30\%, in the focal plane of the reflector 
(see Fig.~\ref{fig:IACTtecnique}).
The camera has a typical diameter of about 1~m, which corresponds to a FoV 
of~$5^{\rm o}$$\times$$5^{\rm o}$.
The signal collected by the camera is analogically transmitted to trigger systems, similar to 
the ones used in high-energy physics.
The events which passed the trigger levels are sent to the data acquisition system, 
which typically operates at a frequency of a few hundreds Hz.
The typical resolution on the arrival time of a signal on a photomultiplier is better 
than 1~ns.

Since, as discussed above, about 10 photons per square meter arrive in the light pool for 
a primary photon of 100 GeV, a light collector of area 100 m$^2$ is sufficient to detect 
gamma-ray showers if placed at mountain-top altitudes.
Due to the faintness of the signal, data can typically be taken only in moonless time, or with moderate moonlight, and without clouds, which limits the total observation time to some 1500 hours per year.

In the GeV-TeV region the background from charged particles is three orders of magnitude larger than the signal.
Hadronic showers, however, have a different topology with respect to electromagnetic showers, being larger and more subject to fluctuations.
One can thus separate 
showers induced by gamma-rays from the hadronic ones on the basis of the shower shape.

\begin{figure}
\centering
\includegraphics[width=.7\textwidth]{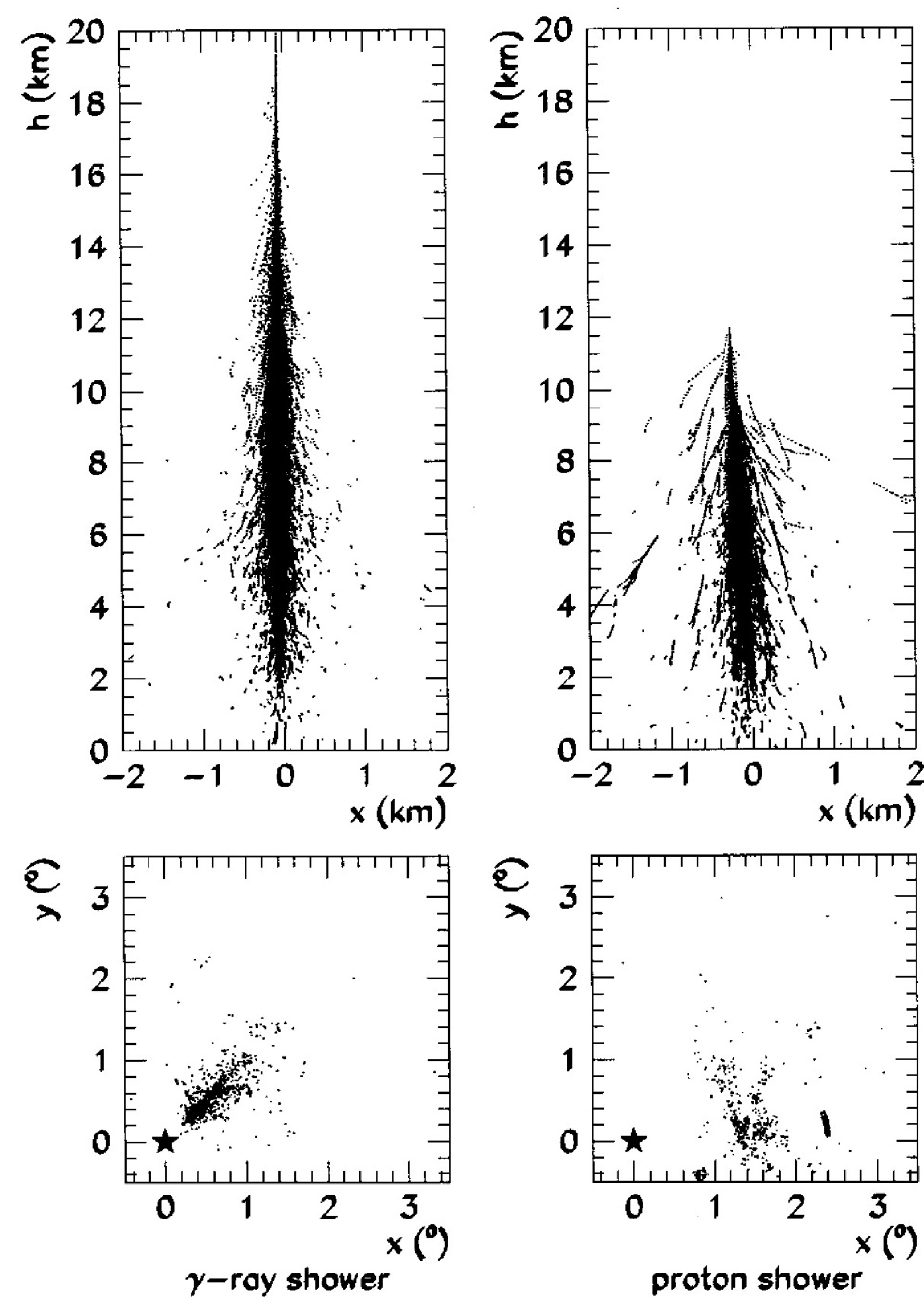}
\caption{\label{fig:hadronphoton}
Development of vertical 1-TeV proton and photon showers in the atmosphere.
The upper panels show the positions in the atmosphere of all shower electrons above the 
Cherenkov threshold;
the lower panels show the resulting Cherenkov images in the focal plane of a 10-m reflecting 
mirror when the showers fall 100 m from the detector (the center of the focal plane is 
indicated by a star)~\cite{fleury}.}
\end{figure}

\begin{figure}
\centering
\includegraphics[width=.49\textwidth]{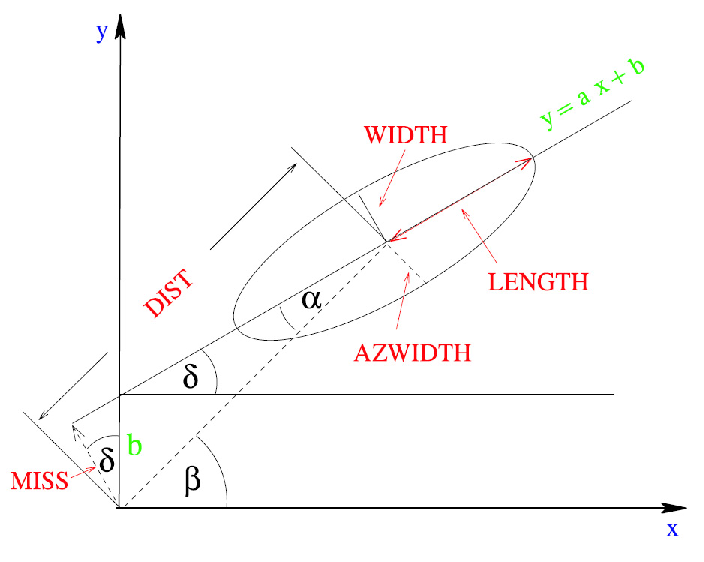}
\includegraphics[width=.49\textwidth]{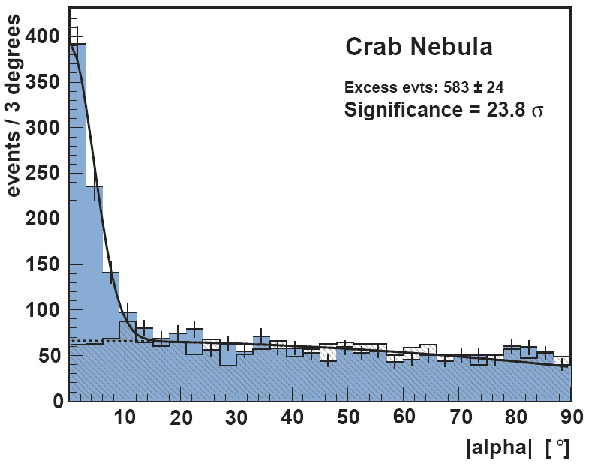}
\caption{\label{fig:alphaplot}
Left: Imaging parameters used by air Cherenkov telescopes to reject the hadronic cosmic-ray 
background: $\alpha$, width, and length.
The ellipse represents the outline of the shower image in the focal plane of the telescope.
Right: An ``$\alpha$-plot'' of the signal from the Crab Nebula, by the MAGIC telescope.}
\end{figure}

Most of the present identification techniques rely on a technique pioneered by Hillas in the 80's~\cite{hillnew}; the discriminating variables are called ``Hillas parameters''.
Several new techniques have been proposed to improve these results~\cite{improvement};
a recent review on the problem of background subtraction is published in~\cite{funnew}.
The intensity (and area) of the image produced is an indication of the shower energy,
while the image orientation is related to the shower direction.
The shape of the image is different for different primary particles, in such a way that its 
characteristics are used to distinguish between events produced by photons and by other 
particles, and to reject the background from charged particles (Fig.~\ref{fig:hadronphoton}).
Since the signal points to a given source, while the background is expected to be uniformly 
distributed, the image is circular for a shower falling directly on the detector, becoming 
more elliptical as the incoming photon displays a nonzero impact parameter.
For a photon coming from the source, the angle $\alpha$ in Fig.~\ref{fig:alphaplot} 
should be close to zero, while it should not peak to zero for hadrons, since their direction
does not come from a single source.

The time structure of Cherenkov images provides an additional discriminator against the 
hadronic background~\cite{tescaro}, which can be used by isochronous detectors (with parabolic 
shape) and with a signal integration time smaller than the duration of the shower (i.e., 
better than 1-2 GHz).

Two pointing modes can be used for the observation: the on-off mode and the wobble mode.
\begin{itemize}
\item
For the on-off mode, in the ``on'' phase the source is located in the camera center; in a 
different time a similar sky region without sources is pointed, and this measurement (called 
``off'') is used to estimate the background.
Since this cannot be done at the same time, conditions (weather, night sky background, etc.) 
can be different: thus an appropriate scaling of the background measurement is needed.
\item
Instead of observing the source in the center of the camera, the telescope can be pointed to 
a sky position slightly off-source.
The background can then be extracted from a so called anti-source position symmetrical 
with respect to the camera center.
This observation mode is called wobble mode.
Wobble mode has in general the disadvantage of a slighly worse sensitivity, but it 
saves observation time.
\end{itemize}

\begin{figure}
\centering
\includegraphics[width=.7\textwidth]{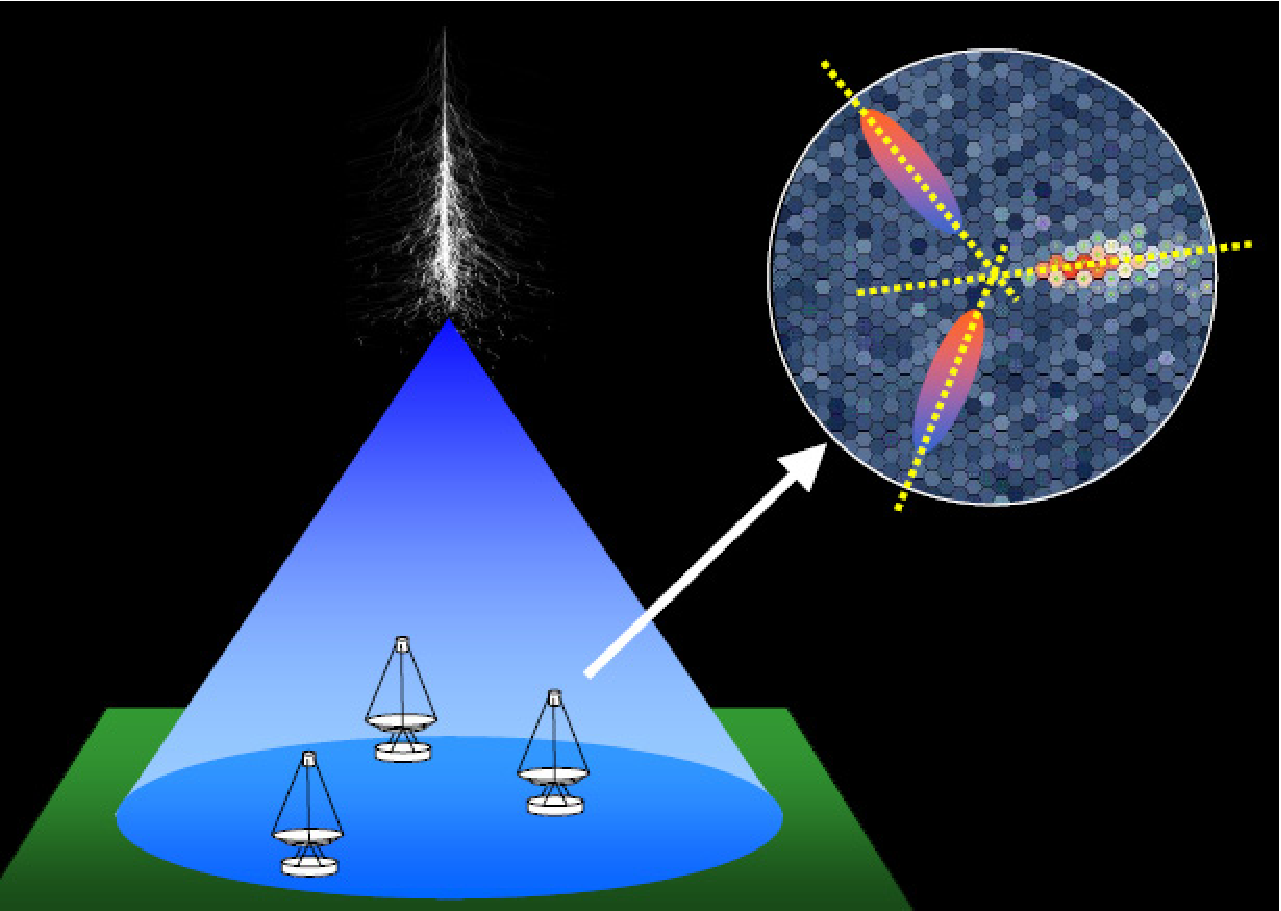}
\caption{\label{fig:multipleIACTs}
A system of more than one Cherenkov telescope~\cite{sitoHess}.}
\end{figure}

Systems of more than one Cherenkov telescope provide a better background rejection, and a 
better angular and energy resolution (see Fig.~\ref{fig:multipleIACTs}) than a single 
telescope.

\begin{figure}
\centering
\includegraphics[width=\textwidth]{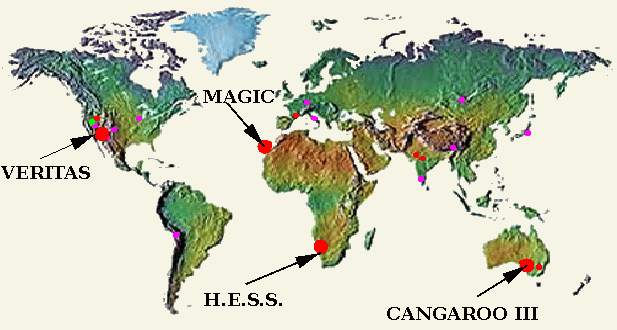}
\caption{\label{fig:IACTs}
Location of the four big IACTs~\cite{sitoHess-desy}.}
\end{figure}

There are four large operating IACTs: CANGAROO III, H.E.S.S., MAGIC and VERITAS (see 
Fig.~\ref{fig:IACTs}), two located in the southern hemisphere and two in the northern 
hemisphere.

\begin{itemize}
\item
CANGAROO (the present setup is called CANGAROO III) is a Japanese and Australian observatory placed near Woomera 
(Australia).
In its final design it will be a system of four telescopes with a surface of 57~m$^2$ each.
The third telescope has been completed in 2004; its precursors, CANGAROO (single telescope) 
and CANGAROO~II (two telescopes), started their activity in 1992 and 1999 respectively.
\item
The H.E.S.S.\ observatory (Fig.~\ref{fig:Hess}) is composed by four telescopes with surface of 
108~m$^2$ each, working since early 2003, while the first of these telescopes is operating 
since summer 2002.
\begin{figure}
\centering
\includegraphics[width=\textwidth]{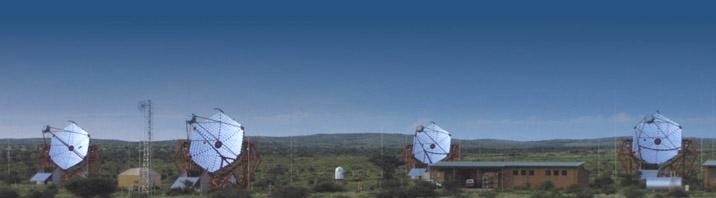}
\caption{\label{fig:Hess}
The H.E.S.S.\ telescopes~\cite{sitoHess-desy}.}
\end{figure}
It is located in the Khomas highlands of Namibia, and it involves several countries, 
Germany and France in particular.
In the future another telescope with a surface of about 600~m$^2$ will be placed in the center 
of the present array.
Among the present detectors, it has the best sensitivity (about 1\% Crab at 5 standard 
deviations in 50 hours of observation) and the best angular resolution (about 0.06 degrees, 
which allows imaging many galactic sources).
The present energy threshold is about 100 GeV at trigger level.
\item
The MAGIC telescope (Fig.~\ref{fig:Magic}) in the Canary Island of La Palma, has a diameter of 
17 m and a reflecting surface of 236~m$^2$, and it is the largest single-dish Cherenkov 
telescope in operation;
due to the largest area it reaches the lowest energy threshold (about 50 GeV).
The collaboration operating MAGIC involves several countries, Germany, Italy, Spain, Finland 
and Switzerland in particular.
Besides the purpose of lowering as much as possible the energy threshold by increasing the 
dish size, the instrument was designed to be able to rapidly slew responding to alerts due to 
transient phenomena (GRBs in particular).
The lightweight construction allows a slewing time of 22~s, three to four times faster than H.E.S.S.
The advantages of a stereoscopic system (as discussed above) motivated the second phase of the 
MAGIC project: the construction of a second 17 m telescope at a distance of about 85 m from 
the first will increase substantially (by a factor of~2) the sensitivity of MAGIC, making it 
similar to the sensitivity of H.E.S.S., and improve the angular resolution from 0.1~degrees to 
about 0.07~degrees.
\begin{figure}
\centering
\includegraphics[width=.9\textwidth]{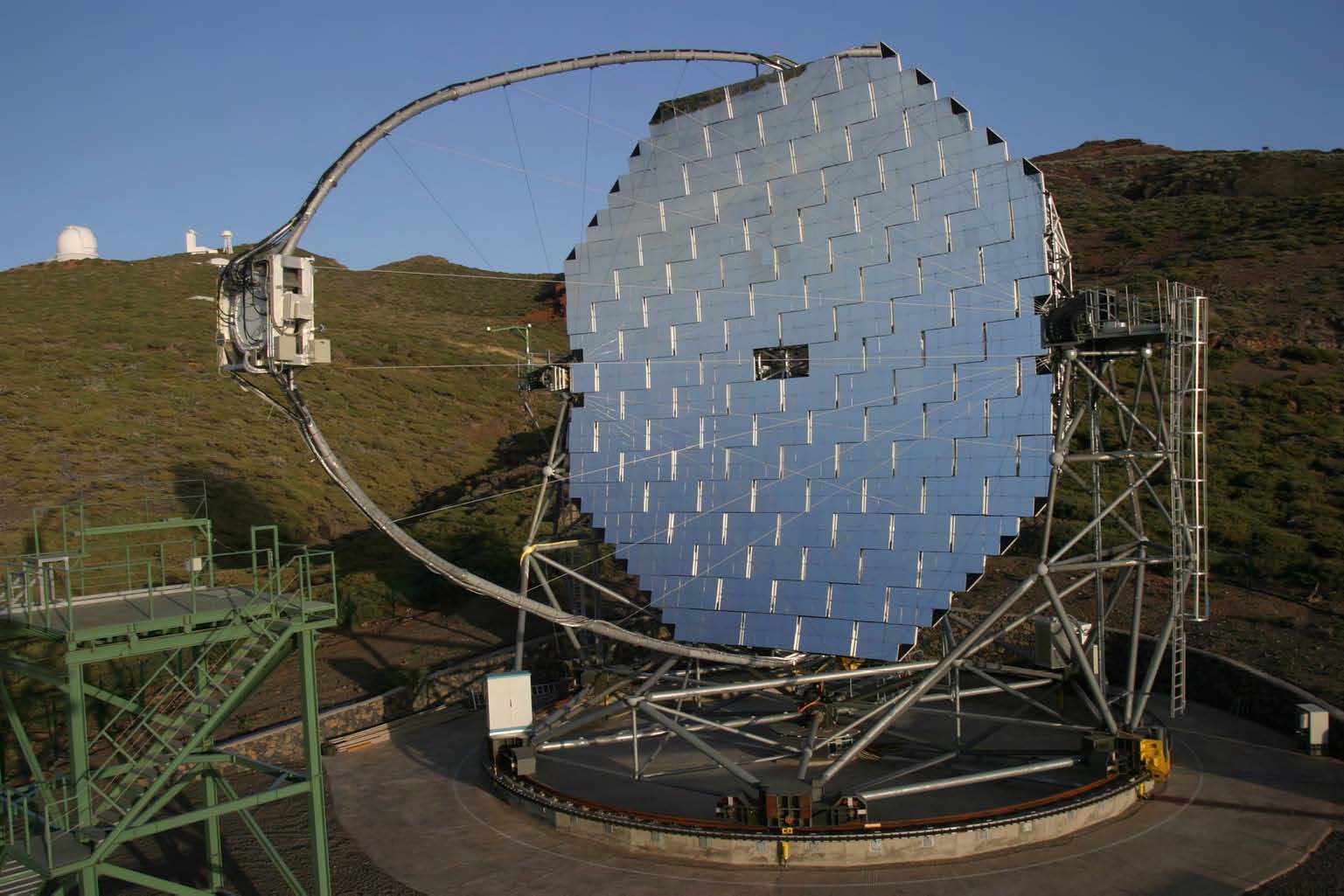}
\caption{\label{fig:Magic}
The MAGIC telescope~\cite{figuraMagic}.}
\end{figure}
\item
VERITAS (\cite{sitoVeritas}; Fig.~\ref{fig:VeritasOperational}) involves Canada, Ireland, the United Kingdom and 
the U.S.A..
The observatory is constituted by an array of four telescopes with a diameter of 12~m and is 
located near Tucson, Arizona.
It is operative since April 2007, but the VERITAS prototype telescope was active since February 2004.
The overall design is rather similar to H.E.S.S..
\begin{figure}
\centering
\includegraphics[width=\textwidth]{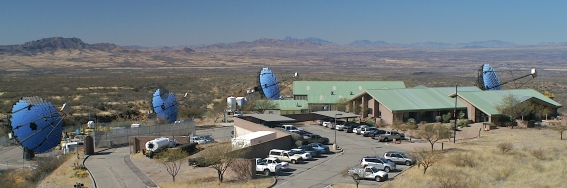}
\caption{\label{fig:VeritasOperational}
The VERITAS site~\cite{sitoVeritas}.}
\end{figure}
\end{itemize}

The main characteristics of the detectors are summarized in Table 1, adapted from 
Ref.~\cite{hinton2}.
Typical sensitivities of H.E.S.S., MAGIC, VERITAS are of about 1\% to 2\% of Crab in 50 hours 
of observation.
\begin{table}
\caption{Main characteristics of currently operating IACTs.
\newline
{\rm The energy threshold given is the approximate trigger-level
threshold for observations close to zenith.
The approximate sensitivity is expressed as the minimum flux (as a percentage of that of the 
Crab Nebula: $\approx 2\times10^{-11}$ photons cm$^{-2}$ s$^{-1}$ above 1 TeV) of a point-like 
source detectable at the $5\sigma$ significance level in a 50~hour observation.}}
\begin{tabular}{|@{\,}l@{\;} | @{}c@{\;} | @{}c@{\;} | @{}c@{\;} | @{}c@{\;} | @{}c@{\;} | @{}c@{\;} | @{}c@{\;} | @{}c@{\;} | @{}c@{\;} |} \hline
Instrument & Lat. & Long. & Alt. & Tels. & Tel. Area & Total A. & FoV
& Thresh. & Sensitivity \\
& ($^{\circ}$) & ($^{\circ}$) & (m) & &
(m$^{2})$ & (m$^{2}$) & ($^{\circ}$) & (TeV) & (\% Crab)
\\\hline
H.E.S.S. & -23 & 16 & 1800 & 4 & 107 & 428 & 5 & 0.1 & 0.7 \\
VERITAS & 32 & -111 & 1275 & 4 & 106 & 424 & 3.5 & 0.1 & 1 \\
MAGIC & 29 & 18 & 2225 & 1 & 236 & 236 (472)$^{\star}$ & 3.5 & 0.05 &
1.6 (0.8) \\
CANGAROO-III & -31 & 137 & 160 & 3 & 57.3 & 172 & 4 & 0.4 & 15 \\
\hline
\end{tabular}
\label{tab:inst}
\begin{tabular}{ @{} l @{\,} l @{} c}
$^{\star}$\ With MAGIC2 (September 2008).     & \hfil    & \hfil
\end{tabular}
\end{table}

The observation at an angle different from the zenith deteriorates the energy threshold.
The dependence of the threshold energy $E_{thr}$ as a function of the angle $\phi$ with 
respect to the zenith can be parametrized as $E_{thr} \simeq E_0 \,\cos^{-2.5}\phi$~\cite{abe} 
where $E_0$ is the threshold for a source at the zenith.
However, due to geometrical factors, the effective area increases when getting away from the 
zenith.

An overlap in the regions of the sky explored by the IACTs allows an almost continuous 
observation of sources placed at mid-latitude;
there is however space for two more installations, one in South America and one (MACE, already 
scheduled for construction) in India.

Negotiations towards a Memorandum of Understanding is ongoing to balance competition and 
cooperation among the present IACTs.

\subsection{Resumé}

\begin{figure} 
\centering
\includegraphics[width=.7\textwidth]{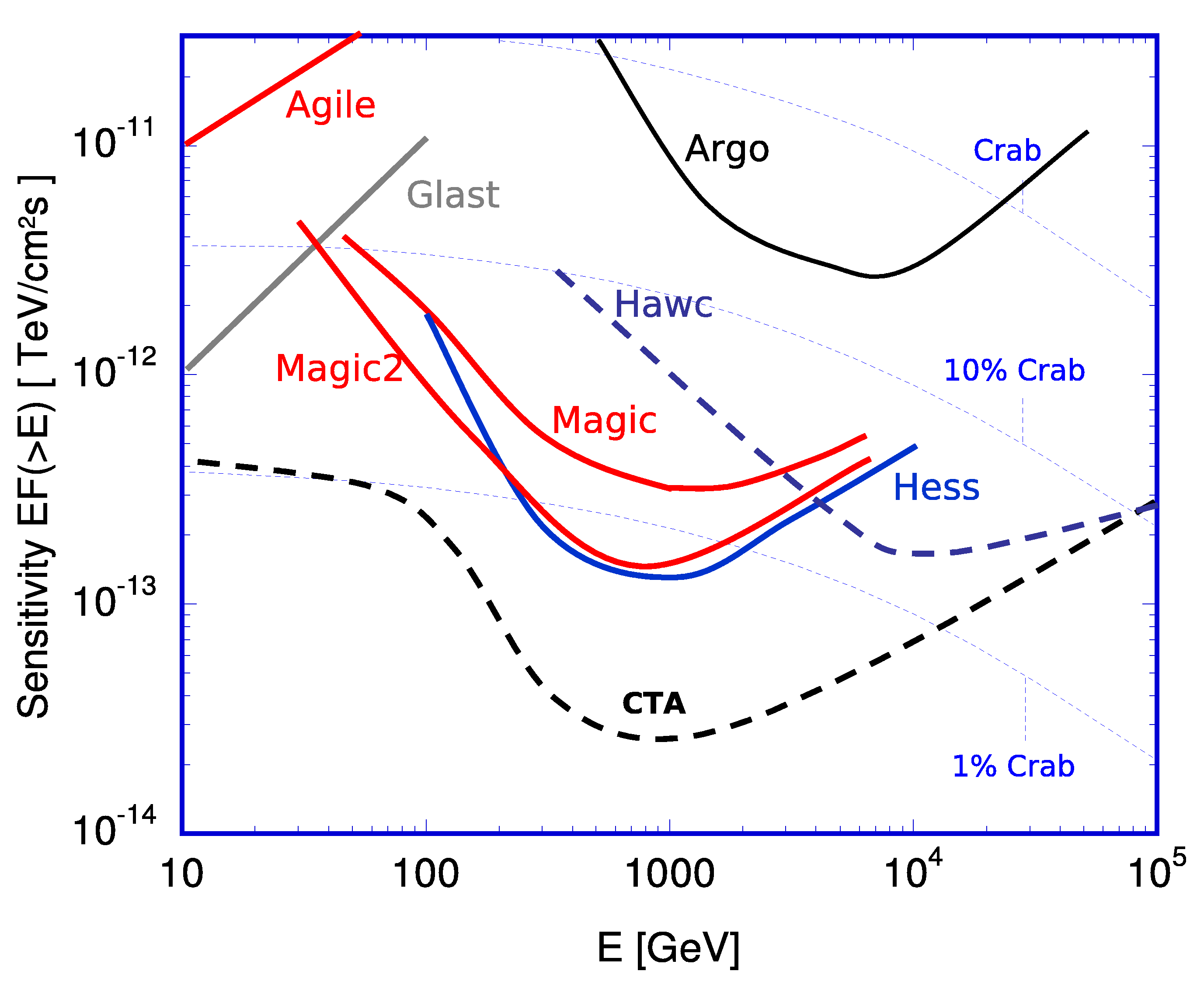}
\caption{\label{fig:Sensitivities}
Sensitivities of some present and future HE gamma detectors~\cite{plotSensitiv}, measured as 
the minimum intensity source detectable at 5\,$\sigma$.
The performance for EAS and satellite detector is based on one year of data taking; for Cherenkov 
telescopes it is based on 50 hours of data.
The sensitivity curve for VERITAS is between MAGIC and MAGIC2.}
\end{figure}

A simplified comparison of the characteristics of the GLAST LAT satellite detector, of the 
IACTs and of the EAS detectors (ground-based) is given in Table~\ref{tab:DetectorsComparison}.
The sensitivities of the above described high-energy detectors are shown in 
Fig.~\ref{fig:Sensitivities}.

\begin{table}
\caption{\label{tab:DetectorsComparison}
A comparison of the characteristics of GLAST, the IACTs and of the EAS particle detector 
arrays.
}
\begin{tabular}{ | l | l | l | l | }
                \hline &&& \\[-.9em]
{\bf Quantity}        & {\bf GLAST}       & {\bf IACTs}          & {\bf EAS}
\\ \hline  &&& \\[-.9em]
Energy range          & 20 MeV -- 200 GeV & 100 GeV -- 50 TeV    & 400 GeV -- 100 TeV
\\         &&& \\[-.9em]
Energy resolution     & 5-10\%            & 15-25\% ($^*$)       & $\sim$ 50\%
\\         &&& \\[-.9em]
Duty Cycle            & 80\%              & 15\%                 & $>$ 90\%
\\         &&& \\[-.9em]
FoV                   & $4 \pi / 5$       & 5 deg $\times$ 5 deg & $4 \pi / 6$
\\         &&& \\[-.9em]
Resolution(PSF)       & 0.1 deg           & 0.07 deg             & 0.5 deg
\\         &&& \\[-.9em]
Sensitivity($^{**}$)  & 1\% Crab (1 GeV)  & 1\% Crab (0.5 TeV)   & 0.5 Crab (5 TeV)
\\        \hline
\end{tabular}
\begin{tabular}{ @{} c @{\,} l @{} c}
($^*$)     & Decreases to 15\% after cross-calibration with GLAST \cite{noimagglast}.                          & \hfil
\\
($^{**}$)  & Computed over one year for GLAST and the EAS, and over 50 hours for the IACTs. & \hfil
\end{tabular}
\end{table}

\section{The emerging HE and VHE gamma-ray sky}

\begin{figure}
\centering
\includegraphics[width=.9\textwidth]{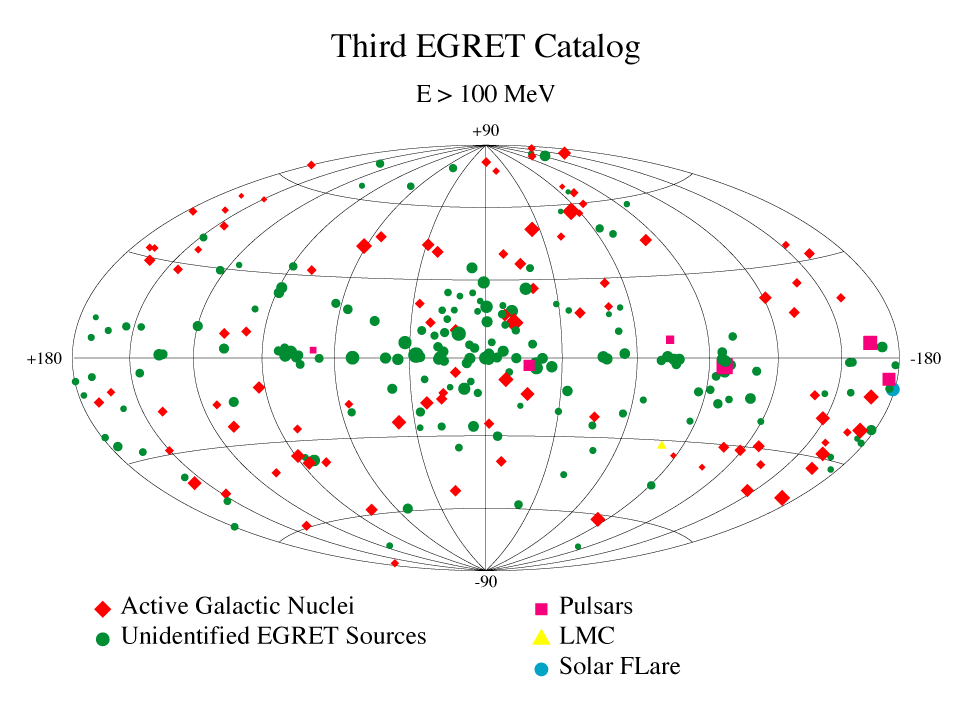}
\caption{\label{fig:EgretCatalog}
Third EGRET catalog (1999) at energies greater than 100~MeV~\cite{3egretCat}.}
\end{figure}

More than one half of the 271 sources detected at energies $>$100~MeV in the third EGRET catalog of~1999 (see Fig.~\ref{fig:EgretCatalog}) are still unidentified.

\begin{figure}
\centering
\includegraphics[width=.95\textwidth]{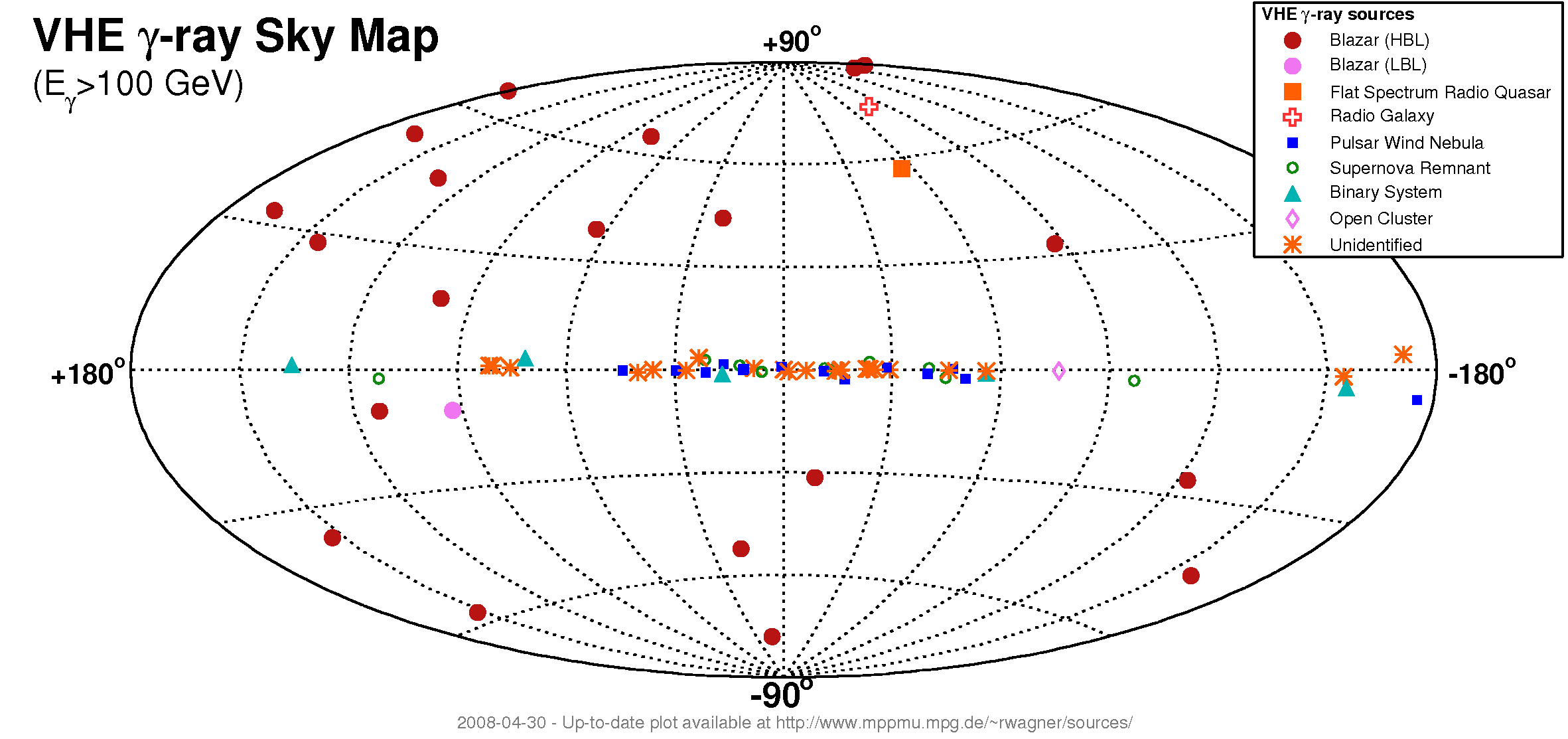}
\caption{\label{fig:GammaSkyComparison}
Known sources in the VHE sky in 2008~\cite{sourcesWagner}.}
\end{figure}

Thanks mostly to Cherenkov telescopes, a large amount of VHE sources has been detected and identified (see Fig.~\ref{fig:GammaSkyComparison}).
When this review has been written 
(June 2008), 76~VHE sources had been detected, which operate as cosmic particle accelerators, and the $\gamma$-rays they emit trace information on the primary electrons or nuclei.
Therefore, it is important to determine the nature of the primary particles and their spatial and momentum distribution.
Among these~sources, 8~are supernova remnants (SNRs), 14~are pulsar wind nebulae (PWNe), one is the galactic centre,
4~are binary systems, one is a VHE pulsar, 25~are unidentified galactic sources, and 23~are active galactic nuclei.

\subsection{Galactic cosmic-ray accelerators}

\begin{figure}
\centering
\includegraphics[width=\textwidth]{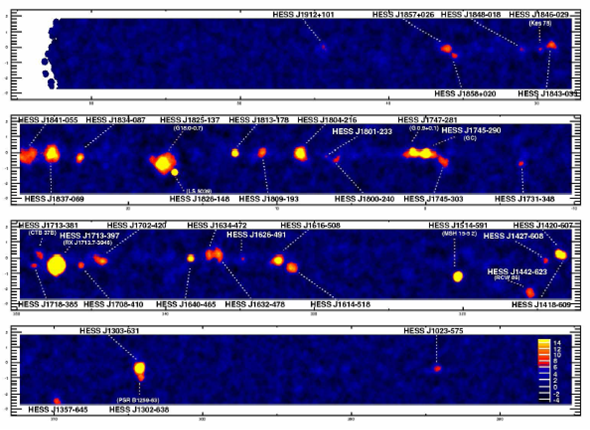}
\caption{\label{fig:galscan}
Known galactic VHE sources from the survey of the galactic plane by the H.E.S.S.\ 
telescope between 2004 and 2007~\cite{HESS:scanicrc}.}
\end{figure}

The recent dramatic growth (by a factor of $\sim$10) in the number of known galactic VHE sources is largely a consequence of the survey of the galactic plane conducted with the southern-located H.E.S.S.\ between 2004 and 2007~\cite{aharonian2006a}.
Fig.~\ref{fig:galscan} shows the current extent of this scan, which now covers essentially the whole inner Galaxy:
$-85^{\circ}$$<$$l$$<$$60^{\circ}$,$-2.5^{\circ}$$<$$b$$<$$2.5^{\circ}$~\cite{HESS:scanicrc}.
Further galactic sources, accessible from the northern hemisphere, were subsequently observed with the MAGIC telescope (e.g., Ref.~\cite{albert2007a}), and even more by H.E.S.S. 
Proposed counterparts of such galactic VHE sources include supernova remnants, PWNe, and accreting binaries.
Whatever their detailed nature, it is expected that galactic VHE sources are related to evolutionary endproducts of massive, bright, short-lived, stellar progenitors.
Hence, these galactic VHE sources are immediate tracers of the current star formation.

About half of the  currently known galactic TeV sources remain unidentified.
This is in part due to the difficulty of identifying extended sources with no clear sub-structure.
Nonetheless, several methods of identification have been successfully applied and the situation is much more favourable than that in the GeV band where only one galactic source class (pulsars) has been unambiguously identified.

\subsubsection{Supernova Remnants}

Galactic cosmic rays have long been suspected to be produced at supernova shock fronts via 
diffusive acceleration.
If the observed VHE $\gamma$-rays were found to be generated through the hadronic channel, 
via $\pi^0$ decay following {\sl pp} interaction with the dense molecular clouds embedding the 
short-lived supernova progenitor, then the acceleration by supernovae of nuclei to energies of 
the order of the knee in the cosmic-ray spectrum would be proven 
(e.g., Ref.~\cite{torres2003}).
However, it is difficult to disentangle the hadronic VHE component from the leptonic one, 
produced by IC scattering of interstellar radiation field photons (in the inner Galaxy) or 
cosmic microwave background photons (in the outer Galaxy) off ultrarelativistic electrons (e.g., 
Ref.~\cite{porter2006}), by measuring $\gamma$-rays over only a decade or so in energy.
The VHE data of RX\,J1713.7-3946 can be explained in terms of either channel, 
leptonic/hadronic if the relevant magnetic field is low/high 
($B$\,$\sim\,$10/100\,$\mu$G~\mbox{\cite{aharonian2006b, berezhko-volk2006}}).
Data in the $\sim\,$0.1--100~GeV band, such as those to be provided by AGILE and GLAST, are 
clearly needed to discriminate between the two channels.
(For Cas~A the high magnetic field, $B\sim\,$1~mG, suggests a mostly 
hadronic~\cite{berezhko2003} VHE emission~\cite{ona2007}.)
The complementarity between AGILE and GLAST on one side and the IACTs on the other side might 
be the key for important discoveries in the future~\cite{tavani}.

Whatever the details, the detection of photons with  $E \gtrsim$\,100~TeV from 
RX\,J1713.7-3946 is a proof of the acceleration of primary particles in supernova shocks to 
energies well above 10$^{14}$~eV.
The differential VHE spectral index is $\sim\,$2.1 all across this SNR, suggesting that the 
emitting particles are ubiquitously strong-shock accelerated, up to energies 
$\sim\,$200/100~TeV for primary cosmic protons/electrons if the hadronic/leptonic channel is 
at work~\cite{aharonian2007}.
This is getting close to the knee of the cosmic-ray spectrum; this fact might signal the high-energy end 
of the galactic cosmic-ray distribution (e.g.\ Ref.~\cite{blasi2005}).

Circumstantial evidence supports a hadronic origin for at least part of the VHE emission.
In several expanding SNRs the X-ray brightness profile behind the forward shock is best 
explained as synchrotron emission from energetic electrons in strong magnetic fields, 
$B$\,$\sim\,$${\mathcal O}$$(10^2)$\,$\mu$G, i.e.~$\sim$\,100 times larger than typical 
interstellar medium values.
Such a large amplified magnetic field disfavors the IC interpretation of the VHE data.
Furthermore, in the remnant HESS\,J1834-087, the maximum of the extended VHE emission 
correlates with a maximum in the density of a nearby molecular cloud~\cite{albert2006a}
-- which suggests hadronic illumination of the target molecular cloud.
A similar situation holds for IC\,443~\cite{ic443} and W\,28~\cite{w28}, both of which appear 
to have emission correlated with available target material rather than with the radio/X-ray 
emission of the SNR itself, suggesting that the VHE emission may arise from hadronic 
interactions in/around the SNR.


\subsubsection{Pulsars and Pulsar Wind Nebulae}

Pulsars are understood as rapidly spinning neutron stars (with periods between
about 1 ms and 1 s) with extremely strong magnetic fields.
(Neutron stars are stellar remnants of supernova explosions.)
Pacini~\cite{pacini1967} argued that such stellar remnants could power SNRs like the Crab nebula, and predicted that they could be observable at radio frequencies.
Rotation generates an induced electric field that overcomes gravity, so charges are pulled out from the neutron star surface (where a layer of plasma still survives), in the form of a relativistic wind that carries most of the pulsar's rotational energy, filling the (corotating) magnetosphere with plasma.
Electron-positron pairs originate in the magnetosphere via interaction with the magnetic field;
they may escape through the polar cap regions in the form of a wind that eventually terminates in the surrounding interstellar medium.

Gamma-ray radiation from rotation-powered pulsars can be produced through several radiation mechanisms in three physically distinct regions: magnetosphere, relativistic wind, and the surrounding nebula.
As for the latter, the pair-wind termination shock into the circumstellar medium establishes a standing reverse shock: this leads to the formation of a nonthermal (synchrotron and Compton) nebula whose spectrum extends from radio to VHE gamma­rays.

To date there is no wide consensus on the physical mechanism for magnetospheric emission.
Objects emitting at X-ray to $\gamma$-ray energies are critical to this scope, because this 
energy band is where most of the radiative luminosity is observed and it is linked with the 
pair wind that is the dominant mode of energy deposition in the circum-pulsar environment.
The most crucial issue about this high-energy emission of pulsars is the location of the 
acceleration region in the magnetosphere:
is it in the outer gap (e.g., Ref.~\cite{cheng1986,romani1996}; review by~\cite{hirotani2005}) 
or in the polar cap (e.g., Ref.~\cite{daug-hard1982-96};
review by~\cite{baring2004}).
Different radiative signatures are predicted in the two cases.

The key difference between the two models is the location of the acceleration zone: near the 
pulsar surface, i.e.\ at the magnetic poles (polar-cap model), or further out from the star 
surface, where the local field is reduced by several orders of magnitudes with respect to 
the field at the star's surface (outer-gap model).
In terms of observables, this distinction translates into the manifestation (or lack thereof) 
of extremely strong magnetic fields.

In polar-cap models, for strong values of the magnetic field (10$^{11}$G~$\lesssim$\,$B$\,$\lesssim$~4$\times$\,\!$10 ^{13}$G),
the conversion of photons into $e^+e^-$ pairs generates cascades that strongly attenuate 
supra-GeV emission in Crab-like or Vela-like (``canonical'') pulsars, generating 
steeper-than-exponential (``super-exponential'') cutoffs in the 10 MeV -- 10 GeV band%
\footnote{For magnetars, pulsars with still higher fields, 
	$B > 4 \times 10^{13}$G, no emission is predicted above 
	$>$100 MeV.}.
In outer-gap models pair creation is obtained through the 
$\gamma\gamma$\,$ \rightarrow$\,$e^+e^-$ process involving surface thermal X-rays as targets, 
and the absorption of $\gamma$-rays in the magnetic field is not sufficient for the 
development of pair cascades and does not significantly distort the emitted spectrum.

Discrimination between different polar-cap versus outer-gap models of pulsar magnetospheric 
emission is one clear goal of VHE astrophysics.

The current observational situation is exemplified by the cases of the PSR\,B1951+32 and 
Vela pulsars~\cite{aharonian2006c,albert2007b}.
In the latter case upper limits to the pulsed emission imply a cutoff energy $E_c\,<\,$32~GeV.
In both, Vela and PSR\,B1951+32, IC emission at TeV energies as predicted by outer-gap models 
is severely constrained, although not all outer-gap models are ruled out.
Deeper sensitivities can test the models further, and certainly no test of these models in the 
range 10--100 GeV has yet been achieved.

Millisecond pulsars, that have lower magnetic fields, in polar-cap models could have a cutoff 
at $\sim$\,100~GeV (i.e., $E_c\propto B^{-1}$), hence their (pulsed) VHE emission could be 
relevant.
Detecting it would be a test of the polar-cap theory~\cite{harding2005}.

\begin{figure}
\centering
\includegraphics[width=.9\textwidth]{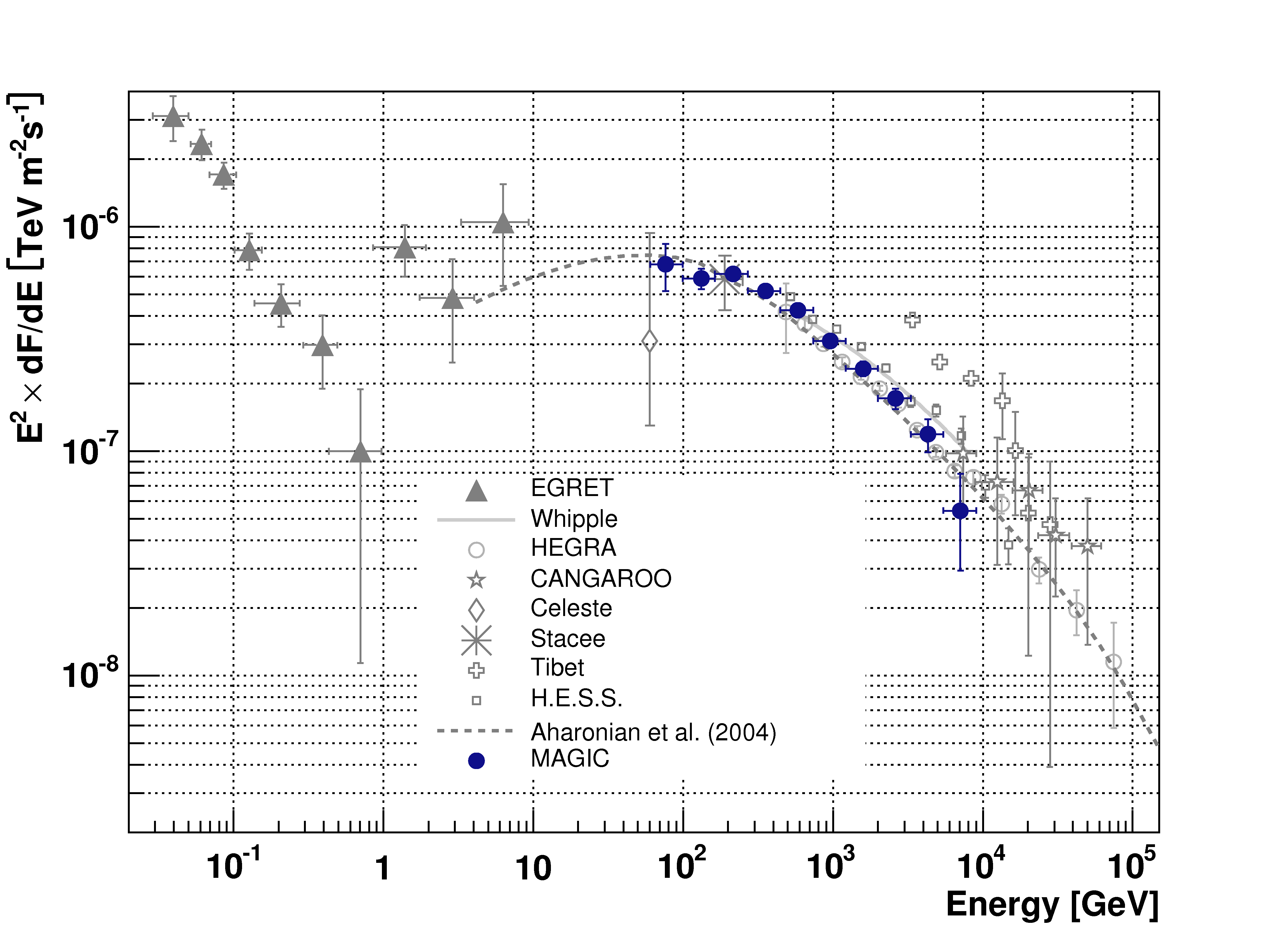}
\caption{\label{fig:Crab}
The spectrum of the Crab Nebula in the HE region~\cite{albert2007c}.}
\end{figure}

The Crab Nebula was the first VHE source to be discovered \cite{weekes1989}.
It still is the brightest steady emitter in the VHE sky, that is used as a calibration candle.
The $\gamma$-ray emission from the Crab is dominated by the pulsed emission from the rotating 
pulsar below GeV energies, and by the steady emission from the nebula above GeV energies.
Its broad-band spectrum is double-peaked (see Fig.~\ref{fig:Crab}), a feature common to all 
pulsar wind nebulae:
the two components are usually attributed to synchrotron radiation and its Compton scattering 
off the parent relativistic electrons that emerge from the termination shock of the 
pulsar wind.
Its spectral energy density decreases in the GeV region, but it has a turn-over and starts to 
be again visible at about 100~GeV.
The Crab Nebula has been observed extensively from the radio energy range up to about 70~TeV.
No pulsed (magnetospheric) VHE emission was originally found in MAGIC data from previous 
observational campaigns, that implies $E_c<$\,50~GeV~\cite{albert2007c}.
The steady nebular spectrum, measured in the about 0.03-30~GeV range by EGRET and in the about 
0.06-70~TeV by several IACTs, shows a bump that starts to dominate at about 1~GeV and peaks at 
about 50~GeV:
this component results from IC scattering, by the synchrotron-emitting electrons, of softer 
photons in the shocked wind region -- i.e., synchrotron, far infrared/millimeter or cosmic microwave background photons.
In spite of its detected IC $\gamma$-ray emission, however, the Crab Nebula is not an 
effective IC emitter as a consequence of its high nebular magnetic field ($B$$\sim$$0.4$mG).

Very recently, the MAGIC collaboration has reported on the detection of the Crab 
pulsar~\cite{atel_crab}.
The source was observed from October 2007 until February 2008 for 40 hours, of which 22 hours 
of data were recorded at optimal weather condition and small zenith angles.
The observation took place after the installation of a new trigger system which lowered the 
threshold from 50-60 GeV down to 25 GeV.
The analysis reveals a pulse profile that shows a very clear signal at a 6.4\,$\sigma$ level 
coinciding also with an optical signal concurrently recorded by a special photosensor in the 
camera center.
Both peaks in the light curve, the main pulse and the interpulse, are of about equal 
intensity.
Both are clearly visible and are in phase with the emission at all other lower energies 
including gamma energies above 100\,MeV as observed by EGRET.

The $\gamma$-ray energy cutoff will be deduced for the first time, allowing a direct 
comparison with emission models~\cite{atel_crab}.

\begin{figure}
\centering
\includegraphics[height=.39\textwidth]{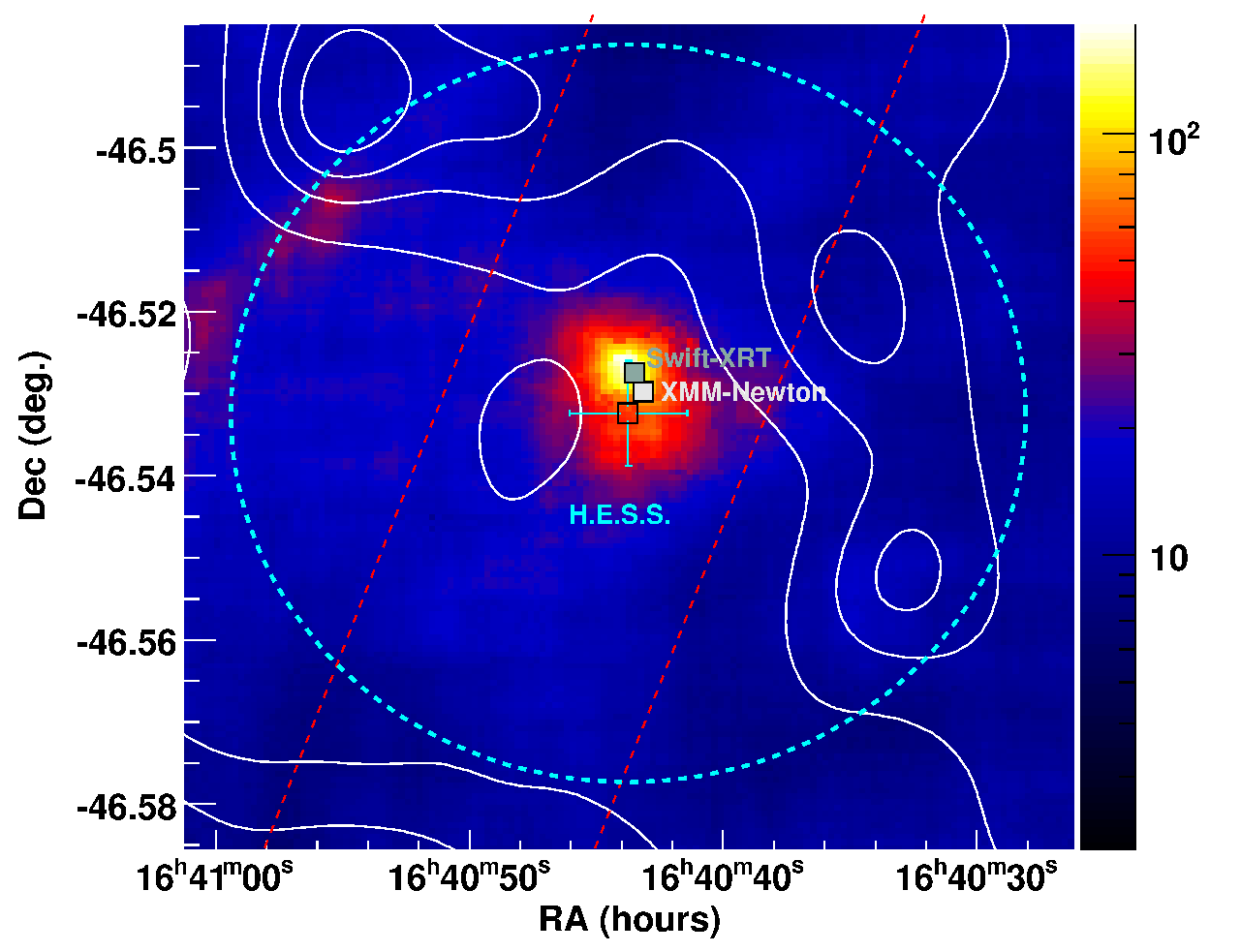}
\includegraphics[height=.39\textwidth]{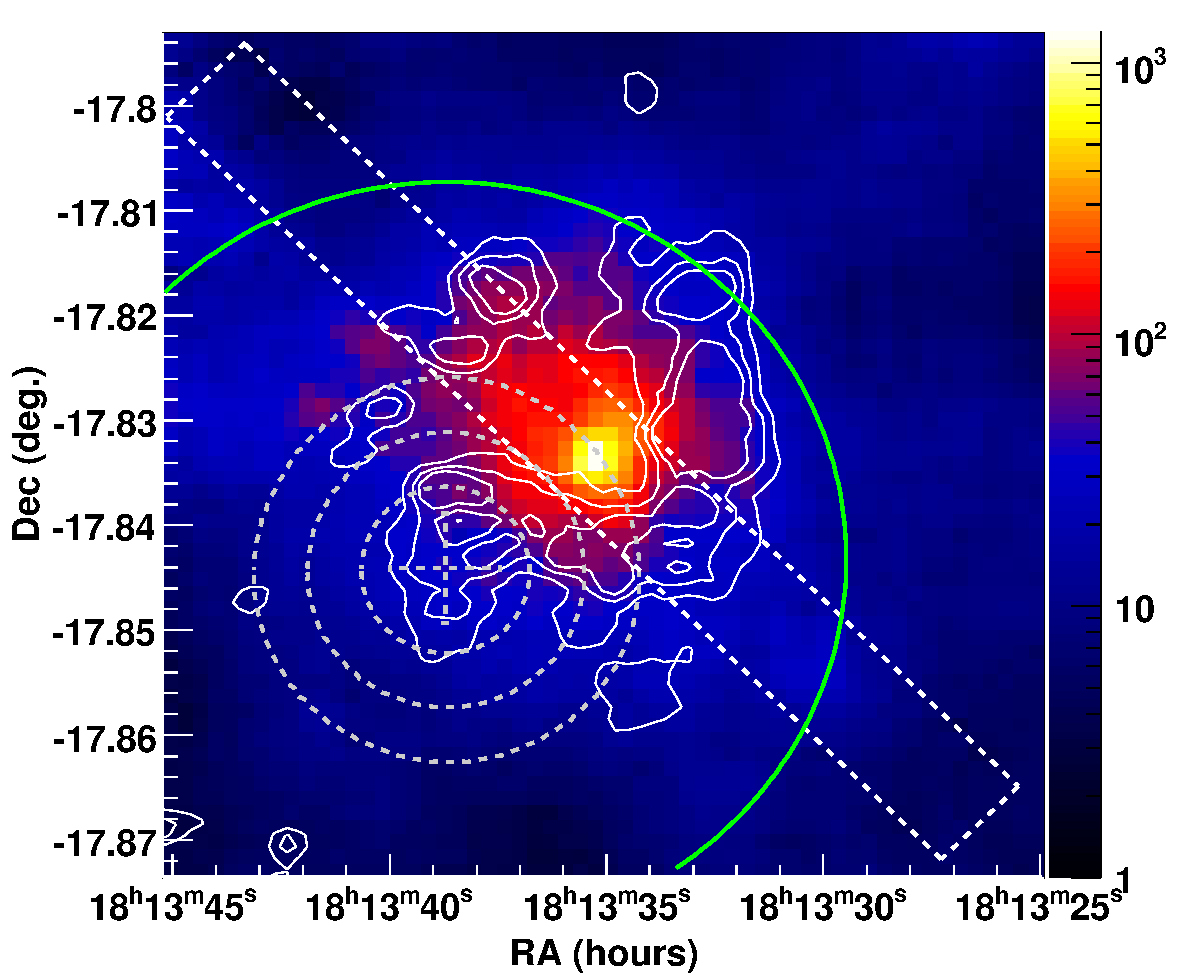}
\caption{\label{fig:pwnHessNew}
Sky map around HESS\,J1640-465~\cite{HESSJ1640} (left) and 
around HESS\,J1813.178~\cite{HESSJ1813} (right).}
\end{figure}

Pulsar wind nebulae -- PWNe are pulsars displaying a prominent nebular emission.
They currently constitute the most populated class of identified galactic VHE sources 
(7 identifications -- see e.g.~Ref.~\cite{gallant2006}).
Even for the unidentified extended VHE sources detected in the H.E.S.S.\ galactic plane 
survey, there is a clear excess of VHE nebulae positionally coincident with high-luminosity 
radio pulsars.
So probably the class of PWN is even more populated than it appears now.
Some PWNe have an X-ray counterpart: HESS\,J1640-465~\cite{HESSJ1640} and HESS\,J1813.178~\cite{HESSJ1813} (see Fig.~\ref{fig:pwnHessNew});
the second source has a radio shell~\cite{HESSJ1813radio}.

Also based on the observed X-ray emission, we can assume that the VHE emission of PWNe is likely of leptonic origin.
Let us consider the case of HESS\,J1825-137, for which spectra have been measured in spatially separated regions~\cite{aharonian2006d}.
In these regions, the VHE spectra become steeper with increasing distance from the pulsar, and the VHE morphology is similar to the X-ray morphology: furthermore the low derived magnetic field (few~$\mu$G) implies that synchrotron X-ray emission is due to electrons of energy higher than the $\gamma$-rays.
This suggests that the observations can be modeled in terms of very energetic electrons that efficiently lose energy via synchrotron losses, aging progressively more rapidly as they are farther away from the acceleration site, and produce VHE $\gamma$-rays via IC~scattering.

The PWNe seen by H.E.S.S.\ are extended sources with dimension of order of 10~pc, in several cases displaced with respect to the pulsar they contain.
This may be due to the supernova explosion occurring in an inhomogeneous medium, and the resulting asymmetric reverse shock displacing the PWN in the direction away from the higher-density medium.
Such an offset may be typical of older PWNe, e.g.\ Vela (spin-down age: 11kyr) versus Crab (1.2 kyr) (e.g.,~\cite{gallant2006}).

\subsubsection{Binary systems}

In both SNRs and PWNe particle acceleration proceeds on the parsec distance scales in the shocks formed in interactions of either the ejecta of supernovae or pulsar winds with the interstellar medium.
A different population of much more compact particle accelerators, which has been revealed by current IACTs, is formed by the TeV binaries.
These systems contain a compact object
-- either a neutron star or a black hole --
that accretes, or interacts with, matter outflowing from a companion star:
hence they are VHE-loud X-ray binaries.
Four TeV binaries have been detected so far: 
PSR\,B1259-63~\cite{aharonian2005a}, LS\,5039~\cite{aharonian2005b, aharonian2006e}, LS\,I\,+61\,303~\cite{albert2006b}, and Cyg\,X-1~\cite{albert2007d}.

PSR\,B1259-63 is powered by the rotation energy of its young 48~ms pulsar, and strong VHE emission is observed from this system in pre- and post-periastron phases when the relativistic pulsar wind collides with the dense equatorial wind blowing from the companion star, which displays a rapid rotation.

LS\,I\,+61\,303 and LS\,5039 may share a similar structure.
The former is composed of a compact object and a rapidly rotating star in a highly eccentric orbit.
Its VHE emission, whose variability constrains the emitting region of LS\,I\,+61\,303 to be larger than the size of the binary system, appears correlated with the radio emission and does not peak at periastron, where the rate $\dot M$ of the transferred mass is expected to be largest.
This picture favors an IC origin of VHE emission, as probably the most efficient at the relatively large scales of the system at peak emission.
LS\,5039 clearly shows periodicity of 3.9 days from the gamma-ray data alone (in agreement with the optical):
the VHE spectrum varies as a function of phase, getting softer when the compact object is hidden by the companion, that may indicate photon-photon absorption or cascading.

The emission from Cyg\,X-1 (for which there is evidence with a significance $\gtrsim$4$\,\sigma$) is point-like and excludes the nearby radio nebula powered by the relativistic jet.
If confirmed, \mbox{Cyg\,X-1} is the first stellar-mass ($\sim$13$\,M_\odot$) black hole, and hence the first established accreting binary, established as a VHE source.

\subsubsection{The centre of the Milky Way}

\begin{figure}
\centering
\includegraphics[width=.9\textwidth]{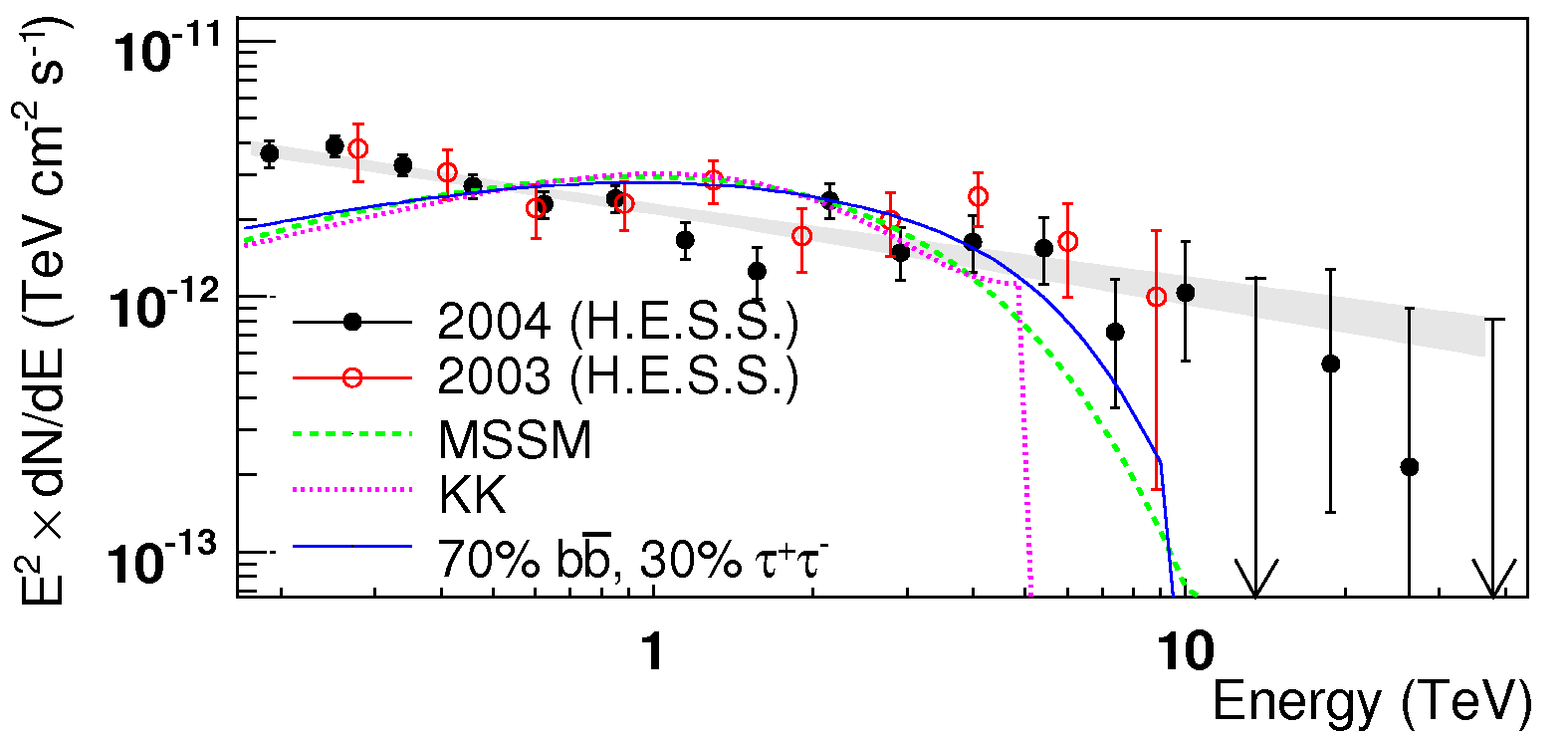}
\caption{\label{fig:GCspectrum}
Differential gamma spectrum from the galactic centre~\cite{aharonian2006f}.}
\end{figure}

The possibility of indirect dark matter detection through its annihilation into VHE 
$\gamma$-rays has aroused interest to observe the centre of the Milky Way.
H.E.S.S.\ and MAGIC observed the galactic centre, measuring a steady flux consistent with 
a differential power-law wiyh a spectral index of about 2.2 (see Fig.~\ref{fig:GCspectrum}), 
up to energies of about 20~TeV with no apparent cutoff~\cite{aharonian2004, albert2006c}.
The $\gamma$-ray source is steady even during X-ray flares and behaves like a point source.
Within the error circle of the measurement of the central source HESS\,J1745-290 
(the position of the centre of the Milky Way is determined within a systematic error of 6'' and 
a statistical error of 6'', too) there are three 
compelling candidates for the origin of the VHE emission: the shell-type SNR Sgr\,A\,East, 
the newly discovered PWN G\,359.95-0.04, and the supermassive black hole Sgr\,A$^{\star}$ 
itself.
Plausible radiation mechanisms include IC scattering of energetic electrons, the decay of 
pions produced in the interactions of energetic hadrons with the interstellar medium or dense 
radiation fields, and curvature radiation of UHE protons close to Sgr\,A$^{\star}$.
These considerations disfavor dark matter annihilation as the main origin of the detected 
flux, whereas a more conventional astrophysical mechanism is likely to be at work (e.g., 
Ref.~\cite{aharonian2006f}).
Furthermore, the lack of flux variability on hour/day/year timescales might suggest that 
particle acceleration occurs in a steady object, such as a SNR or a PWN, and not in the 
central black hole.

The diffuse emission correlates with molecular clouds and suggests an enhanced cosmic-ray 
spectrum in the galactic centre~\cite{aharonian2006g}.
Its morphology and spectrum suggest recent in situ cosmic-ray acceleration:
because the photon indexes of the diffuse emission and of the central source HESS\,J1745-290 
are similar, the latter source could be the accelerator in question.

\subsubsection{Mystery sources and {Cygnus region}}

\begin{figure}
\centering
\includegraphics[width=.9\textwidth]{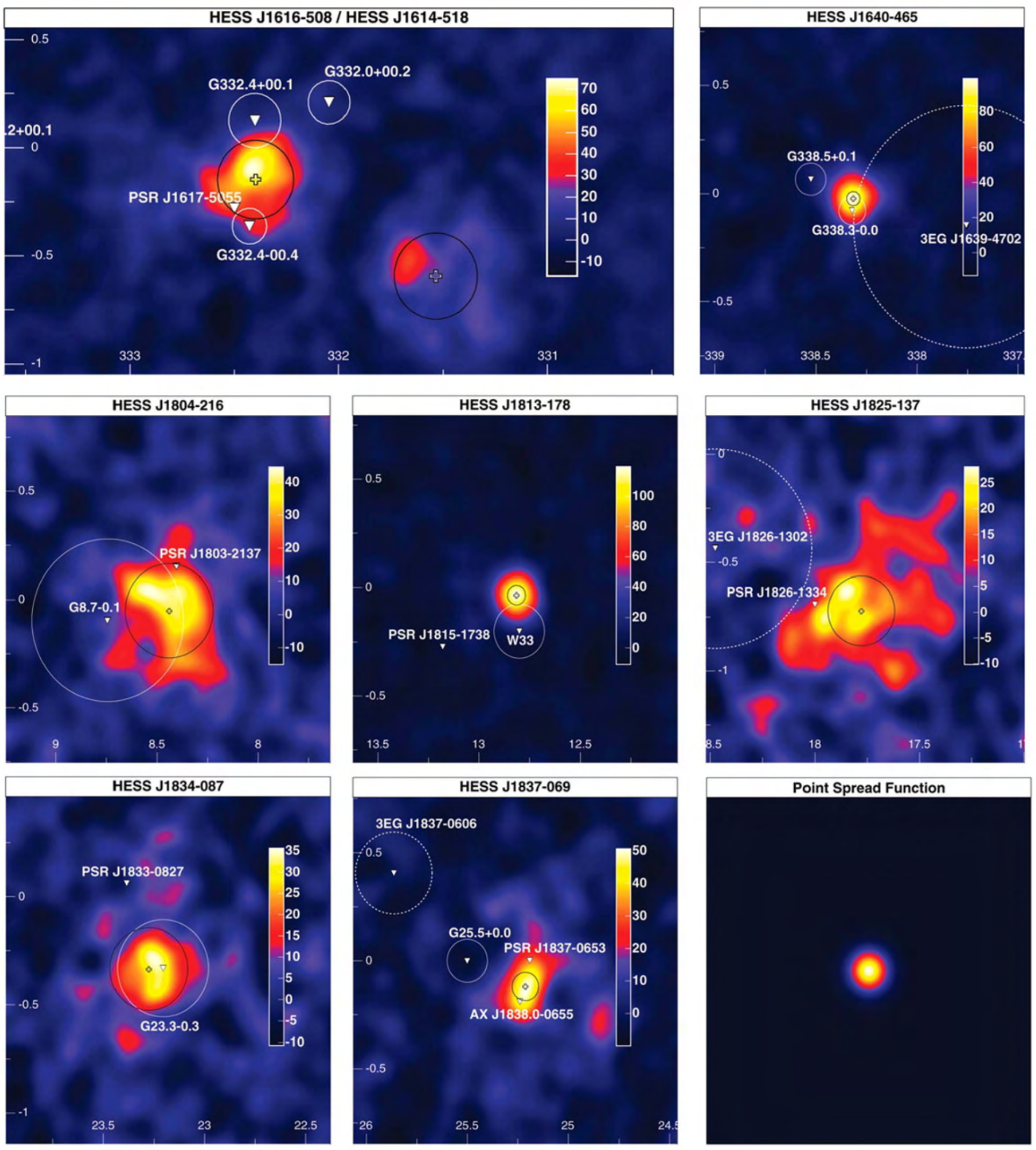}
\caption{\label{fig:VHEw/oXradio}
Some H.E.S.S.\ sources without a X or radio counterpart~\cite{noXhess}.}
\end{figure}

There are examples of galactic VHE sources which do not have a counterpart in the X or radio 
energy bands (see Fig.~\ref{fig:VHEw/oXradio}).
This situation is the likely outcome of experimental and physical considerations.
As an example of the former, extended sources may have different morphologies at different 
frequencies;
as for the latter, sources my emit substantially in the GeV-TeV range and negligibly in other 
spectral bands (``dark accelerators"), e.g.\ TeV J2023+4130.
Whatever their detailed nature, a reasonable expectation is that these sources, too, are 
related to compact stellar endproducts and descend from massive, bright, short-lived, 
progenitor stars.
Hence, these dark accelerators would still be linked to current star-formation activity.

\begin{figure}
\centering
\includegraphics[width=.7\textwidth]{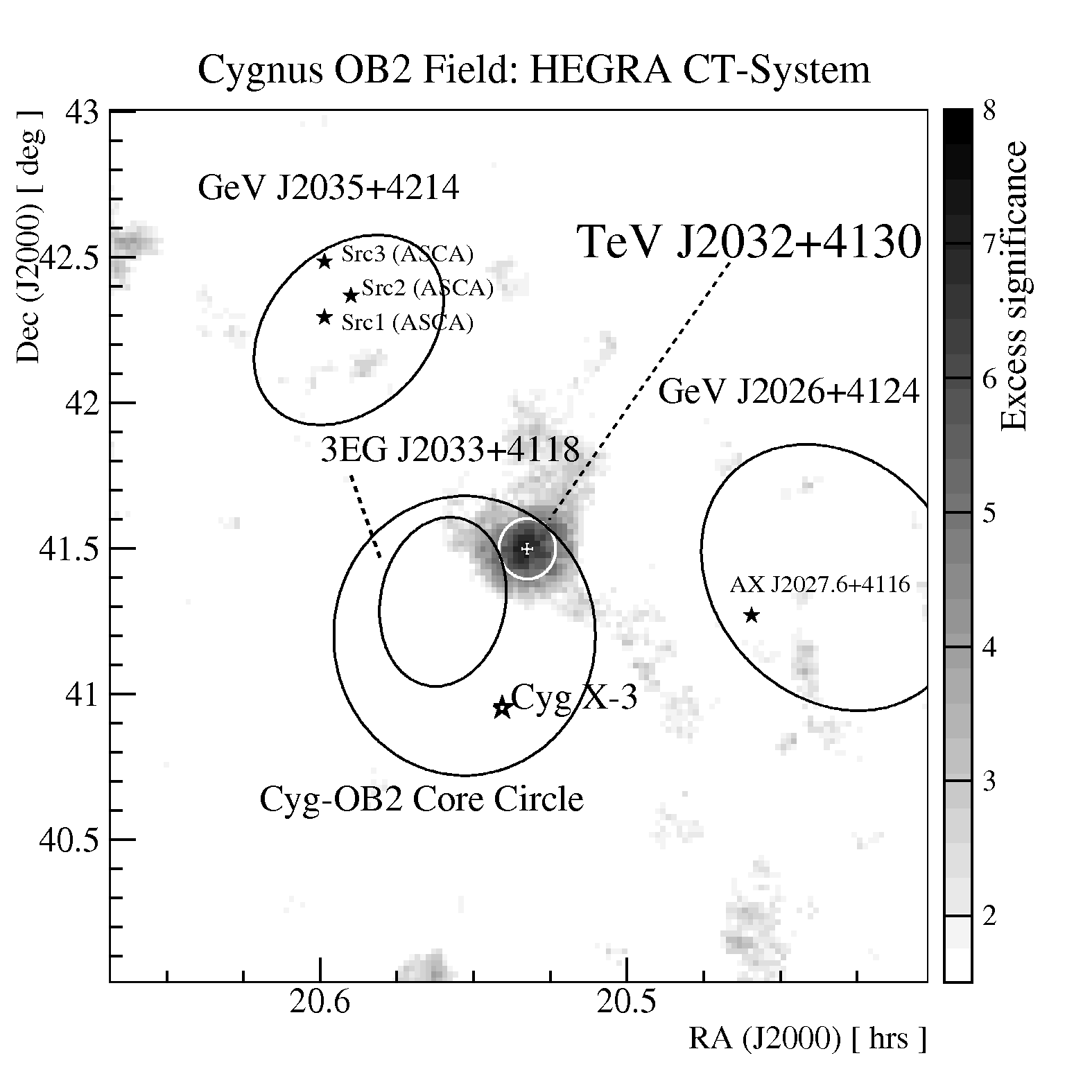}
\caption{\label{fig:J2023}
The source TeV\,J2023+4130 as seen by HEGRA~\cite{HegraJ2023}.}
\end{figure}

MAGIC, HEGRA and Whipple observed the source TeV\,J2023+4130 (see Fig.~\ref{fig:J2023}).
Whipple detected it at 6.1\,$\sigma$ with a flux corresponding to 8\% of the Crab 
flux~\cite{WhippleJ2023};
MAGIC observed it at 6\,$\sigma$ with a flux 4\% of the Crab flux, a size of 6' and a spectral index 
of 1.8~\cite{MAGICJ2023}, while HEGRA detected it at 7\,$\sigma$ with a flux 5\% of the Crab 
flux, a size of 6' and a spectral index of 1.9~\cite{HegraJ2023}.

\begin{figure}
\centering
\includegraphics[width=.9\textwidth]{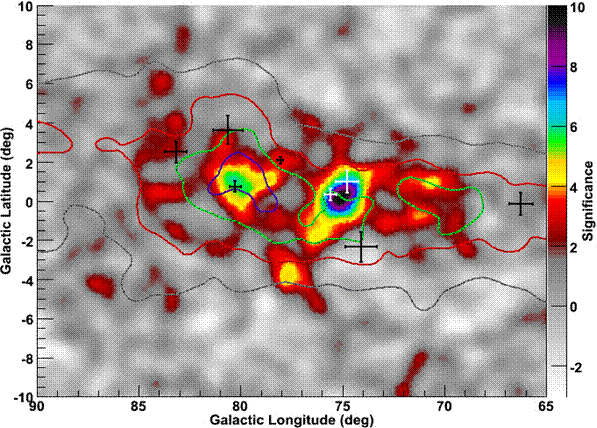}
\caption{\label{fig:MilagroVHEs}
The sources MGRO\,J2031+41 and MGRO\,J2019+37 as seen by MILAGRO (picture from the first paper 
cited in Ref.~\cite{MilagroVHEs}).}
\end{figure}

\begin{figure}
\centering
\includegraphics[width=.7\textwidth]{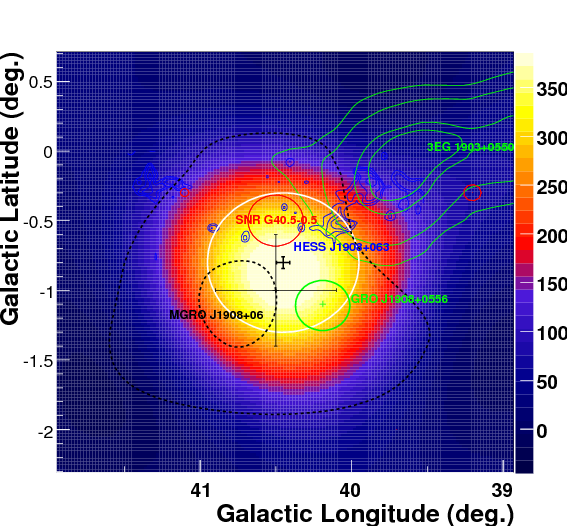}
\caption{\label{fig:MilagroHESS}
The source MGRO\,J1908+06~\cite{j1908-hess}.}
\end{figure}

The MILAGRO telescope discovered the three sources MGRO\,J2031+41, MGRO J2019+37, and 
MGRO\,J1908+06~\cite{MilagroVHEs} (see Fig.~\ref{fig:MilagroVHEs}).
The second source was seen also by the EAS detector TIBET AS-$\gamma$~\cite{TibetICRC}.
The source MGRO\,J1908+06 is still unidentified;
the H.E.S.S.\ observations in its direction are shown in Fig.~\ref{fig:MilagroHESS}.

Explanations for these unidentified sources range from exotic to conventional.
One of the former type is that these sources originate from the annihilation of dark matter
in localised clumps, i.e.\ subhalos within the Galaxy's halo.
A more conventional explanation is that they originate from the collision of cosmic hadrons 
($pp$ interaction), which entails emission mostly in the VHE band - contrary to electrons which 
typically produce comparable fluxes by synchrotron emission.
However, even in this conventional scenario, the acceleration site for these hadrons remains 
unknown (see~\cite{salsa2008, druah2008}).

\subsection{Extragalactic sources of VHE $\gamma$-rays}

\subsubsection{Satellites of the Milky Way}

Outlooks for WIMP annihilation detection in Draco by current IACTs are not very promising:
for a neutralino mass $m_\chi$=100~GeV and a variety of annihilation modes, and in the 
favorable case of a maximal (cuspy) inner halo profile, VHE detection can occur with 40h by a 
IACT if the average value of the neutralino cross section times the velocity is 
$<$\,\!$\sigma$$v$$>$\,$\gtrsim$10$^{-25}$~cm$^3$s$^{-1}$, which is somewhat larger than the 
maximum value for a thermal relic with a density equal to the measured (cold) dark matter 
density.
The prospects are better in the HE range with GLAST (100~MeV--10~GeV):
for a maximal (cuspy) halo, 1~yr of GLAST observation should be able to yield a detection if 
$m_\chi$\,$\lesssim$\,500~GeV and 
$<$$\sigma$\,$v$$>$\ $\sim$\ 3$\times$10$^{-25}$~cm$^3$s$^{-1}$~\cite{bergrstrom-hooper2006}.

No evidence of dark matter annihilation $\gamma$-rays has been unambiguously claimed so 
far~\cite{dracomagic}.
An apparently extended signal from the direction of NGC\,253 has been definitely interpreted 
as due to hardware malfunction~\cite{itoh2007}.

\subsubsection{Star-forming galaxies}

Diffuse $\gamma$-ray emission from {\sl pp} interactions of cosmic-ray nuclei with target 
interstellar medium photons makes up about 90\% of the $>$\,100~MeV luminosity of the Milky 
Way~\cite{strong2000}.
However, the VHE flux from a galaxy like the Milky Way located 1~Mpc away would be well below 
current IACT sensitivities.
Indeed, only loose upper limits on the VHE flux from normal galaxies have been obtained, even 
for local galaxies and for the VHE-bright starburst galaxies (e.g., Ref.~\cite{torres2004}).
Detailed models of VHE emission from 
NGC\,253~\cite{volk1996,paglione1996,domingo-sanamaia-torres2005} and for 
Arp\,220~\cite{torres2004} are only slackly constrained by current upper 
limits~\cite{aharonian2005c, arbert2007e}.
The increasingly deep upper limit, even for the more prominent star-forming galaxies, suggest 
that non-beamed emission from extragalactic distances may be beyond detection for current 
IACTs and possibly within reach for upcoming and future instruments~\cite{perear2008}.

\subsubsection{Active Galactic Nuclei}

Synchrotron-Compton models have long provided the mainstream theoretical framework to 
model AGN emission~\cite{maraschi1992}.
VHE data have been crucial to close the model, initially introduced to explain the 
radio-to-X-ray data that were interpreted as synchrotron radiation.
For example, let us consider the simplest scheme of blazar emission, where one blob of 
magnetized nonthermal plasma, moving relativistically towards the observer, emits via synchrotron 
and Compton mechanisms (the latter, scattering synchrotron off their parent electrons).
The observed synchrotron emission 
shows a degeneracy between the value of the magnetic field and the electron number density;
the Compton emission
provides the second condition needed to break the degeneracy.
In general, knowledge of the whole spectral energy distribution is required for a complete 
description of the emitting electrons' distribution and environment (e.g., 
Ref.~\cite{tavecchio1998}).
As is clear from basic knowledge of the synchrotron and Compton radiation mechanism, the 
parameters that specify the properties of the emitting plasma in the SSC model are 
the electron spectrum, 
the magnetic field, and the size and Lorentz factor of the plasma itself.
Truly, a simultaneous spectral energy distribution provides us with a snapshot of the emitting population of particles at a given time.

The VHE emission of blazars has also been interpreted as due to accelerated hadrons.
This possibility is particularly exciting in view of the recently claimed association of the arrival directions ultra-high-energy cosmic rays with the sky position of nearby AGNs~\cite{auger07}.

At high energy, the 3$^{\rm rd}$ EGRET catalogue~\cite{hartman1999} includes more than 130 sources known to be extragalactic (e.g., Ref.~\cite{padovani2007}),
all of which are AGN (except the Large Magellanic Cloud), and about 97$\%$ of which are blazars.
Of these, about 93$\%$ are low-frequency-peaked blazars (LBLs) and only about 3$\%$ are high-frequency-peaked blazars (HBLs).
Therefore, blazars -- mostly of the LBL type -- dominate EGRET's extragalactic HE sky.
Blazars dominate the extragalactic VHE sky, too.
However, these are mostly HBLs -- at present, only 1/16 is of the LBL type.
The HBL dominance descends from a selection bias: for a given HE flux, HBLs have a higher VHE 
flux than LBLs, because both spectral humps are shifted to higher frequencies.

Blazar observations have been a top priority for VHE astrophysics ever since the discovery 
of TeV emission from Mrk\,421~\cite{punch1992}.
To date, firm VHE detections of AGN include about 20 blazars 
(see Ref.~\cite{persic-deangelis2007}), the 
quasar 
3C\,279, and the radio galaxy M\,87 
(Ref.~\cite{m87_ahar03}).

After concluding that the extragalactic HE $\gamma$-ray sky is dominated by blazars, 
one comment is in order.
Given their peculiar orientation, blazars are rare.
Assuming that the maximum angle with respect to the line of sight an AGN jet can have for a 
source to be called a blazar is about 15$^{\circ}$, only about 3$\%$ of all radio-loud AGN, 
and therefore about 0.3$\%$ of all AGN, are blazars.
Assuming that about 1$\%$ of the galaxies host an AGN, this implies that only 1 out of about 
30~000 galaxies is a blazar.
Hence, the fact that the GeV and TeV skies are dominated by blazars is surprising.
The explanation stems of course from the blazars' defining characteristics:
\begin{itemize}
\item
{\it high-energy electrons}: in some blazars the synchrotron emission peaks in the X-ray 
range: this suggests the presence of high-energy electrons that can produce HE/VHE radiation 
via Compton scattering;
\item
{\it strong non-thermal (jet) component}: HE/VHE emission is clearly non-thermal and 
related to the jet: the stronger the latter, the stronger the former; and
\item
{\it relativistic beaming}: in sources as compact as blazars (as suggested by their 
short variability timescales) all GeV photons, would be absorbed through pair-producing 
$\gamma\gamma$ collisions with target X-ray photons.
\end{itemize}
Beaming ensures the intrinsic radiation density to be much smaller than the observed one, 
so that VHE photons encounter a much lower $\gamma\gamma$ opacity and hence manage to leave 
the source.
Furthermore, relativistic beaming causes a strong amplification of the jet's 
observed flux, so providing a powerful bias toward 
blazar detection.

The known TeV blazars are variable in flux in all wavebands.
Even simple one-zone homogeneous SSC modeling predicts the X-ray and TeV flux variability to 
be closely correlated, both emissions being linked to the same electron population.
Observational evidence, although still statistically limited, supports this prediction (e.g., 
Ref.~\cite{pian1998}).
Blazar variability, both in flux and spectrum, has been observed at VHE frequencies down to minute 
timescales.
For Mkn\,501, observed with the MAGIC telescope at $>$\,100~GeV during 24 nights between May 
and July 2005, the integrated flux and the differential photon spectra could be measured on a 
night-by-night basis~\cite{albert2007f}.
During the observational campaign, the flux variations (from about 0.4 to about 4 Crab units) 
were correlated with the spectral changes (i.e., harder spectra for higher fluxes), and a 
spectral peak showed up during the most active phases.
A rapid flare occurred on the night of July 10$^{\rm th}$, 2005, showing a doubling time as 
short as about 2 minutes and a delay of about 3 minutes as a function of energy of the emitted 
photons.

One further aspect of TeV spectra of blazars is that they can be used as probes of the EBL.
The TeV photons emitted by a blazar interact with the EBL photons and are likely absorbed 
via pair production.
Whatever its intrinsic shape at emission, after traveling through the EBL-filled space, a 
blazar spectrum will reach the observer distorted by absorption.
The strength of the absorption is measured by the optical depth $\tau(E,z)$ for the
attenuation between the blazar, located at a distance redshift $z$, 
and the Earth~\cite{fazio-stecker1970, stecker1992}.
Usually, either
({\it i})
the shape and intensity of EBL($z$) is assumed, and the TeV spectrum is corrected before the 
SSC modeling is performed (e.g., Ref.~\cite{dejager-stecker,kneiske2004}); or
({\it ii})
based on assumptions on the intrinsic VHE spectrum, EBL($z$) is solved for: e.g., based on 
analysis of the observed hard VHE spectra of the distant blazars 1ES\,1101-232 and 
H\,2359-309, a low EBL energy density at $z \lesssim$\,0.2 has been 
derived~\cite{aharonian2006i}.
The two approaches can be used in combination to estimate the distance to the VHE 
source~\cite{mazin-goebel2007}.
Based on the inferred attenuation of blazar VHE emission by the EBL, and in particular on the 
detection of the distant ($z$=0.536) quasar 3C\,279 (see below), the transparency of the 
Universe at VHE $\gamma$-rays is deduced to be maximal, at the level implied by the known 
cosmic evolution of the stellar populations of galaxies.

\begin{figure}
\centering
\includegraphics[width=.7\textwidth]{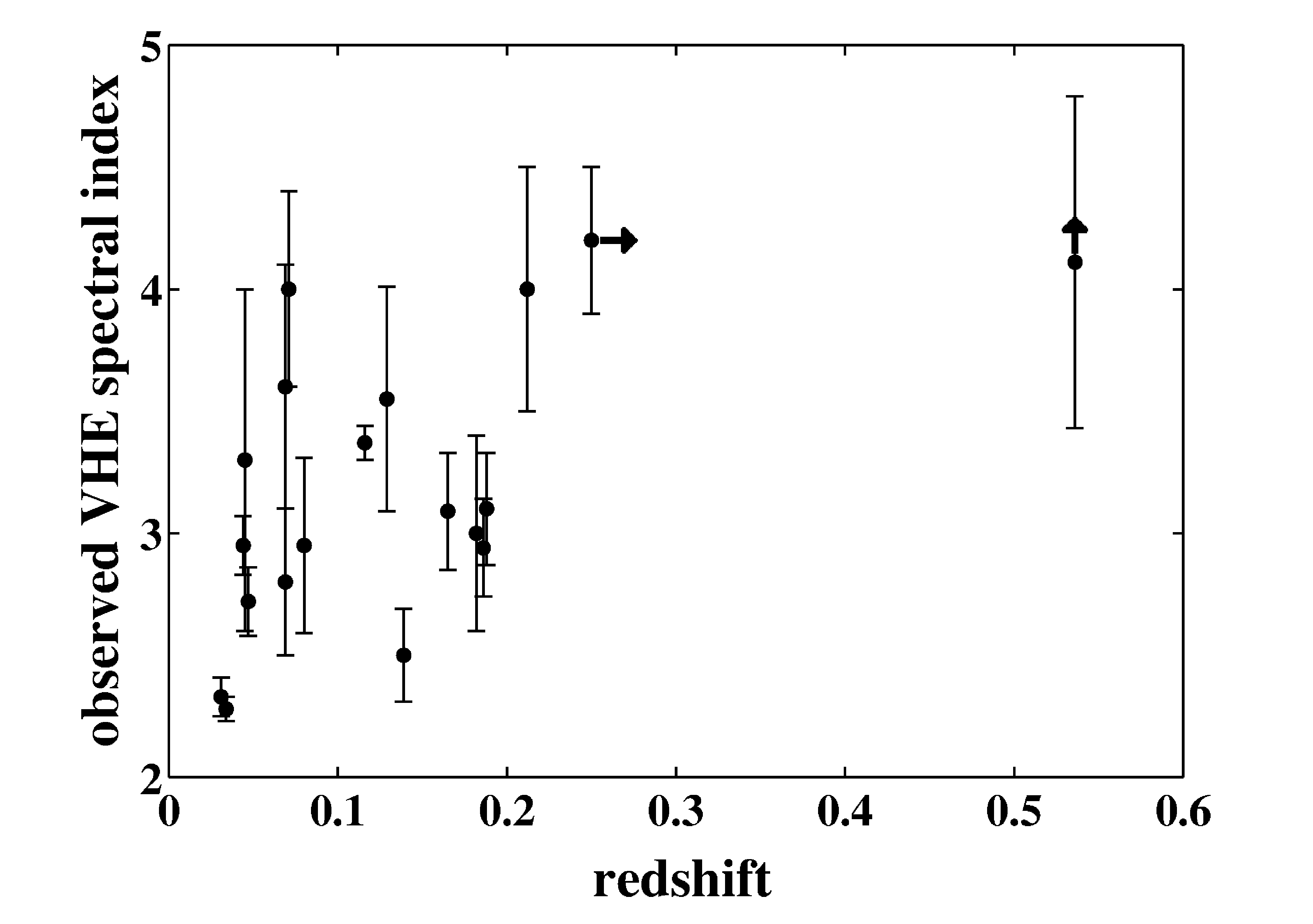}
\caption{\label{fig:spectralindex}
Observed spectral indexes for AGN in the VHE region.}
\end{figure}

More than 20 AGN have been detected as VHE sources (see Table~\ref{fig:AGNtable}) by the time this 
review is written (May 2008).
The AGN observed at VHE are uniformly distributed in the high galactic latitude sky.
Measured spectral indices are plotted versus redshift in Fig.~\ref{fig:spectralindex}.

\begin{table}
\caption{\label{fig:AGNtable}
Detected AGN at TeV energies~\cite{persic-deangelis2007}.}
\begin{tabular}{ l l l r l l l l }
\hline
\noalign{\smallskip}
Source    & ~~~~~~$z$ & ~~~~~~~~$\alpha_\gamma$ & ~~~~~~~~F$_\gamma$~~~~~~~~ & IACT & References\\
\noalign{\smallskip}
\hline
\noalign{\smallskip}
M\,87          &  ~~~0.0044 &                 &                     & H   & \cite{m87_ahar03} \\
Mrk\,421       &  ~~~0.031 & $2.33 \pm 0.08$ & 1.03($\pm$0.03)$\times10^{-10}$ & M   & \cite{albert2007q} \\
Mrk\,501       &  ~~~0.034 & $2.28 \pm 0.05$ & 1.71($\pm$0.11)$\times10^{-11}$ & M   & \cite{albert2007f} \\
               &           & $2.45 \pm 0.07$ & 3.84($\pm$1.00)$\times10^{-12}$ & M   & \\
1ES\,2344+514  &  ~~~0.044 & $2.95 \pm 0.12$ & 1.21($\pm$0.10)$\times10^{-11}$ & M   & \cite{albert2007e} \\
Mrk\,180       &  ~~~0.045 & $3.30 \pm 0.70$ & 8.46($\pm$3.38)$\times10^{-12}$ & M   & \cite{albert2006e} \\
1ES\,1959+650  &  ~~~0.047 & $2.72 \pm 0.14$ & 3.04($\pm$0.35)$\times10^{-11}$ & M   & \cite{albert2006f} \\
BL\,Lacertae   &  ~~~0.069 & $3.60 \pm 0.50$ & 3.28($\pm$0.26)$\times10^{-12}$ & M   & \cite{albert2007h} \\
 PKS 0548-322  &  ~~~0.069 & $2.80 \pm 0.30$ &   3.3($\pm$0.7)$\times10^{-12}$  & H   & \cite{superina}\\
PKS\,2005-489  &  ~~~0.071 & $4.00 \pm 0.40$ & 3.32($\pm$0.48)$\times10^{-12}$ & H   & \cite{HESS-PKS2155,Perlman99} \\
RGB\,J0152+017 &  ~~~0.080 & $2.95 \pm 0.36$ & 4.43($\pm$1.24)$\times10^{-12}$ & H   & \cite{aharonian2008a,donato01} \\
W\,Comae       &  ~~~0.102 &                 &                 & V   & \cite{atel1422} \\
PKS\,2155-304  &  ~~~0.116 & $3.37 \pm 0.07$ & 2.89($\pm$0.18)$\times10^{-11}$ & H   & \cite{aharonian2005l} \\
1ES\,1426+428  &  ~~~0.129 & $3.55 \pm 0.46$ & 2.53($\pm$0.43)$\times10^{-11}$ & W   & \cite{horan02,sambruna97} \\
1ES 0806+524	 &  ~~~0.138 &                 &                      & V   & \cite{swo2}\\ 
1ES\,0229+200  &  ~~~0.139 & $2.50 \pm 0.19$ & 4.46($\pm$0.71)$\times10^{-12}$ & H   & \cite{aharonian2005m,Perlman96,donato05} \\
H\,2356-309    &  ~~~0.165 & $3.09 \pm 0.24$ & 2.55($\pm$0.68)$\times10^{-12}$ & H   & \cite{aharonian2006m,wood84} \\
1ES\,1218+304  &  ~~~0.182 & $3.00 \pm 0.40$ & 1.01($\pm$0.26)$\times10^{-11}$ & M   & \cite{albert2006i} \\
1ES\,1101-232  &  ~~~0.186 & $2.94 \pm 0.20$ & 4.35($\pm$0.69)$\times10^{-12}$ & H   & \cite{aharonian2007d} \\
1ES\,0347-121  &  ~~~0.188 & $3.10 \pm 0.23$ & 3.86($\pm$0.73)$\times10^{-12}$ & H   & \cite{aharonian2007e} \\
1ES\,1011+496  &  ~~~0.212 & $4.00 \pm 0.50$ & 6.40($\pm$0.32)$\times10^{-12}$ & M   & \cite{albert2007l} \\
PG\,1553+113   &  \,$>$0.25& $4.20 \pm 0.30$ & 5.24($\pm$0.87)$\times10^{-12}$ & M,H & \cite{albert2007m} \\
3C\,279        &  ~~~0.536 &  $4.1 \pm 0.7$  &   2.9($\pm$0.5)$\times10^{-11}$ & M   & \cite{albert_3c279} \\
S5\,0716+714   &   ~~~?    &                 &            $\sim\times10^{-11}$ & M   & \cite{atel1500} \\
\noalign{\smallskip}
\hline
\end{tabular}
\noindent
The observed photon spectral index $\alpha_\gamma$ 
between 0.2 TeV and 2 TeV (0.6 TeV for 3C279: the absolute value of the spectral index between 0.2 TeV and 2 TeV for this source might be larger)
and the flux $F_\gamma$ observed at energy $>$0.2 TeV (in erg cm$^{-2}$ s$^{-1}$), for blazars observed at different redshift $z$.
The errors indicate the statistical uncertainty; the corresponding systematic uncertainties on the spectral index are typically about 0.1 for H.E.S.S.\ and about 0.2 for MAGIC.
In the column IACT, the Cherenkov telescope with which the data have been collected:
symbols stand for H$=$H.E.S.S., M$=$MAGIC, V$=$VERITAS, W$=$Whipple.
In the last column, the references.
\end{table}

\begin{figure}
\centering
\hspace*{-2.5mm}
\includegraphics[height=.49\textwidth]{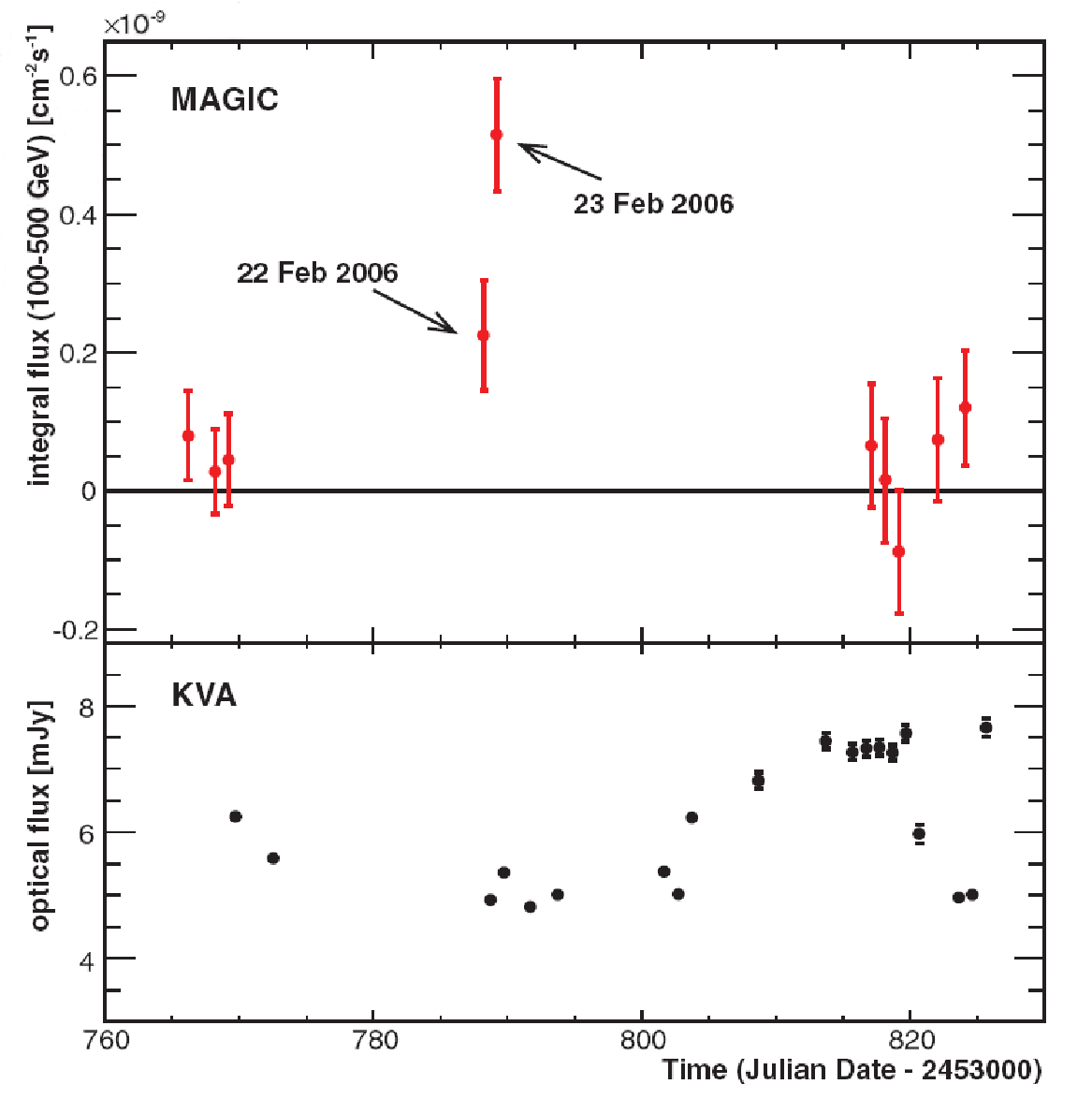}
\hspace*{-3mm}
\includegraphics[height=.48\textwidth]{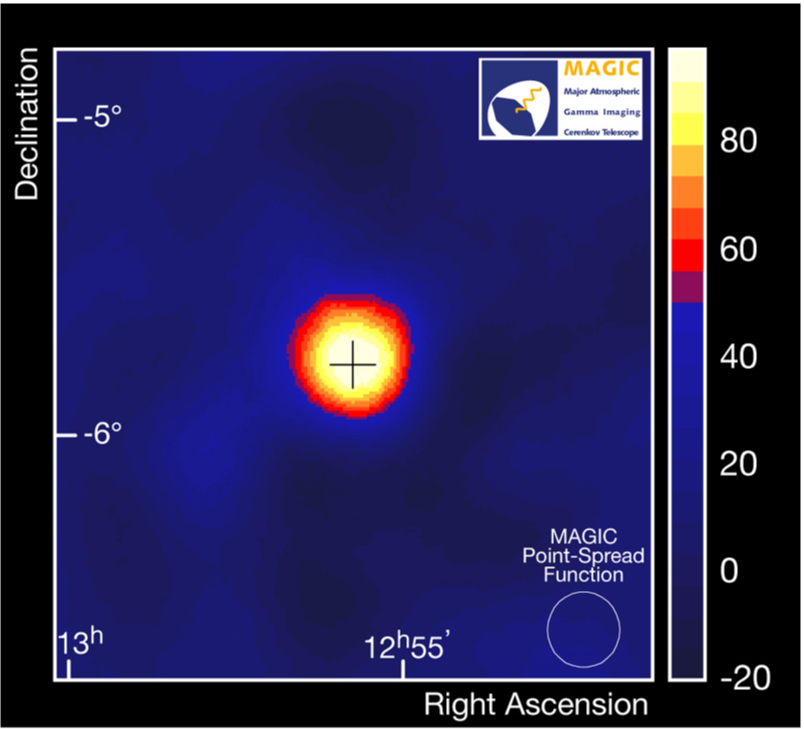}
\caption{\label{fig:3c279lightcurve}
Lighcurve (left) and skymap on February 23$^{\rm rd}$,~2006 (right) of 3C\,279~\cite{albert_3c279}.}
\end{figure}

Currently, the most distant established VHE source is 3C\,279, with $z = 0.536$.
This AGN is a Flat Spectrum Radio Quasar (FSRQ) that was observed by EGRET to be strongly 
flaring (see Fig.~\ref{fig:3c279lightcurve}).
The VHE spectrum of 3C\,279 has been observed by MAGIC at 6.1\,$\sigma$ for energies lower than 220~GeV and at 5.2\,$\sigma$ within 220 and 600~GeV~\cite{teshimaICRC2007}.

\begin{figure}
\centering
\includegraphics[width=.44\textwidth]{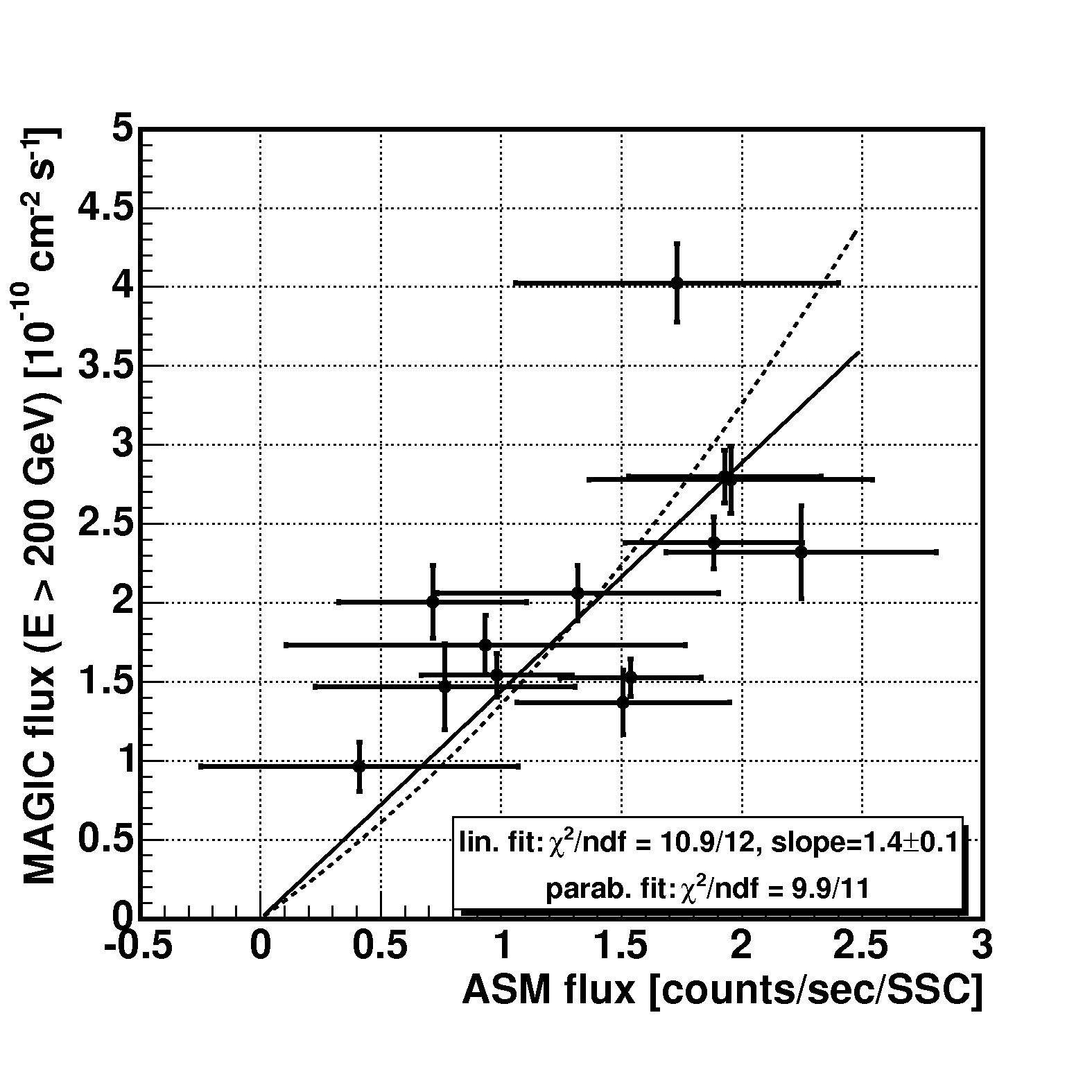}
\includegraphics[width=.55\textwidth]{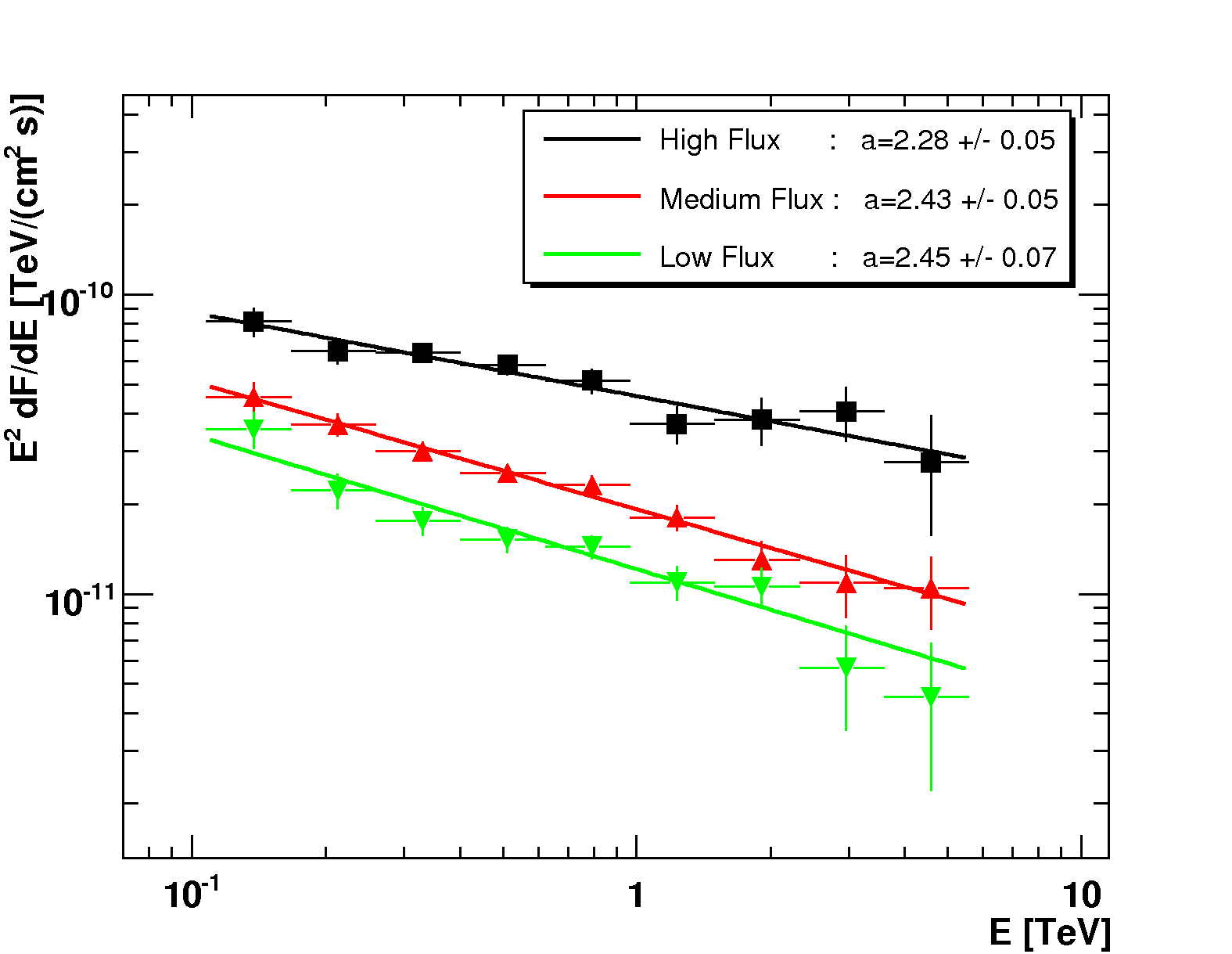}
\caption{\label{fig:Mkn421-501}
Left: TeV-X ray correlation for from Mkn\,421~\cite{albert2007q}.
Right: the spectral energy distribution from Mkn\,501~\cite{albert2007f} during different 
flares.}
\end{figure}

The very well studied sources M\,87, Mkn\,421, and Mkn\,501 (Fig.~\ref{fig:Mkn421-501}) show a 
strong temporal variability.
These sources, which are monitored by all IACTs, show correlations between TeV and X-ray 
emission.
During the flares, the spectral features change, and the IC peak shifts to higher frequencies 
(Fig.~\ref{fig:Mkn421-501}).
The highlight of these observations is the detection of very fast (few minutes timescale) 
variability with some time delay between photons of different energy.
Such lags are a potentially powerful diagnostics of acceleration and energy loss processes, 
whereas the short timescales involved can place tight limits to the size of the emitting 
regions and the Lorentz factor of the jet.
The H.E.S.S.\ observations of PKS\,2155-304 (located at $z$=0.116), showed a very fast flux 
variability: on the night of July 28~${\rm th}$,~2006 it had a peak flux about 50 times its 
average flux (and about 15 times the Crab flux), and rapidly doubled it in four successive 
episodes (in 67$\pm50$~s, 116$\pm50$~s, 173$\pm50$~s, and 178$\pm50$~s respectively).
As for M\,87, this source is not a blazar but a radio galaxy that harbors the most massive 
known black hole in the nearby universe and whose jet is $\sim$30$^o$ from the line of sight 
to the observer: hence the two-day timescale measured by H.E.S.S.~\cite{m87variab} is remarkable.

\begin{figure}
\centering
\includegraphics[width=.95\textwidth]{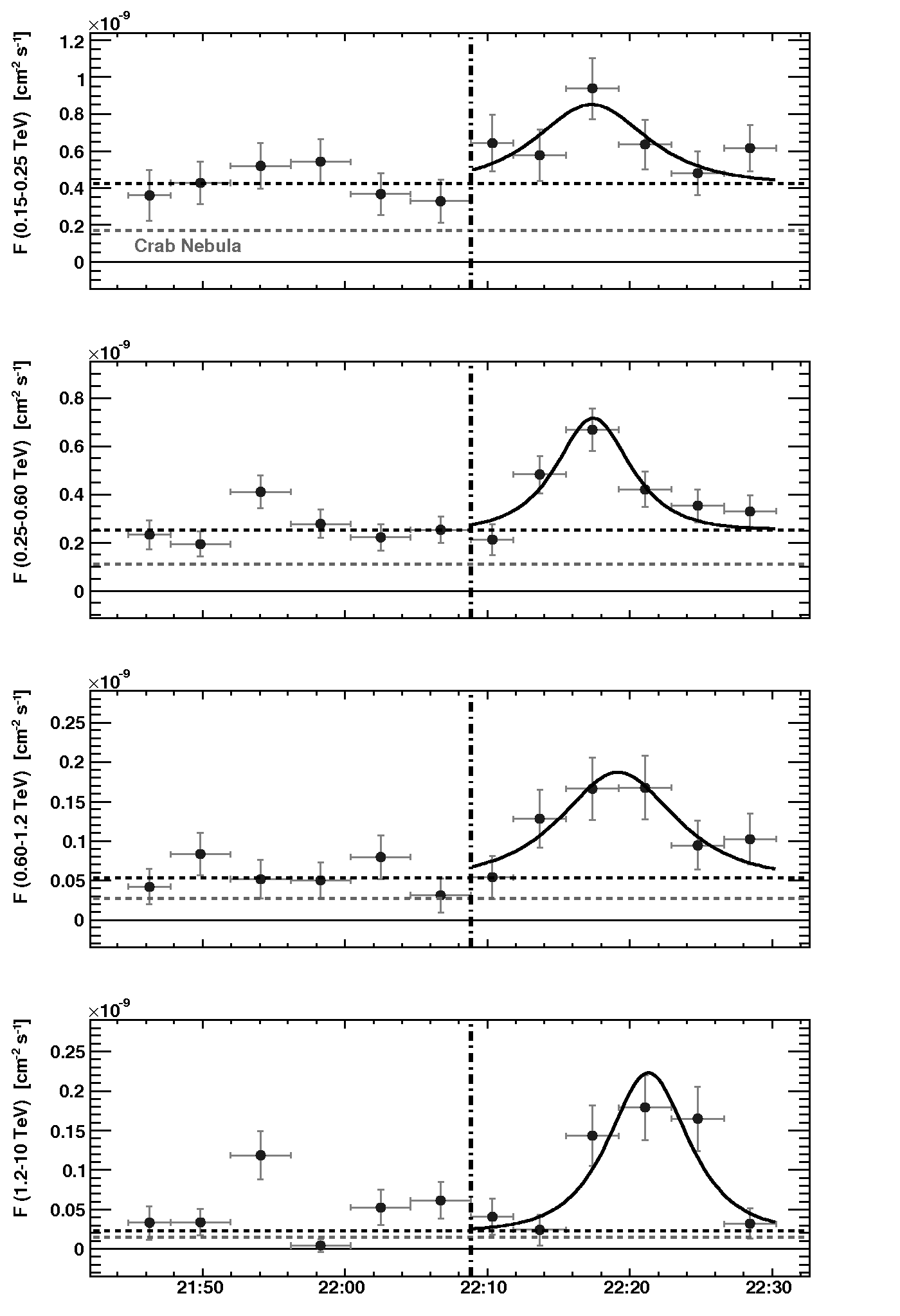}
\caption{\label{fig:Mkn501Magic2}
Integral flux of Mkn\,501 detected by MAGIC in four different energy 
ranges~\cite{albert2007f}.}
\end{figure}

The variability of the AGN in the VHE region can provide information about possible violations 
of the Lorentz invariance by means of the light dispersion expected in some quantum gravity 
models~\cite{amelino}.
The velocity of light in such models is given by the formula:
$$
V = c \left[
1 + \xi \left(\frac{E}{E_{QG}}\right) + \xi_2 \left(\frac{E}{E_{QG}}\right)^2 + \ldots
\right]
$$
where $E_{QG}$ is the energy scale where Lorentz invariance is violated, and the $\xi$s are 
parameters of order unity which can be positive or negative.
At first order photons of different energies emitted at the same time are detected with a time 
delay $\Delta$$t$\,$\simeq$\,$\xi$$\frac{E}{E_{QG}}$\,\!$\frac{z}{H_0}$\,$ = $\,$\xi$\,\!$\frac{E}{E_{QG}}$\,\!$\frac{L}{c}$.
The MAGIC data about Mkn\,501~\cite{MAGIC-Mkn501} showed a negative correlation between the 
arrival time of photons and their energy (Fig.~\ref{fig:Mkn501Magic2}), yielding, if one 
assumes that the delay is due to quantum gravity effects, to an evaluation of $E_{QG}$$\sim$$0.03~M_P$ 
($E_{QG}$$>$$0.02~M_P$)%
\footnote{Quantum Gravity effects are usually parametrized as a function of the reduced 
	Planck mass $M_P$, where violations are expected to become sizeable.
	The reduced Planck mass, $[(\hbar c)(8 \pi G)]^{1/2}$, is, apart from factors of 
	order~1, the mass of a black hole whose Schwarzschild radius equals its 
	Compton wavelength.}.
H.E.S.S.\ observations of PKS\,2155~\cite{HESS-PKS2155} evidenced no effect, allowing to set a 
lower limit $E_{QG} > 0.04~M_P$;
in 1999 Whipple data~\cite{WhippleQG} lead to $E_{QG} > 0.005~M_P$ and the X-ray limits 
from Gamma-Ray Bursts give $E_{QG} > 0.01~M_P$.

In most quantum gravity scenarios violations to the universality of the speed of light happen at order 
larger than 1:
$\Delta t \simeq \left( {E}/{E_{QG}} \right) ^\nu$ with $\nu > 1$.
In this case the VHE detectors are even more sensitive with respect to other instruments;
for $\nu = 2$ the data from PKS\,2155 give $E_{QG} > 3 \cdot 10^{-9}~M_P$.

\subsubsection{Gamma-Ray Bursts}

Gamma-Ray Bursts (GRBs) are widely interpreted as originating in relativistic ``fireballs" 
following the core-collapse of massive stars and/or the coalescence of two compact objects.
GRBs are interesting both for astrophysical reasons and related to the physics of photon 
propagation.
At present, no VHE $\gamma$-ray emissions from GRBs have been positively detected; the current record of a HE photon is set by EGRET at 18~GeV, two hours after the primary burst.

\begin{figure}
\centering
\includegraphics[width=.9\textwidth]{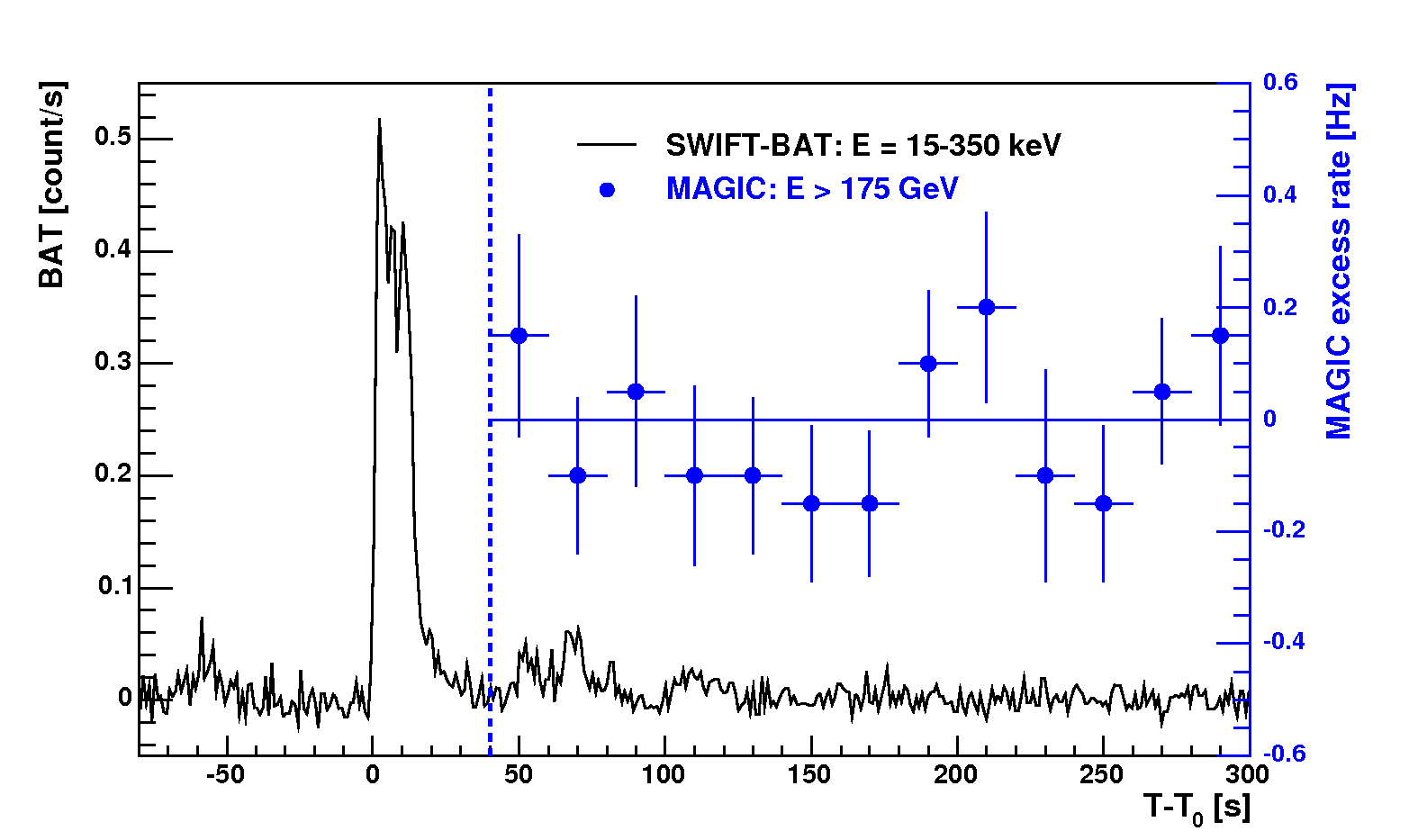}
\\
\includegraphics[width=.9\textwidth]{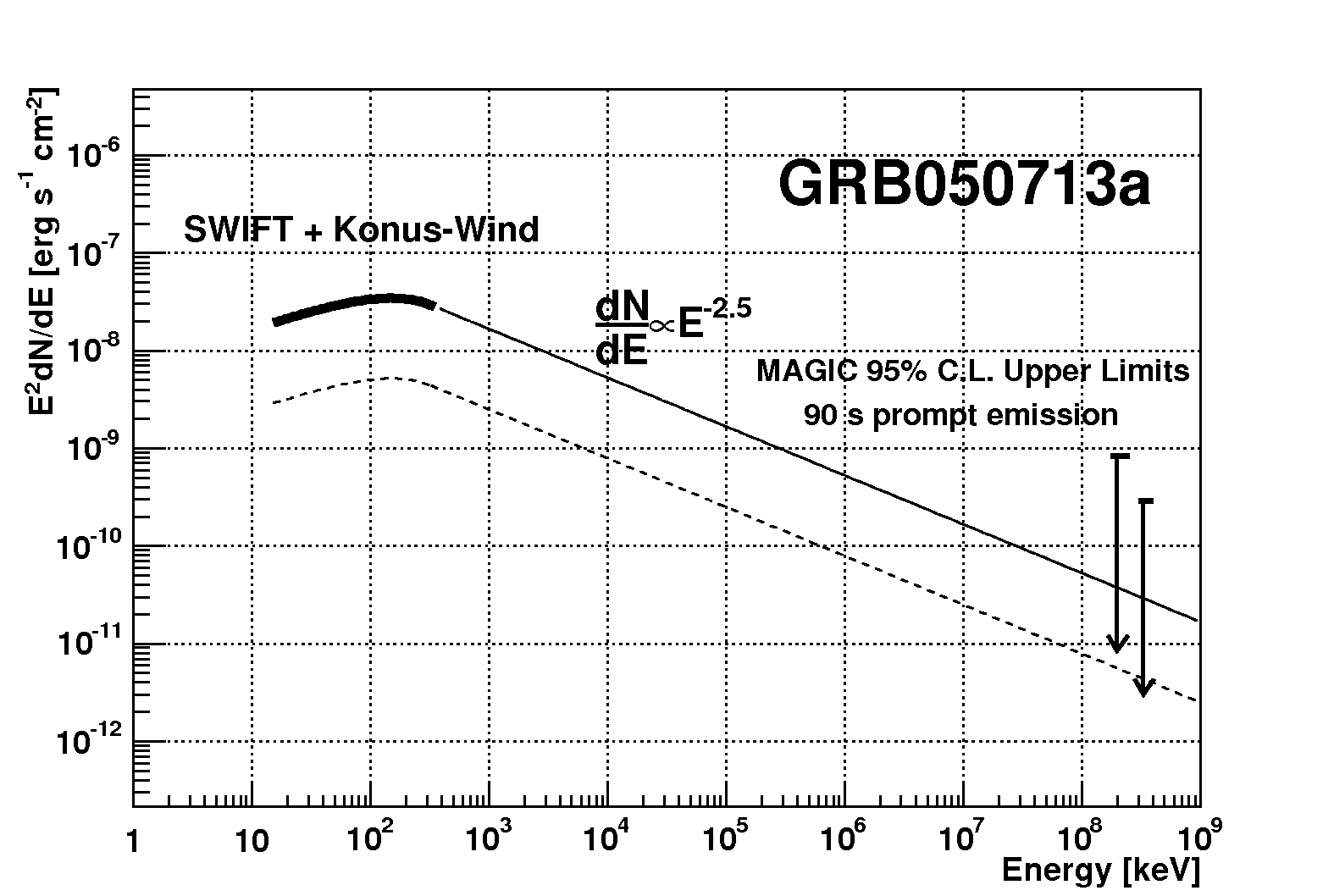}
\caption{\label{fig:MAGICgrb}
Plots of the MAGIC data about the GRB\,050713a~\cite{albert2006d}:
up, excess event rate compared with SWIFT-BAT observations;
down: upper limits on the spectrum.}
\end{figure}

MAGIC observed part of the prompt-emission phase of GRB\,050713a (Fig.~\ref{fig:MAGICgrb}) as 
a response to an alert by the Swift satellite~\cite{albert2006d}.
However, no excess at energies larger than 175~GeV was detected, neither during the prompt emission phase nor 
later -- but the upper limits to the MAGIC flux are compatible with simple extrapolations of 
the power-law spectrum measured by Swift, with spectral index of about 1.6, to hundreds of GeV.
In general, however, the cosmological distances of these sources prevent VHE 
detection~\cite{albert2007g}: the average redshift of the GRBs for which MAGIC was alerted 
(and whose $z$ are known) is $\langle z \rangle =3.22$, whereas at 70~GeV the cosmological 
$\gamma$-ray horizon is $z \sim 1$.
Complementary Whipple data~\cite{horan2007} and MILAGRO data~\cite{abdo2007} provide upper 
limits on, respectively, the late VHE emission (about 4~hours after the burst) from several 
long-duration GRBs, and on the prompt/delayed emission from several, reputedly nearby 
($z\!\lesssim$\,0.5), short-duration GRBs.
The ``naked-eye" event GRB080319B~\cite{bloom2008} was unfortunately not observed with MAGIC, 
in spite of its optimal sky location and relative proximity ($z$=0.937), due to its occurring 
at twilight, nor with AGILE, due to its being occulted by the Earth at the time of its 
occurrence.

\begin{figure}
\centering
\includegraphics[width=.7\textwidth]{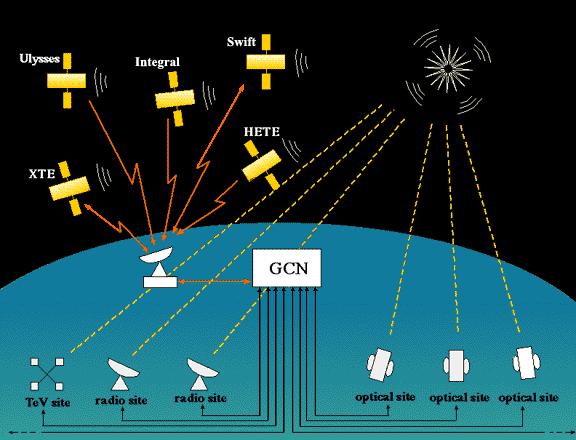}
\caption{\label{fig:GCNnetwork}
The GCN Network for GRB alerts~\cite{schemaGCN}.}
\end{figure}

The MAGIC telescope is at present the best Cherenkov telescope to observe GRBs, due to its fast 
movement and its low energy threshold; furthermore, MAGIC is in the GCN Network for GRB alerts (see 
Fig.~\ref{fig:GCNnetwork}) which is active since April 2005.
However, VHE observations of GRBs are severely limited by EBL absorption: only a small 
fraction will likely occur at $z<$1, and these will have to be observed at energies 
substantially below 100\,GeV.
The MILAGRO and HAWC telescopes has characteristics suitable to the GRB observation, owing to 
their very large field of view and $\sim$100\% duty cycle.

\section{The future}

\subsection{The future of present detectors}

\begin{figure}
\centering
\includegraphics[width=.7\textwidth]{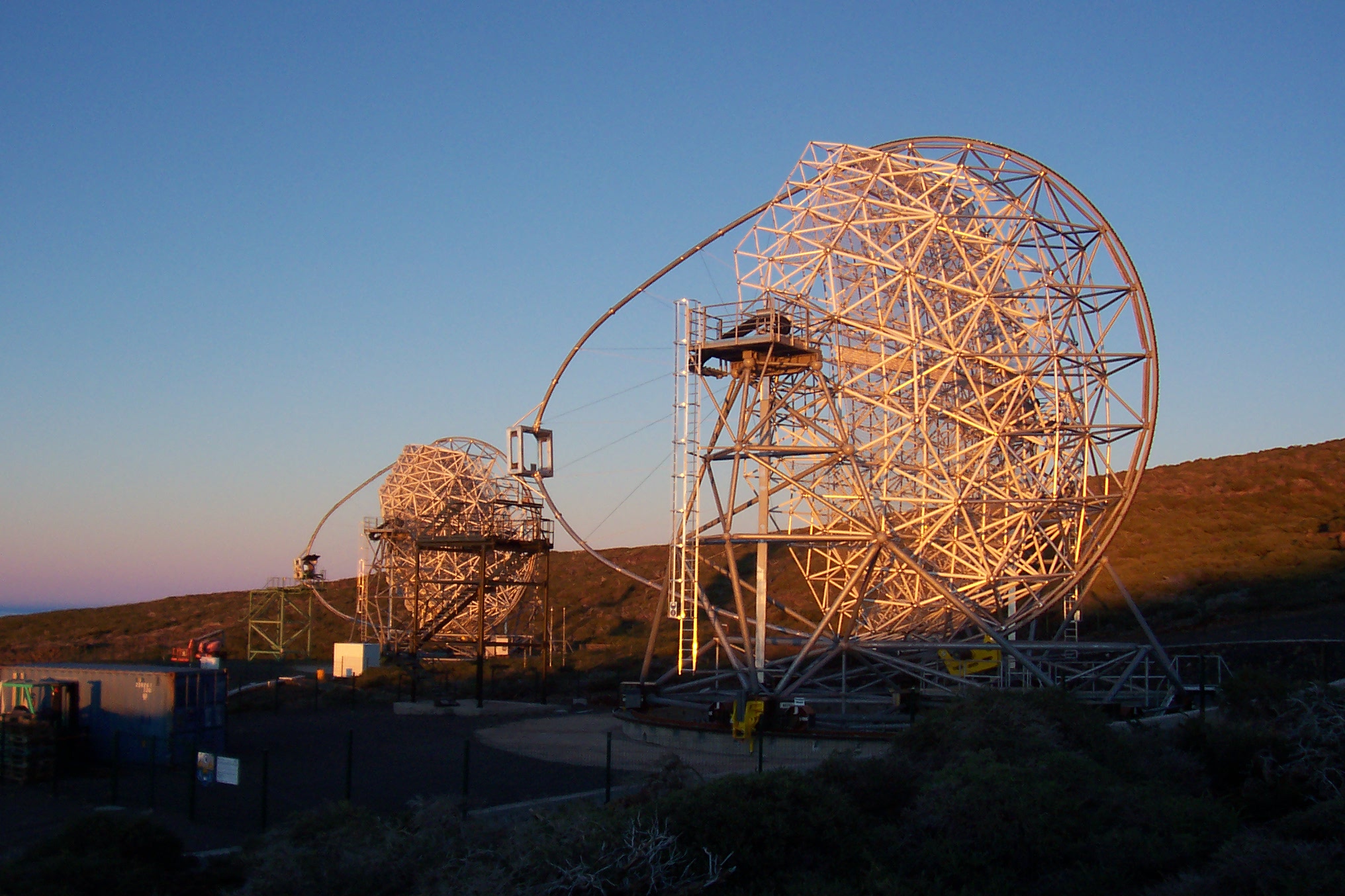}
\\[1mm]
\includegraphics[width=.7\textwidth]{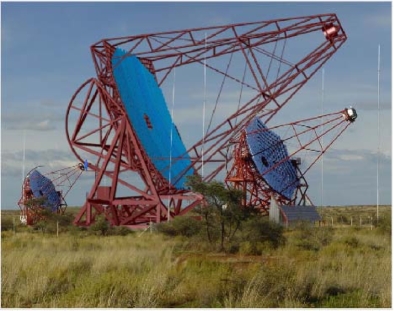}
\caption{\label{fig:nextTelescopes}
The telescopes MAGIC~2 (real view)~\cite{magic-sito} and
H.E.S.S.~2 (artist view)~\cite{hess2figura}.}
\end{figure}

The next years will be very exciting for the observations of the $\gamma$-ray sky.
The GLAST satellite telescope is in orbit since June 2008.
A second MAGIC telescope, at a distance of 85~m from the first one, starts operating in September 2008 (see Fig.~\ref{fig:nextTelescopes}); it will increase substantially the sensitivity of MAGIC and improve the angular resolution to about 0.07 degrees.
With this new telescope, MAGIC enters in phase 2 (MAGIC~2).
The H.E.S.S.\ collaboration has started the construction of a large telescope, which will be inaugurated after 2009 (see Fig.~\ref{fig:nextTelescopes}) and will lead the instrument into its phase 2 (H.E.S.S.~2).
With its diameter of 28~m, the new telescope, located in the middle of the four existing telescopes, will be the largest Cherenkov telescope in the world, and it should decrease the trigger threshold to some 30 GeV.

What next?
Given the present space technology and the progress that one can reasonably foresee, space instruments beyond GLAST are forbidden for the next some 30 years - the cost of space technology forbidding enterprises at a scale one-two orders of magnitude larger than GLAST.
The future will then be for ground-based instruments.

When designing a new instrument to detect VHE sources, we must learn from experience.
H.E.S.S.\ and MAGIC have explored with enormous success most of the sky, including the galactic plane and most EGRET sources, and we learned~that:
\begin{itemize}
\item[-]
Extensive Air Shower detectors need to increase their sensitivity by a factor of about 10 to stay in the market;
they should become larger and collect a flux of about 1\% Crab per year or less;
\item[-]
to further improve the IACTs, it is needed to build instruments with a larger field of view by using new geometries or by replicating the present Cherenkov instruments, maybe with some improvements such as adopting silicon photomultipliers (cheaper, lighter and smaller, with advantages on the weight and the engineering of the camera) or adopting new technologies for mirrors.
The cost of telescopes is presently dominated by mechanics, mirrors, and camera.
\end{itemize}

Two longer-term projects for ground telescopes are under discussion:
the Cherenkov Telescope Array (CTA)~\cite{CTA-figura}, and an EAS detector called High Altitude Water Cherenkov (HAWC)~\cite{HAWC}.

\subsection{The Cherenkov Telescope Array}

The CTA facility is a European project meant to explore the sky in the energy range
from 10~GeV to 100~TeV  and it is designed to combine guaranteed science with significant discovery potential.
It is designed to be built using demonstrated technologies, i.e., being a replica of present Cherenkov telescopes.
The CTA is a cornerstone towards a multi-messenger exploration of the universe.

\begin{figure}
\centering
\includegraphics[width=.7\textwidth]{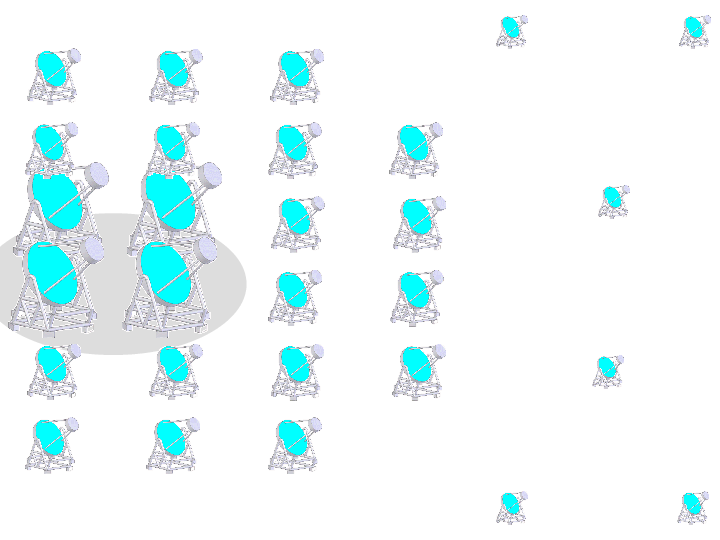}
\caption{\label{fig:CTAzones}
A possible layout for the CTA ~\cite{CTA-figura}.}
\end{figure}

In the most ambitious and expensive scheme, for which the cost foreseen is of the order of 100-150 million euros, the array layout will be composed by 3 zones (see Fig.~\ref{fig:CTAzones}), with different types of telescopes.
\begin{itemize}
\item[-]
The low-energy section, which is the innermost part, is covered with a sampling of 10\% by large telescopes with medium-field cameras; its radius will be about 70~m and its energy threshold will be of about 10 GeV.
\item[-]
The medium-energy section will be covered with a sampling of 1\% by mid-size telescopes; its radius will be 250~m in addition to the low-energy section and its energy threshold will be of about 100~GeV.
\item[-]
The high-energy section will be covered at 0.05\% by small telescopes with wide-field cameras; its radius will be few kilometers in addition to the medium-energy section and its energy threshold will be of about 1~TeV.
\end{itemize}
The field of view increases from 4-5 degrees in the inner region to 8-10 degrees in the 
outer sections.

In such a scenario, the CTA will operate in four different modes:
\begin{itemize}
\item[-]
deep wide-band mode, when all the telescopes track the same source;
\item[-]
survey mode, when the telescopes survey the sky using a staggered pattern of fields of view;
\item[-]
search and monitoring mode, when subclusters of telescopes track different sources;
\item[-]
narrow-band mode, when halo telescopes accumulate high-energy data, while the core telescopes hunt for pulsars.
\end{itemize}

Two sites are foreseen for the CTA: one in the northern and one in the southern hemisphere.

At the 30$^{\rm th}$ International Cosmic Ray Conference (2007), a board of scientists 
affiliated to institutions from U.S.A.\ presented a white paper~\cite{WhitePaperTeam} about 
the status and the future of ground-based $\gamma$-ray astronomy, in which they propose some 
baselines for the development of the future instruments.
In particular, they identify some requirements that the future instruments should meet, among 
which are the following:
\begin{itemize}
\item[-]
one order of magnitude better sensitivity by means of a footprint area on the order of 
1~km$^2$;
\item[-]
enlargement of the explored energy range;
\item[-]
improvement of the angular resolution;
\item[-]
a factor 2 of increase of the field of view.
\end{itemize}
The technology they propose is an array of mid-size IACTs (5-15~m diameter) coupled to an 
array of water Cherenkov detectors, the latter to provide a large field of view and a high 
duty-cycle.
For what concerns the IACTs themselves, they suggest to:
\begin{itemize}
\item[-]
reduce the construction and operational costs;
\item[-]
move from the present Davies-Cotton or parabolic optics to Cassegrain optics, in order to 
increase the field of view;
\item[-]
design the readout and triggering electronics, choose the photodetector and the mirror 
fabrication technique in such a way to improve the sensitivity and cut down the expenses.
\end{itemize}

The future Cherenkov arrays will observe many new sources at low energies, but maybe very 
exciting new discoveries will come from the few expected new sources in the high energy 
region, which could generate important new physics.

Finally, research has started for large field-of-view imaging atmospheric Cherenkov telescopes 
with Fresnel optics~\cite{GAW}, which could combine the precision and the sensitivity of IACTs 
with a field of view as large as 40 degrees.

An intermediate model for the CTA, for which the cost could be dropped to some 20 million 
euros or less, would be to stage the final detector by building a third MAGIC telescope and 
several H.E.S.S.\,1 detectors, all with advanced high-efficiency cameras.
In this case the overall sensitivity would improve by a factor of 2 to 3 instead than the 
final design factor of 10, but this would give the time to explore the potential of the 
observations by the joint collaboration with a subsstancial increase in sensitivity and with 
limited funding.

\subsection{Large EAS}

The High Altitude Water Cherenkov (HAWC) observatory, the successor of MILAGRO, will be 
located at the extreme altitude of 4~100~m in the Sierra Negra (Mexico).
Such an altitude will allow lowering the threshold, by intercepting a larger amount of charged 
secondaries.
It will incorporate new design solutions like placing the photomultipliers in isolated tanks, 
and adopting a larger spacing between them.
It will have a 22~500~m$^2$ sensitive area (to be compared with the 4~000~m$^2$ of MILAGRO) 
and since it will reuse in the beginning the MILAGRO photomultipliers and electronics it will 
cost only about 10 to 20~million dollars to complete the detector
(to be compared to the about 150 million euros foreseen for the ``full version'' of the CTA).
HAWC will improve the sensitivity of MILAGRO by at least one order of magnitude: it will 
detect the Crab Nebula in one day at 5\,$\sigma$ and collect four times the Crab flux in 15 
minutes, being suitable for $\gamma$-ray bursts observations.
HAWC will probably have a lower energy threshold at 200~GeV.

The capability of the HAWC observatory to perform a highly sensitive all-sky survey will 
enable the monitoring of the known sources and yield to the discovery of new sources of known 
types and new classes of TeV sources.
Furthermore, HAWC can investigate the origin and propagation of cosmic rays above the 
TeV - tens~of~TeV region.

\section{Conclusions}

High-energy photons are a powerful probe of fundamental physics under extreme conditions, 
since they are produced in the highest energy phenomena, they often travel through large 
distances, and their interactions display large boosts towards the center of mass.

Observation of X- and $\gamma$-rays gives an exciting view of the HE universe
thanks to satellite-based telescopes (AGILE, GLAST) and to ground-based detectors
like the IACTs, which discovered more than 60 new VHE sources in the 
last 3 years and are going on this way.
This large population of VHE-$\gamma$-ray sources, which are often unknown sources, poses 
questions on the transparency of the Universe at these energy ranges; this might indicate 
the existence of new physics.

\begin{figure}
\centering
\includegraphics[width=.9\textwidth]{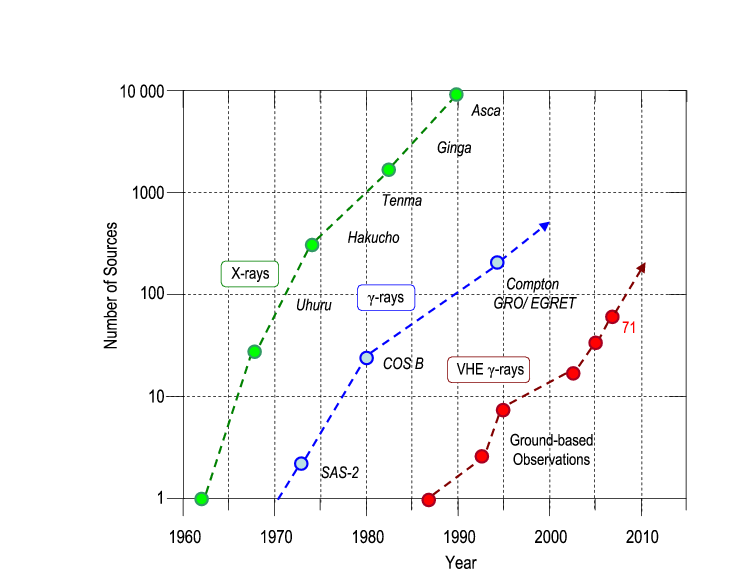} \hspace*{5mm}
\caption{\label{fig:KifunePlot}
The ``Kifune plot'' which shows the number of detected sources as a function of time from year 
1960, with an extrapolation to the next future.
The plot has been updated in~\cite{hinton}.}
\end{figure}
The plot in Fig.~\ref{fig:KifunePlot}, usually referred to as ``Kifune plot'', shows how the 
amount of detected X-ray, $\gamma$-ray and VHE-$\gamma$-ray sources varied and an 
extrapolation to the future; the contribution from the different observatories is highlighted.
In this plot we can see that the advent of the IACTs like MAGIC and H.E.S.S.\ increased by a 
factor of 10 the known VHE-$\gamma$-ray source population.
The progress achieved with the latest generation of IACTs is comparable with the one drawn by 
EGRET with respect to the previous $\gamma$-ray satellite detectors.

This exciting scenario gives handles for the study of new mechanisms about the 
VHE-$\gamma$-ray origin and propagation, and many astrophysical constraints are feeding the 
theories.
So far no clear sources above about 50 TeV have been detected, but it is still uncertain 
whether they do not exist or there is just a technological limit on detecting them with the 
current facilities; the future instruments will tell.
The exploration of the VHE sources has just started and in the next three years (2008/2010) 
a factor of 10 improvement in the GeV range is expected by the AGILE and GLAST observations, 
while a factor 2-3 improvement in the TeV range will be reached by H.E.S.S.~2, MAGIC~2 and 
VERITAS.

To go beyond the TeV energies the best solution comes from the combination provided by the CTA 
and HAWC observatories, and possibly from new concepts of large-field of view facilities.

\section*{Acknowledgements} The Authors thank 
Abelardo Moralejo,
Barbara De Lotto,
Carlotta Pittori,
Fabrizio Tavecchio,
Florian G\"obel,
Marco Roncadelli, 
Robert Wagner
for comments and suggestions.
A special acknowledgement goes to the anonymous referee who helped in improving the article.


\begin{thebibliography}{111}

\bibitem{vhess}
\BY{V.~Hess} \IN{Phys.\ Z.}{13}{1913}{1084}.

\bibitem{pacini}
\BY{D.~Pacini} \IN{Nuovo Cim.}{3}{1912}{93}.

\bibitem{rossi}
\BY{B.~Rossi} \TITLE{High-energy Particles} (Prentice-Hall, New York)~1952;\\
\BY{B.~Rossi} \TITLE{Cosmic Rays} (McGraw-Hill, New York)~1964.

\bibitem{gzk}
\BY{K.~Greisen} \IN{Phys.\ Rev.\ Lett.}{16}{1966}{748};\\
\BY{G.T.~Zatsepin \atque V.A.~Kuzmin} \IN{J.\ Exp.\ Theor.\ Phys.\ Lett.}{4}{1966}{78}.

\bibitem{auger0}
\BY{T.~Yamamoto, for the Pierre Auger Collaboration}
Proc.\ 30$^{\rm th}$ Intl.\ Cosmic Ray Conf.\ (ICRC 2007), in press (arXiv:0707.2638).

\bibitem{fermiacc}
\BY{E.~Fermi} \IN{Phys.\ Rev.}{75}{1949}{1169}.

\bibitem{magfield}
See, for example,
\BY{E.~Zwiebel \atque C.~Heiles} \IN{Nature}{385}{1997}{131}.

\bibitem{auger07}
\BY{The Pierre Auger Collaboration} \IN{Science}{318}{2007}{938}.

\bibitem{emagfield}
\BY{A.~De Angelis, M.~Roncadelli \atque M.~Persic} 
\IN{Mod.\ Phys.\ Lett.\ A}{23}{2008}{315}.


\bibitem{ressel-turner1989}
\BY{M.T.~Ressell \atque M.S.~Turner} \IN{Comm.\ on Astrophys.}{14}{1990}{323}.

\bibitem{aharobook}
\BY{F.A.~Aharonian} \TITLE{Very High Energy Cosmic Gamma Radiation} (World Scientific, Singapore)~2004.

\bibitem{fleury}
\BY{C.M.~Hoffman, C.~Sinnis, P.~Fleury, \atque M.~Punch} \IN{Rev.~Mod.~Phys.}{71}{1999}{897}.

\bibitem{hinton}
\BY{J.~Hinton} 
Proc.\ 30$^{\rm th}$ Intl.\ Cosmic Ray Conf.\ (ICRC 2007), in press (arXiv:0712.3352).

\bibitem{hinton2}
\BY{J.~Hinton} 
New J.\ Phys., in press (arXiv:0803.1609).

\bibitem{weekes1989}
\BY{T.C.~Weekes et al.} \IN{Astrophys.~J.}{342}{1989}{379}.

\bibitem{mannheim1993}
\BY{K.~Mannheim} \IN{Astron.\ Astrophys.}{269}{1993}{76}.

\bibitem{aharonian2000}
\BY{F.A.~Aharonian} \IN{New Astron.}{5}{2000}{377}.

\bibitem{urry-padovani1995}
\BY{C.M.~Urry \atque P.~Padovani} \IN{Publ.~Astron.~Soc.~Pacific}{107}{1995}{803}.

\bibitem{padovani2007}
\BY{P.~Padovani} \IN{Am.~Inst.~Phys.~Conf.~Proc.}{921}{2007}{19}.

\bibitem{ulrich1997}
\BY{M.H.~Ulrich et al.} \IN{Ann.~Rev.~Astron.~Astrophys.}{35}{1997}{445}.

\bibitem{fossati1998}
\BY{G.~Fossati et al.} \IN{Mon.~Not.~R.~Astron.~Soc.}{299}{1998}{433}.

\bibitem{ghisellini1998}
\BY{G.~Ghisellini et al.} \IN{Mon.~Not.~R.~Astron.~Soc.}{301}{1998}{451}.

\bibitem{dermer-schlickeiser1993}
\BY{C.~Dermer \atque R.~Schlickeiser} \IN{Astrophys.~J.}{416}{1993}{458}.

\bibitem{rees1967}
\BY{M.J.~Rees} \IN{Mon.~Not.~R.~Astron.~Soc.}{137}{1967}{429}.

\bibitem{maraschi1992}
\BY{L.~Maraschi, G.~Ghisellini, \atque A.~Celotti} \IN{Astrophys.~J.}{397}{1992}{L5}.

\bibitem{rees-meszaros1992}
\BY{M.J.~Rees \atque P.~Meszaros} \IN{Mon.~Not.~R.~Astron.~Soc.}{258}{1992}{41}.

\bibitem{meszaros-rees1993}
\BY{P.~Meszaros \atque M.J.~Rees} \IN{Astrophys.~J.}{405}{1993}{278}.

\bibitem{sari1998}
\BY{R.~Sari et al.} \IN{Astrophys.~J.}{497}{1998}{L17}.

\bibitem{meszaros2006}
\BY{P.~Meszaros} \IN{Rep.~Prog.~Phys.}{69}{2006}{2259}.

\bibitem{vanAlbada1985}
\BY{T.S.~van Albada et al.} \IN{Astrophys.~J.}{295}{1985}{305}.

\bibitem{sarazin1986}
\BY{C.~Sarazin} \IN{Rev.~Mod.~Phys.}{58}{1986}{1}.

\bibitem{spergel2003}
\BY{D.N.~Spergel et al.} \IN{Astrophys.~J.~Suppl.}{148}{2003}{175}.

\bibitem{pdg}
\BY{W.-M.~Yao et al.} \IN{J.\ Phys.}{G33}{2006}{1} and 2007 partial update for~2008.

\bibitem{GLASTscienceBrochure}
\TITLE{GLAST Science Brochure (March 2001)}, \texttt{http://glast.gsfc.nasa.gov/science}.

\bibitem{recentberg}
\BY{T.~Bringmann, L.~Bergstr\"om, \atque J.~Edsj\"o} 
\IN{J.~High Energy Phys.}{01}{2007}{049}.

\bibitem{aharonian2006g}
\BY{F.A.~Aharonian et al.} \IN{Nature}{439}{2006}{695}.

\bibitem{CoppiAharonian}
\BY{P.~Coppi \atque F.A.~Aharonian} \IN{Astrophys.~J.}{487}{1997}{L9}.

\bibitem{heitler}
\BY{W.~Heitler} \TITLE{The quantum theory of radiation} (Oxford University Press, Oxford)~1960.

\bibitem{lee98}
\BY{S.~Lee} \IN{Phys.\ Rev.\ D}{58}{1998}{043004}.

\bibitem{gould:1967a}
\BY{R.J.~Gould \atque G.P.~Schr\'eder} \IN{Phys.\ Rev.}{155}{1967}{1408}.

\bibitem{stecker2001}
\BY{F.W.~Stecker} \IN{Int.\ Astron.\ Union Symp.}{204}{2001}{135}.

\bibitem{stecker1992}
\BY{F.W.~Stecker, O.C.~De Jager, \atque M.H.~Salomon} \IN{Astrophys.~J.}{390}{1992}{L49}.

\bibitem{aharonian2006i}
\BY{F.A.~Aharonian et al.} \IN{Nature}{440}{2006}{1018}.

\bibitem{mazin}
\BY{D.~Mazin \atque M.~Raue} \IN{Astron.\ Astrophys.}{471}{2007}{439}.

\bibitem{peebles}
\BY{P.J.E.~Peebles} \TITLE{Principles of Physical Cosmology} (Princeton University Press, Princeton)~1993.

\bibitem{padmanabhan}
\BY{T.~Padmanabhan} \TITLE{Theoretical Astrophysics, Volume III: Galaxies and Cosmology} (Cambridge University Press, Cambridge)~2002.

\bibitem{blanch-vari}
\BY{O.~Blanch, J.~Lopez, \atque M.~Martinez} \IN{Astropart.\ Phys.}{19}{2003}{245};\\
\BY{O.~Blanch \atque M.~Martinez} \IN{Astropart.\ Phys.}{23}{2005}{588};\\
\BY{O.~Blanch \atque M.~Martinez} \IN{Astropart.\ Phys.}{23}{2005}{598};\\
\BY{O.~Blanch \atque M.~Martinez} \IN{Astropart.\ Phys.}{23}{2005}{608}.

\bibitem{mannheim1999}
\BY{K.~Mannheim} \IN{Rev.\ Mod.\ Astron.}{12}{1999}{167}.

\bibitem{primack2001}
\BY{J.R.~Primack et al.} \IN{Astropart.\ Phys.}{11}{1999}{93};\\
\BY{J.R.~Primack et al.} \IN{AIP Conference Proceedings}{558}{2001}{463}.

\bibitem{noi1}
\BY{A.~De Angelis, O.~Mansutti, \atque M.~Roncadelli} \IN{Phys.\ Lett.~B}{659}{2008}{847}.

\bibitem{noi2}
\BY{A.~De Angelis, M.~Roncadelli, \atque O.~Mansutti} \IN{Phys.\ Rev.~D}{76}{2007}{121301}.

\bibitem{hooperserpico}
\BY{M.~Simet, D.~Hooper, P.~Serpico} \IN{Phys.\ Rev.~D}{77}{2008}{063001}.

\bibitem{kneiske2004}
\BY{T.M.~Kneiske et al.} \IN{Astron.\ Astrophys.}{413}{2004}{807}.

\bibitem{kifune}
\BY{T.~Kifune} \IN{Astrophys.~J.}{518}{1999}{L21}.

\bibitem{amelino}
\BY{G.~Amelino-Camelia et al.} \IN{Nature}{393}{1998}{763}.

\bibitem{SuW}
\BY{A.~De Angelis \atque L.~Peruzzo} \IN{Sterne Weltraum}{8}{2007}{26}.

\bibitem{rossi1941}
\BY{B.~Rossi, K.~Greisen} \IN{Rev.\ Mod.\ Phys.}{13}{1941}{240}.

\bibitem{rossigreisen}
\BY{J.~Nishimura, K.~Kamata} \IN{Prog.\ Theor.\ Phys.}{7}{1952}{185};\\
\BY{K.~Greisen} \IN{Rev.\ Mod.\ Phys.}{13}{1960}{240}.

\bibitem{wagnertesi}
\BY{R.M.~Wagner} Ph.D.\ thesis, 2006,
Technische Universit\"at M\"unchen, MPP-2006-245.

\bibitem{GLASTimage}
\TITLE{GLAST Gallery},
\texttt{http://www-glast.sonoma.edu/resources/multimedia/gallery/}.

\bibitem{satelliteDetector}
\TITLE{CGRO Science Support Center Image Gallery},\\
\texttt{http://cossc.gsfc.nasa.gov/docs/cgro/images/epo/gallery/glast/}.

\bibitem{agile-sito}
\TITLE{AGILE, Astro-rivelatore Gamma a Immagini Leggero},
\texttt{http://agile.asdc.asi.it/}.

\bibitem{lat-gmb}
\BY{Goddard Space Flight Center, NASA} \TITLE{The GLAST Instruments}, \\
\texttt{http://glast.gsfc.nasa.gov/public/instruments.html}.

\bibitem{GLASTintero}
\BY{Goddard Space Flight Center news, NASA}\\
\texttt{http://www.nasa.gov/centers/goddard/news/topstory/2007/GlastLaunch\_prt.htm}.

\bibitem{GlastPerformance}
\BY{Stanford Linear Accelerator Center} \TITLE{GLAST LAT Performance}, \\
\texttt{http://www-glast.slac.stanford.edu/software/IS/}.

\bibitem{sciamiVeritas}
\BY{The MILAGRO Gamma-Ray Observatory, LANL} \TITLE{The Detection of Cosmic Rays}, \\
\texttt{http://www.lanl.gov/milagro/detecting.shtml}

\bibitem{sitoArgo}
\BY{INFN Roma~3} \TITLE{The Argo-YBJ Experiment}, \\
\texttt{http://193.204.162.110/$\sim$nucleare/argo/argo.html}.

\bibitem{sitoMilagro}
\BY{The MILAGRO Gamma-Ray Observatory, LANL}
\texttt{http://www.lanl.gov/milagro/}.

\bibitem{hillnew}
\BY{A.M.~Hillas} 
Proc. 19$^{\rm th}$ Intl.\ Cosmic Ray Conf.\ (ICRC 1985);\\
\BY{A.M.~Hillas} \IN{Space Sci.\ Rev.}{75}{1996}{17}.

\bibitem{improvement}
\BY{F.~Piron et al.} \IN{Astron.\ Astrophys.}{374}{2001}{895};\\
\BY{M.~de Naurois} arXiv:astro-ph/0607247v1;\\
\BY{M.~Lemoine-Goumard, B.~Degrange, \atque M.~Tluczykont} \IN{Astropart.\ Phys.}{25}{2006}{195}.

\bibitem{funnew}
\BY{D.~Berge, S.~Funk, \atque J.~Hinton} \IN{Astron.\ Astrophys.}{466}{2007}{1219}.

\bibitem{tescaro}
\BY{V.R.~Chitnis \atque P.N.~Bhat} \IN{Astropart.\ Phys.}{15}{2001}{29};\\
\BY{S.~Raducci} Thesis at the Universit\`a di Udine,~2004;\\
\BY{D.~Tescaro} Thesis at the Universit\`a di Padova,~2005;\\
\BY{R.~Mirzoyan et al.} \IN{Astropart.\ Phys.}{25}{2006}{342};\\
\BY{D.~Tescaro et al.} arXiv:0709.1410v1~[astro-ph];\\
\BY{G.~Cabras et al.} arXiv:0804.3896v1~[astro-ph].

\bibitem{sitoHess}
\BY{Max Plank Institute} \TITLE{The H.E.S.S.\ Telescopes}, \\
\texttt{http://www.mpi-hd.mpg.de/hfm/HESS/public/telescope/hn\_telescopes.htm}.

\bibitem{sitoHess-desy}
\BY{University of Hamburg} \TITLE{H.E.S.S.\ overview}, \\
\texttt{http://www-hess.desy.de/pages/hes\_overview.html}.

\bibitem{figuraMagic}
\BY{Max-Plank-Institut f\"ur Physik, Munich} \TITLE{MAGIC Picture Gallery}, \\
\texttt{http://wwwmagic.mppmu.mpg.de/gallery/}.

\bibitem{sitoVeritas}
\TITLE{VERITAS Homepage},
\texttt{http://veritas.sao.arizona.edu}.

\bibitem{abe} 
\BY{A.~Moralejo} private communication.

\bibitem{noimagglast} 
\BY{D. Bastieri et al.} \IN{Astropart. Phys.}{23}{2005}{572}  

\bibitem{plotSensitiv}
\BY{Institut f\"ur Theoretische Physik und Astrophysik, Fakult\"at f\"ur Physik und Astronomie, Universit\"at W\"urzburg}
\TITLE{Observations with the MAGIC Telescope}, \\
\texttt{http://www.astro.uni-wuerzburg.de/mphysics/}.

\bibitem{3egretCat}
\texttt{http://cossc.gsfc.nasa.gov/docs/cgro/cossc/egret/3rd\_EGRET\_Cat.html}\\
or Ref.~\cite{hartman1999}.

\bibitem{sourcesWagner}
\texttt{http://www.mppmu.mpg.de/$\sim$rwagner/sources/}.

\bibitem{aharonian2006a}
\BY{F.A.~Aharonian et al.} \IN{Astrophys.~J.}{636}{2006}{777}.

\bibitem{HESS:scanicrc}
\BY{S.~Hoppe et al.}
Proc.\ 30$^{\rm th}$ Intl.\ Cosmic Ray Conf.\ (ICRC 2007), in press (arXiv:0710.3528).

\bibitem{albert2007a}
\BY{J.~Albert et al.} \IN{Astrophys.~J.}{664}{2006}{L87}.

\bibitem{torres2003}
\BY{D.F.~Torres et al.} \IN{Phys.~Rep.}{382}{2003}{303}.

\bibitem{porter2006}
\BY{T.A.~Porter et al.} \IN{Astrophys.~J.}{648}{2006}{L29}.

\bibitem{aharonian2006b}
\BY{F.A.~Aharonian et al.} \IN{Astron.\ Astrophys.}{449}{2006}{223}.

\bibitem{berezhko-volk2006}
\BY{E.G.~Berezhko \atque H.J.~V\"olk} \IN{Astron.\ Astrophys.}{451}{2006}{98}.

\bibitem{berezhko2003}
\BY{E.G.~Berezhko et al.} \IN{Astron.\ Astrophys.}{400}{2003}{971}.

\bibitem{ona2007}
\BY{J.~Albert et al.} \IN{Astron.\ Astrophys.}{474}{2007}{937}.

\bibitem{tavani}
\BY{M.~Persic, A.~De Angelis, F.~Longo, \atque M.~Tavani}
Proc.\ 30$^{\rm th}$ Intl.\ Cosmic Ray Conf.\ (ICRC 2007), in press (arXiv:0709.1881).

\bibitem{aharonian2007}
\BY{F.A.~Aharonian et al.} \IN{Astron.\ Astrophys.}{464}{2007}{235}.

\bibitem{blasi2005}
\BY{P.~Blasi} \IN{Mod.~Phys.~Lett.}{A20}{2005}{3055}.

\bibitem{albert2006a}
\BY{J.~Albert et al.} \IN{Astrophys.~J.}{643}{2006}{L53}.

\bibitem{ic443}
\BY{J.~Albert et al.} \IN{Astrophys.~J.}{664}{2007}{L87}.

\bibitem{w28}
\BY{F.~Aharonian et al.} \IN{Astron.\ Astrophys.}{481}{2008}{401}.

\bibitem{pacini1967}
\BY{F.~Pacini} \IN{Nature}{216}{1967}{567}.

\bibitem{cheng1986}
\BY{K.S.~Cheng et al.} \IN{Astrophys.~J.}{300}{1986}{500}.

\bibitem{romani1996}
\BY{R.W.~Romani} \IN{Astrophys.~J.}{470}{1996}{469}.

\bibitem{hirotani2005}
\BY{K.~Hirotani} \IN{Adv.\ Space Res.}{35}{2005}{1085}.

\bibitem{daug-hard1982-96}
\BY{J.K.~Daugherty \atque A.~Harding} \IN{Astrophys.~J.}{252}{1982}{337}; \\
\BY{J.K.~Daugherty \atque A.~Harding} \IN{Astrophys.~J.}{458}{1996}{278}.

\bibitem{baring2004}
\BY{M.G.~Baring} \IN{Adv.\ Space Res.}{33}{2004}{552}.

\bibitem{aharonian2006c}
\BY{F.A.~Aharonian et al.} \IN{Astron.\ Astrophys}{448}{2006}{L43}.

\bibitem{albert2007b}
\BY{J.~Albert et al.} \IN{Astrophys.~J.}{669}{2007}{1143}.

\bibitem{harding2005}
\BY{A.K.~Harding, V.V.~Usov, \atque A.~G.~Muslimov} \IN{Astrophys.~J.}{622}{2005}{531}.

\bibitem{albert2007c}
\BY{J.~Albert et al.} \IN{Astrophys.~J.}{674}{2008}{1046}.

\bibitem{atel_crab}
\BY{M.~Teshima for the MAGIC Collaboration} \IN{ATel}{1491}{2008}\!\!.

\bibitem{gallant2006}
\BY{Y.~Gallant} \IN{Astrophys.\ Space Sci.}{309}{2007}{197}.

\bibitem{HESSJ1640}
\BY{S.~Funk et al.} \IN{Astrophys.~J.}{662}{2007}{517}.

\bibitem{HESSJ1813}
\BY{S.~Funk et al.} \IN{Astron.\ Astrophys.}{470}{2007}{249}.

\bibitem{HESSJ1813radio}
\BY{D.J.~Helfand, R.H.~ Becker, \atque R.L.~White} arXiv:astro-ph/0505392; \\
\BY{C.L.~Brogan et al.} \IN{Astrophys.~J.}{629}{2005}{L105}; \\
\BY{D.J.~Helfand et al.} \IN{Astrophys.~J.}{665}{2007}{1297}.

\bibitem{aharonian2006d}
\BY{F.A.~Aharonian et al.} \IN{Astron.\ Astrophys.}{460}{2006}{365}.

\bibitem{aharonian2005a}
\BY{F.A.~Aharonian et al.} \IN{Astron.\ Astrophys.}{442}{2005}{1}.

\bibitem{aharonian2005b}
\BY{F.A.~Aharonian et al.} \IN{Science}{309}{2005}{746}.

\bibitem{aharonian2006e}
\BY{F.A.~Aharonian et al.} \IN{Astron.\ Astrophys.}{460}{2006}{743}.

\bibitem{albert2006b}
\BY{J.~Albert et al.} \IN{Science}{312}{2006}{1771}.

\bibitem{albert2007d}
\BY{J.~Albert et al.} \IN{Astrophys.~J.}{665}{2007}{L51}.

\bibitem{aharonian2004}
\BY{F.A.~Aharonian et al.} \IN{Astron.\ Astrophys.}{425}{2004}{L13}.

\bibitem{albert2006c}
\BY{J.~Albert et al.} \IN{Astrophys.~J.}{638}{2006}{L101}.

\bibitem{aharonian2006f}
\BY{F.A.~Aharonian et al.} \IN{Phys.~Rev.~Lett.}{97}{2006}{221102}.

\bibitem{noXhess}
\BY{F.A.~Aharonian et al.} \IN{Science}{307}{2005}{1938}.

\bibitem{WhippleJ2023}
\BY{A.~Konopelko et al.} \IN{Astrophys.~J.}{658}{2007}{1062}.

\bibitem{MAGICJ2023}
\BY{J.~Albert et al.} \IN{Astrophys.~J.}{675}{2008}{L25}.

\bibitem{HegraJ2023}
\BY{F.A.~Aharonian et al.} \IN{Astron.\ Astrophys.}{431}{2005}{197}.

\bibitem{MilagroVHEs}
\BY{A.~Abdo et al.} \IN{Astrophys.~J.}{658}{2007}{L33}; \\
\BY{A.~Abdo et al.} \IN{Astrophys.~J.}{664}{2007}{L91}.

\bibitem{TibetICRC}
\BY{M.~Amenomori et al.}
Proc.\ 30$^{\rm th}$ Intl.\ Cosmic Ray Conf.\ (ICRC 2007), in press (arXiv:0710.2757).

\bibitem{j1908-hess}
\TITLE{H.E.S.S.\ Source of the Month, August 2007: Confirming MGRO\,1908+06},\\ \texttt{http://www.mpi-hd.mpg.de/hfm/HESS/public/som/Som\_8\_07.htm}.

\bibitem{salsa2008}
\BY{M.~Salvati, B.~Sacco} 
\IN{Astron.\ Astrophys.}{485}{2006}{2008}.

\bibitem{druah2008}
\BY{L.~Drury, F.A.~Aharonian} \TITLE{Astropart.\ Phys.} (2008), in press (arXiv:0802.4403).

\bibitem{bergrstrom-hooper2006}
\BY{L.~Bergstr\"om \atque D.~Hooper} \IN{Phys.~Rev.~D}{73}{2006}{063510}.

\bibitem{dracomagic}
\BY{J.~Albert et al.} 
\IN{Astrophys.\ J.}{679}{2008}{428}.

\bibitem{itoh2007}
\BY{C.~Itoh et al.} \IN{Astron.\ Astrophys.}{462}{2007}{67}.

\bibitem{strong2000}
\BY{A.W.~Strong et al.} \IN{Astrophys.~J.}{537}{2000}{763}.

\bibitem{torres2004}
\BY{D.F.~Torres} \IN{Astrophys.~J.}{617}{2004}{966}.

\bibitem{volk1996}
\BY{H.~V\"olk et al.} \IN{Space Sci.~Rev.}{75}{1996}{279}.

\bibitem{paglione1996}
\BY{T.A.D.~Paglione et al.} \IN{Astrophys.~J.}{460}{1996}{295}.

\bibitem{domingo-sanamaia-torres2005}
\BY{E.~Domingo-Santamar\'{\i}a \atque D.~Torres} \IN{Astron.\ Astrophys.}{444}{2005}{403}.

\bibitem{aharonian2005c}
\BY{F.A.~Aharonian et al.} \IN{Astron.\ Astrophys.}{442}{2005}{177}.

\bibitem{arbert2007e}
\BY{J.~Albert et al.} \IN{Astrophys.~J.}{658}{2007}{245}.

\bibitem{tavecchio1998}
\BY{F.~Tavecchio et al.} \IN{Astrophys.~J.}{509}{1998}{608}.

\bibitem{perear2008}
\BY{M.~Persic, Y.~Rephaeli, \atque Y.~Arieli} \TITLE{Astron.~Astrophys.} (2008), in press (arXiv:0802.0818).

\bibitem{hartman1999}
\BY{R.C.~Hartman} \IN{Astrophys.~J.~Suppl.}{123}{1999}{79}.

\bibitem{punch1992}
\BY{M.~Punch et al.} \IN{Nature}{358}{1992}{477}.

\bibitem{persic-deangelis2007}
\BY{M.~Persic \atque A.~De Angelis} 
\IN{Astron.~Astrophys.}{483}{2008}{1}.


\bibitem{pian1998}
\BY{E.~Pian et al.} \IN{Astrophys.~J.}{492}{1998}{L17}.

\bibitem{albert2007f}
\BY{J.~Albert et al.} \IN{Astrophys.~J.}{669}{2007}{862}.

\bibitem{fazio-stecker1970}
\BY{G.G.~Fazio \atque F.~W.~Stecker} \IN{Nature}{226}{1970}{135}.

\bibitem{dejager-stecker}
\BY{O.C.~de~Jager \atque F.~W.~Stecker} \IN{Astrophys.~J.}{566}{2002}{738}.

\bibitem{mazin-goebel2007}
\BY{D.~Mazin \atque F.~Goebel} \IN{Astrophys.~J.}{655}{2007}{L13}.

\bibitem{m87_ahar03} 
\BY{F.~Aharonian et al.} \IN{Astron.\ Astrophys.}{403}{2003}{L1}.

\bibitem{albert2007q}
\BY{J.~Albert et al.} \IN{Astrophys.~J.}{663}{2007}{125}.

\bibitem{albert2007e}
\BY{J.~Albert et al.} \IN{Astrophys.~J.}{662}{2007}{892}.

\bibitem{albert2006e}
\BY{J.~Albert et al.} \IN{Astrophys.~J.}{648}{2006}{L105}.

\bibitem{albert2006f}
\BY{J.~Albert et al.} \IN{Astrophys.~J.}{639}{2006}{761}.

\bibitem{albert2007h}
\BY{J.~Albert et al.} \IN{Astrophys.~J.}{666}{2007}{L17}.

\bibitem{superina}
\BY{G. Superina et al.}
Proc.\ 30$^{\rm th}$ Intl.\ Cosmic Ray Conf.\ (ICRC 2007), in press (arXiv:0710.4057, 138).

\bibitem{HESS-PKS2155}
\BY{F.A.~Aharonian et al.} \IN{Astron.\ Astrophys.}{442}{2005}{895}.

\bibitem{Perlman99}
\BY{E.S.~Perlman et al.} \IN{Astrophys.~J.}{523}{1999}{L11}.

\bibitem{aharonian2008a}
\BY{F.A.~Aharonian et al.} 
\TITLE{Astron.~Astrophys.} (2008), in press (arXiv:0802.4021v2).

\bibitem{donato01}
\BY{D.~Donato et al.} \IN{Astron.\ Astrophys.}{375}{2001}{739}.

\bibitem{atel1422} 
\BY{S.~Swordy for the VERITAS Collaboration} \IN{ATel}{1422}{2008}\!\!.

\bibitem{aharonian2005l}
\BY{F.A.~Aharonian et al.} \IN{Astron.\ Astrophys.}{436}{2005}{L17}.

\bibitem{horan02}
\BY{D.~Horan et al.} \IN{Astrophys.~J.}{571}{2002}{753}.

\bibitem{sambruna97}
\BY{R.M.~Sambruna et al.} \IN{Astrophys.~J.}{483}{1997}{774}.

\bibitem{swo2}
\BY{S.~Swordy for the VERITAS Collaboration} \IN{ATel}{1415}{2008}\!\!.

\bibitem{aharonian2005m}
\BY{F.A.~Aharonian et al.} \IN{Astron.\ Astrophys.}{475}{2007}{L9}.

\bibitem{Perlman96}
\BY{E.S.~Perlman et al.} \IN{strophys.~J.~S.}{104}{1996}{251}.

\bibitem{donato05}
\BY{D.~Donato, R.M.~Sambruna, \atque M.~Gliozzi} \IN{Astron.\ Astrophys.}{433}{2005}{1163}.

\bibitem{aharonian2006m}
\BY{F.A.~Aharonian et al.} \IN{Astron.\ Astrophys.}{455}{2006}{461}.

\bibitem{wood84}
\BY{K.S.~Wood et al.} \IN{Astrophys.~J.~S.}{56}{1984}{507}.

\bibitem{albert2006i}
\BY{J.~Albert et al.} \IN{Astrophys.~J.}{642}{2006}{L119}.

\bibitem{aharonian2007d}
\BY{F.A.~Aharonian et al.} \IN{Astron.\ Astrophys.}{470}{2007}{475}.

\bibitem{aharonian2007e}
\BY{F.A.~Aharonian et al.} \IN{Astron.\ Astrophys.}{473}{2007}{L25}.

\bibitem{albert2007l}
\BY{J.~Albert et al.} \IN{Astrophys.~J.}{667}{2007}{L21}.

\bibitem{albert2007m}
\BY{J.~Albert et al.} \IN{Astrophys.~J.}{654}{2007}{L119}.

\bibitem{albert_3c279} 
\BY{J.~Albert et al.} 
\IN{Science}{320}{2008}{1752}.

\bibitem{atel1500} 
\BY{M.~Teshima for the MAGIC Collaboration} \IN{ATel}{1500}{2008}\!\!.

\bibitem{m87variab}
\BY{J.-Ph.~Lenain}
\TITLE{Rev.\ Mex.\ A.\ A.}, in press (arXiv:0709.1366).

\bibitem{teshimaICRC2007}
\BY{M.~Teshima et al.} Proc.\ of the 30$^{\rm th}$ International Cosmic Ray Conference (ICRC 2007), arXiv:0709.1475v1~[astro-ph].

\bibitem{MAGIC-Mkn501}
\BY{J.~Albert et al.} 
\TITLE{Phys.\ Lett.\ B}, submitted (arXiv:0708.2889).

\bibitem{WhippleQG}
\BY{S.D.~Biller et al.} \IN{Phys.\ Rev.\ Lett.}{83}{1999}{2108}.

\bibitem{albert2006d}
\BY{J.~Albert et al.} \IN{Astrophys.~J.}{641}{2006}{L9}.

\bibitem{albert2007g}
\BY{J.~Albert et al.} \IN{Astrophys.~J.}{667}{2007}{358}.

\bibitem{horan2007}
\BY{D.~Horan et al.} \IN{Astrophys.~J.}{655}{2007}{396}.

\bibitem{abdo2007}
\BY{A.A.~Abdo et al.} \IN{Astrophys.~J.}{666}{2007}{361}.

\bibitem{bloom2008}
\BY{J.S.~Bloom et al.} 
\TITLE{Astrophys.\ J.} (2008), in press (arXiv:0803.3215).

\bibitem{schemaGCN}
\BY{Goddard Space Flight Center, NASA} \TITLE{GCN: The Gamma-ray bursts Coordinates Network},
\texttt{http://gcn.gsfc.nasa.gov/}.

\bibitem{magic-sito}
\TITLE{MAGIC picture gallery}, \texttt{http://wwwmagic.mppmu.mpg.de/gallery/pictures/}.

\bibitem{hess2figura}
\BY{D.~Horns} \IN{J.~Phys.: Conf.~Ser.}{60}{2007}{119}.

\bibitem{CTA-figura}
\texttt{http://www.mpi-hd.mpg.de/hfm/CTA/}.

\bibitem{HAWC}
See for example:
\BY{A.J.~Smith} \IN{J.\ Phys.}{60}{2007}{131}.

\bibitem{WhitePaperTeam}
\BY{H.~Krawczynski et al.\ for the White Paper Team}
Proc.\ of the 30$^{\rm th}$ Intl.\ Cosmic Ray Conf.\ (ICRC 2007), in press (arXiv:0709.0704).

\bibitem{GAW}
\BY{G.~Cusumano et al.} 
Proc.\ of the 30$^{\rm th}$ Intl.\ Cosmic Ray Conf.\ (ICRC 2007), in press (arXiv:0707.4541).

%
%
%
%
%
%

\end{thebibliography}
\end{document}